\documentclass[preprint,prd,tightenlines,nofootinbib,eqsecnum,superscriptaddress]{revtex4-2}

\pdfoutput=1

\usepackage{amsmath}    
\usepackage{amsfonts}   
\usepackage{amssymb}    
\usepackage{mathtools}  

\usepackage[T1]{fontenc}   
\usepackage{mathrsfs}      
\DeclareSymbolFontAlphabet{\mathrsfs}{rsfs}

\usepackage{graphicx}   
\usepackage{overpic}    

\usepackage{array}      
\usepackage{booktabs}   
\usepackage[table]{xcolor}  

\usepackage{orcidlink}  
\usepackage{hyperref}
\hypersetup{
    colorlinks = true,
    linktocpage,
    unicode,
    linkcolor  = blue,
    filecolor  = magenta,
    urlcolor   = cyan,
}

\definecolor{CiteColor}{rgb}{0, 0.5, 0}       
\definecolor{RefColor} {rgb}{0.55, 0, 0}      
\definecolor{URLColor} {rgb}{0, 0, 0.35}      

\hypersetup{citecolor=CiteColor, linkcolor=RefColor, urlcolor=URLColor}

\newcolumntype{C}{>{$}c<{$}}

\AtBeginDocument{%
  \heavyrulewidth  = .08em
  \lightrulewidth  = .05em
  \cmidrulewidth   = .03em
  \belowrulesep    = .65ex
  \belowbottomsep  = 0pt
  \aboverulesep    = .4ex
  \abovetopsep     = 0pt
  \cmidrulesep     = \doublerulesep
  \cmidrulekern    = .5em
  \defaultaddspace = .5em
}

\newcommand{\PBdone}[1]{\cellcolor{green!12}#1}   
\newcommand{\PBind} [1]{\cellcolor{blue!10}#1}    
\newcommand{\PBtodo}[1]{\cellcolor{orange!18}#1}  
\newcommand{\PBnz}  [1]{\cellcolor{gray!18}#1}    
\newcommand{\PBna}     {\cellcolor{black!4}\textendash}  

\newcommand{\quand}{\quad\text{and}\quad}   
\newcommand{\tTD}  {\text{\tiny{TD}}}       
\newcommand{\ud}   {\mathrm{d}}             
\newcommand{\ui}   {\mathrm{i}}             

\newcommand{\scE}{\mathscr{E}}   
\newcommand{\scL}{\mathscr{L}}   

\newcommand{\RR}{\mathbb{R}}
\newcommand{\CC}{\mathbb{C}}
\newcommand{\NN}{\mathbb{N}}

\newcommand{\mcM}{\mathcal{M}}   
\newcommand{\mcN}{\mathcal{N}}   
\newcommand{\mcC}{\mathcal{C}}   
\newcommand{\mcA}{\mathcal{A}}   
\newcommand{\mcB}{\mathcal{B}}   
\newcommand{\mcP}{\mathcal{P}}   
\newcommand{\mcT}{\mathcal{T}}   

\newcommand{\lambdaH}{\lambda_{\small{H}}}   
\newcommand{\ph}{\phantom{o}}                
\newcommand{\fX}{\mathfrak{X}}               

\begin{document}

\title{Symplectic mechanics of relativistic spinning compact bodies. I.\texorpdfstring{\\}{} 
linear‑in‑spin integrability under Killing–Yano symmetry
}

\author{Paul Ramond, \orcidlink{0000-0001-7123-0039}}\,\email{ramond@lpccaen.in2p3.fr}
\affiliation{Universit\'{e} de Caen Normandie, ENSICAEN, CNRS/IN2P3, LPC Caen UMR6534, F-14000 Caen, France}

\author{Soichiro Isoyama, \orcidlink{0000-0001-6247-2642}}\,\email{isoyama@yukawa.kyoto-u.ac.jp}
\affiliation{Department of Physics, National University of Singapore, 21 Lower Kent Ridge Rd, 119077 Singapore}

\begin{abstract}
We study the Hamiltonian dynamics of a neutral, massive spinning test body at linear order in spin, as governed by the Mathisson--Papapetrou--Tulczyjew--Dixon equations in a 4-dimensional spacetime admitting a non-degenerate Killing--Yano tensor.
The Tulczyjew--Dixon spin supplementary condition is imposed as a set of algebraic constraints, and the resulting constraint surface is equipped with a Poisson--Dirac bracket, yielding a non-degenerate, 10-dimensional physical phase space.
On this reduced space we identify five functionally independent first integrals in involution: the autonomous Hamiltonian, two constants of motion associated with two commuting Killing vectors, a generalised Carter constant, and the R\"udiger constant; all four descending from the Killing--Yano tensor.
All calculations are covariant, and the result is a purely geometric statement: it requires no background field equations and holds for any metric admitting a non-degenerate Killing--Yano tensor, beyond Kerr, beyond vacuum and beyond general relativity.
Our results identify Killing--Yano symmetry as the sole geometric source of linear-in-spin integrability. Extensions to quadratic-in-spin dynamics, including the spin-induced quadrupole, and to tidally-induced quadrupolar effects for non-spinning bodies, are treated in companion papers.
\end{abstract}

\maketitle
\tableofcontents

\section{Introduction}

\subsection{Context: The relativistic motion of test bodies}

The dynamics of compact binary systems remain at the forefront of gravitational wave astronomy \cite{Ab.al2.16,GWTC1.19,GWTC2.20,GWTC3.21,GWTC4.25,LIGOScientific:2026wfs}. 
Of particular interest are extreme and intermediate mass ratio inspirals (EMRIs/IMRIs), which are premier targets for the next generation of space-based and ground-based gravitational wave observatories. These include the Laser Interferometer Space Antenna (LISA) \cite{LISAproposal, LISA:2024hlh}, TianQin/Taiji \cite{Gong:2021gvw}, deci-hertz detectors like DECIGO \cite{Kawamura:2020pcg}, and third-generation ground-based facilities such as the Einstein Telescope (ET) \cite{Punturo.al.10, Hild:2010id} and Cosmic Explorer (CE) \cite{Reitze:2019iox, Evans:2021gyd}. Our ability to predict the motion of such systems with great accuracy directly impacts the likelihood of detecting them and extracting their physical parameters through their gravitational-wave emission. 

Fortunately, the orbital dynamics of compact binaries exhibit a remarkable ``universality.'' 
The complex internal structure of an extended compact object does not influence its motion arbitrarily, rather, it is captured through a multipolar expansion. One models the real, extended object as a point particle endowed with a finite set of multipole moments—--monopole (mass), dipole (spin), quadrupole, and so on---with the induced moments determined by explicit models typically involving deformability coefficients (e.g., Love numbers).
The higher the number of multipoles retained, the finer and more realistic the description of the object. The dynamics are then governed by a coupled problem: an equation of motion determining the representative worldline of the body, and evolution equations for the multipole moments along that worldline. Crucially, this very general result holds for any sufficiently small object with a conserved stress-energy-momentum tensor \cite{Di.74, Ha.15}.

In such highly asymmetric binaries, this multipolar description is rigorously formalized by treating the secondary as an extended test body moving in the curved background spacetime of the primary. For such a small object, the foundations of the multipolar schemes were laid in 1937 by Mathisson \cite{Ma.37}, with key contributions by Papapetrou and Tulczyjew  \cite{Pa.51,Tu.59}, culminating in the 1970s with Dixon's comprehensive framework \cite{Di.74} (
see also \cite{Di.15} and Chap.~2 of \cite{PhDHal.21} for a historical review. For other approaches to the formulation, see, e.g., \cite{Frolov.17,Kim:2025xka,Kim:2026bqi}).
A modern and enlightening formulation is Harte's ``generalized Killing fields'' approach \cite{Ha.12, Ha.15, Ha.20}, which makes explicit the distinction between kinematical and dynamical effects. The relevant evolution equations, now known as the Mathisson--Papapetrou--Tulczyjew--Dixon (MPTD) equations, are natural manifestations of the attempt to maintain Poincar\'e invariance along the body's representative worldline \cite{Ha.08}. These equations have been thoroughly studied across a variety of background spacetimes---including exact and hairy black holes, as well as cosmological metrics \cite{Ha.07, ObuPue.11, SaHoAs.14}---and are recovered in various approximation schemes \cite{Po.16, LoomisBrown.17, MiPo.21, ScheopnerVines.24}.

\subsection{Context: Relativistic Hamiltonian formulation}

Because the MPTD equations are first-order ordinary differential equations (ODEs) for the linear momentum and spin, casting them into a Hamiltonian framework\footnote{In this context, by a ``Hamiltonian formulation'' we refer broadly to a system of ODEs generated by a scalar function on a phase space endowed with a specific geometric (Poisson or symplectic) structure, rather than strictly requiring canonical coordinates. See App.~\ref{app:hamsystems} for details.} provides a natural geometric foundation for analyzing their conservation laws and integrability. 
In general relativity, two distinct approaches have historically been pursued. 
The first extends the Lagrangian formulation of a free spinning particle to curved spacetime \cite{HaRe.74}. 
Inspired by this work, the authors of Ref.\cite{Ba.al.09} (see also ~\cite{KuLeLuSe.16}) provided a Hamiltonian formulation of the dipolar MPTD system via a singular Legendre transformation of this Lagrangian, 
with subsequent extensions to quadrupolar \cite{ViKuStHi.16} and octupolar orders \cite{Ma.15}. Similar developments have been extensively pursued within the effective field theory (EFT) approach (see, e.g., \cite{Po.06, Porto:2016pyg, Levi:2018nxp}). 
While these Lagrangian and EFT-based formulations are well-established and highly effective for dynamical evolution, they typically rely on a $3+1$ spacetime split and gauge-dependent spin supplementary conditions (SSCs). Although powerful for their respective purposes, this loss of manifest covariance makes it less straightforward to directly exploit four-dimensional geometric symmetries, such as KY tensors.

The second approach goes back to Souriau's symplectic formulation \cite{Souri.70, Kun.72} (cf. the recent note on Souriau's pioneering ideas \cite{DamSou.24}). Although our treatment does not follow Souriau's construction directly, it is likewise anchored in symplectic geometry and adopts its guiding philosophy: treat the Hamiltonian system as the primary object, not merely as the Legendre transform of a Lagrangian. The general symplectic/Poisson structure employed here is already present---though somewhat implicit---in Souriau's \cite{Souri.70} and K{\"u}nzle's work \cite{Kun.72}; to our knowledge, its first explicit appearance in this setting is in \cite{dAKuvHo.15}.
Naturally, our formulation shares features with covariant approaches in this vein 
\cite{dAKuvHo.15,DaKu.al.16,WiStLu.19,WitzHJ.19,Kim:2026wrs,Witzany:2026eqc}. 
These works operate at the level of the equations of motion with a degenerate Poisson structure \cite{dAKuvHo.15, DaKu.al.16} \footnote{While Ref.~\cite{dAKuvHo.15} refers to these brackets as \emph{symplectic}, they are degenerate and thus, within standard conventions (see App.~\ref{app:hamsystems}), are not strictly so.}, or employ extended phase spaces with auxiliary variables designed for specific applications such as Hamilton--Jacobi separability \cite{WiStLu.19, WitzHJ.19,Witzany:2026eqc}.

Our approach is complementary, forming part of a broad programme towards a fully covariant, symplectic mechanics of spinning compact bodies~\cite{Ra.CQG.24,
Ra.Iso.PapII.24, Ra.Iso.Dru.IntegO2.26}.
Establishing Liouville--Arnold integrability on the physical constraint surface requires a non-degenerate symplectic structure; this necessity leads us to a different construction from those earlier works. 
Because the MPTD equations alone are under determined, an SSC is required to supply the missing algebraic relations by fixing a representative worldline. We adopt the Tulczyjew-Dixon (TD) SSC, interpretable as setting the mass dipole to zero in the frame aligned with the body's momentum, which yields a unique center-of-mass worldline. 
In the Lagrangian formalism, such constraints are typically treated using the Dirac-Bergmann algorithm \cite{Ba.al.09, Bro.22}. Here, however, we handle the TD SSC entirely within the geometric framework of constrained Hamiltonian systems \cite{Arn, Derigl.22}. 
In this symplectic context, algebraic constraints are treated simply as geometric projections (pull backs) onto a hypersurface of the phase space—analogous to projecting a tensor onto a spatial hypersurface in general relativity. By applying the Dirac bracket, we systematically reduce the initially 14-dimensional phase space to a non-degenerate, 10-dimensional physical phase space. This geometric reduction provides the exact symplectic structure required to rigorously establish Liouville-Arnold integrability on the physical constraint surface.

\subsection{Motivation: From geodesic integrability to spinning bodies around black holes
}

While the Hamiltonian framework developed here applies to any background geometry, the question of integrability naturally restricts our focus to spacetimes endowed with special symmetries. At the monopolar (geodesic) order, the equations of motion describe the most general setup for ``free systems'' \cite{Arn.66}. Owing to their purely kinematical evolution, such systems can sometimes possess enough constants of motion to be solvable by quadrature, i.e., using algebraic manipulations and standard calculus. Indeed, the integrability of geodesics across a broad class of black hole and symmetric spacetimes is now well understood and attributed to unique symmetry features, as thoroughly detailed in \cite{Yasui:2011pr,Frolov.17}.

Among these, the Kerr spacetime plays a central role. However, its integrability is highly non-trivial due to a lack of sufficient isometries \footnote{For readers approaching this work from a dynamical systems perspective, we recall that in general relativity, an isometry (a symmetry of the metric tensor) is generated by a Killing vector field; see, e.g., App.~C.1 of \cite{Wald} for precise definitions.}. Remarkably, Carter's original proof of integrability \cite{Carter.68} relied purely on the symplectic mechanics of the geodesic Hamiltonian, predating the discovery of the underlying Killing-St\"ackel tensor by two years \cite{WalkPen.70}.

This profound connection between hidden symmetries and symplectic geometry has enabled the extensive use of advanced analytical and numerical tools inherent to Hamiltonian mechanics: Hamilton-Jacobi equation \cite{Carter.68}, action-angle variables \cite{Schm.02, WitzAA.22}, phase space analysis \cite{Dean.99}, Hamiltonian frequencies and resonances \cite{Schm.02, BrGeHiPRL.15, ZeLuWi.20, Lynch.al.24}, perturbation (KAM-type) theory \cite{HinFle.08, Xue.20}, Birkhoff normal forms via Lie series \cite{Kera.al.23}, symplectic integrators \cite{Wu.al.21, Wang.al.21}, and Poincar\'e-Melnikov theory \cite{WitSemSuk.15}.

However, the inclusion of the secondary's spin is widely expected to break this integrability and induce chaos, as suggested by numerical simulations in both Schwarzschild \cite{Bomb.al.92, VerHir.10,ZeLuWi.20} and Kerr \cite{SuzMae.97, Se.99, KySe.07, Han.08,KuLeLuSe.16} backgrounds. 
Yet, when limiting to effects that are strictly linear in the particle's spin, constants of motion constructed from the Killing-Yano (KY) tensor \cite{Yano.52} have long been known to exist \cite{Rudiger.I.81, Rudiger.II.83, ComDru.22,Andersson:2025bhq}. Leveraging these constants, recent works have explored linear-in-spin integrability in the Kerr spacetime (and beyond), including Hamilton--Jacobi separation \cite{WitzHJ.19,Witzany:2026eqc} and explicit analytical solutions \cite{WiPio.23,SkouWitz.25}.

Despite this considerable progress, the most classical method for proving Liouville-Arnold integrability (i.e., exhibiting a full set of Poisson-commuting first integrals on a reduced physical phase space) has not appeared in the literature even for the Kerr spacetime; partial results were obtained in, e.g., ~\cite{Apos.96,KuCa.12,KuLeLuSe.16,Frolov.17,WitzHJ.19}.
Our work fills this gap. Unlike $3+1$ splits or coordinate-based calculations that can obscure underlying spacetime symmetries, a fully covariant Hamiltonian approach preserves the four-dimensional geometric structure. This allows us to directly link the existence of a non-degenerate Killing-Yano tensor to the emergence of Poisson-commuting invariants.
By providing a rigorous, covariant proof of integrability for any spacetime admitting a non-degenerate Killing-Yano tensor, we demonstrate that the powerful Hamiltonian toolkit developed for geodesics can be safely and rigorously extended to linear-in-spin dynamics.

\subsection{An overview of main results}

The objective of this paper is threefold: (i) to place the linear-in-spin MPTD dynamics under the TD SSC on a robust and Hamiltonian footing; (ii) to establish its integrability in spacetimes with KY symmetry; and (iii) to prepare for subsequent works where integrability results are exploited, and additional extended-body effects are added to the formalism.

\subsubsection{Covariant symplectic formulation (this work)}

Our first result is the systematic construction of the physical phase space.
Starting from the natural 14-dimensional Poisson manifold spanned by spacetime position, momentum, and spin, we fix the two spin-sector Casimirs to reach a 12-dimensional symplectic leaf, then enforce the TD SSC via a Dirac-bracket reduction to obtain the final 10-dimensional physical phase space.
On this reduced space, our Hamiltonian formulation of the linear-in-spin MPTD dynamics satisfies five essential criteria. It is at once:
\begin{itemize}\itemsep0em
    \item symplectic : equipped with a non-degenerate Poisson structure;
    \item covariant : independent of any specific choice of spacetime coordinates or tetrad fields;
    \item self-consistent : strictly truncated to avoid retaining terms of the same spin order as those omitted;
    \item a first integral : the autonomous Hamiltonian is a true constant of motion, not a vanishing phase-space constraint;
    \item 10-dimensional : matching the exact number of physical degrees of freedom required for the MPTD system under the TD SSC.
\end{itemize}

\subsubsection{Proof of Liouville-Arnold Integrability at linear order (this work)}

\begin{figure}[ht]
    \centering
        \begin{overpic}[width=\linewidth]{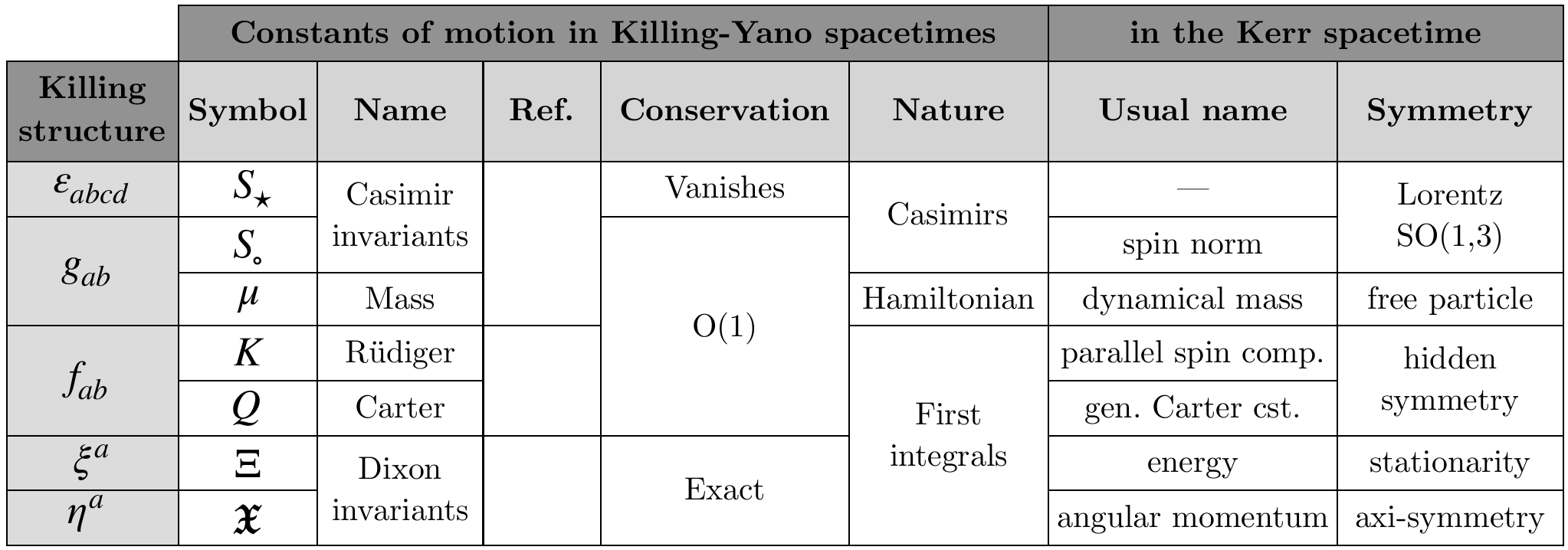}
        \put(32.1,18.5){\eqref{eq:def-norms}}
        \put(31.6,10.2){\eqref{defKQ}}
        \put(32.1,3.4){\eqref{defFk}}
    \end{overpic}
    \caption{
    Constants of motion in a spacetime with Killing-Yano symmetry and in Kerr. 
    The table summarizes the five independent constants of motion 
    $(H, \Xi, \fX, K, Q)$ that underpin the Liouville-Arnold integrability of the linear-in-spin MPTD dynamics under the TD SSC. 
    In the first column, the Levi-Civita tensor $\varepsilon_{abcd}$, the metric and $g_{ab}$ act as \textit{trivial} Killing objects (a valence-4 KY tensor and a valence-2 KS tensor, respectively). 
    In the Kerr limit, the invariant $\fX$ reduces to $a (L - a E)$ where $a$, $E$ and $L$ 
    are the Kerr spin parameter, orbital energy, and axial angular momentum.  
    }
    \label{fig:tableconstant}
\end{figure}

Building on this 10-dimensional formulation, our second result is the proof of Liouville-Arnold integrability. We demonstrate that in any spacetime admitting a non-degenerate KY tensor, the linear-in-spin dynamics exhibit five functionally independent first integrals. 
Remarkably, aside from the Hamiltonian, all these constants of motion are generated entirely by the KY symmetry. 
As summarized in Figure \ref{fig:tableconstant}, they consist of: 
\begin{itemize}\itemsep0em
    \item The Hamiltonian $H$, numerically equal to $-\mu^2/2$, encoding the body's conserved (dynamical) mass $\mu$. It is conserved because the particle evolves freely at dipolar order (no self-induced force or torque from higher-order multipoles). 
    \item Two Dixon invariants $(\Xi,\fX)$, associated with the two Killing vectors generated by the KY tensor, typically linked to (a linear combination of) the small body's energy and angular momentum in stationary, axisymmetric spacetimes. 
    \item The (linear-in-spin) R\"udiger constant $K$, constructed directly from the KY tensor, representing the projection of the small body's spin 4-vector onto a covariant notion of total angular momentum.
    \item The (generalized) Carter constant $Q$, also built solely from the KY tensor and its associated Killing-St\"ackel (KS) tensor.
\end{itemize}

These constants of motion have a rich history in the literature. The spinning corrections to the geodesic Killing invariants ($\Xi$ and $\fX$) were already known to the pioneers of the MPTD equations (see, e.g., the historical account by Dixon \cite{Di.15}), while the existence of the R\"udiger and generalized Carter constant ($K$ and $Q$) in spacetimes admitting a non-degenerate KY tensor has been established in the 1980s by R\"udiger \cite{Rudiger.I.81, Rudiger.II.83}.
However, their mutual involution (and thus the formal integrability of the system) had only been recently established for the specific case of the Kerr metric \cite{WitzHJ.19,ComDru.22,Ra.CQG.24, Witzany:2026eqc}. 
By performing fully covariant calculations, we demonstrate that these five invariants are in pairwise involution with respect to the Dirac bracket for any spacetime admitting a non-degenerate KY tensor. This rigorously establishes Liouville-Arnold integrability on the physical constraint surface, ensuring that phase-space trajectories are confined to invariant tori at linear order in spin.

\subsubsection{Geometric assumptions and admissible spacetimes}

Our results rely on a single geometric assumption: the background spacetime must admit a non-degenerate KY tensor. We make no \textit{a priori} assumptions about the spacetime curvature. Yet, the existence of a KY tensor strongly constrains the geometry. By the classification of Dietz and R\"udiger \cite{DiRu.I.81, DiRu.II.82}, the Weyl tensor is necessarily of Petrov type D, and the Ricci tensor is restricted to a pure-trace term plus a single trace-free component.

At a purely geometric level (without imposing Einstein field equations), this defines an infinite-dimensional family of metrics. Since the KY tensor squares to an irreducible Killing-St\"ackel (KS) tensor, these geometries naturally belong to the separability structures of Benenti and Francaviglia \cite{Benenti:1979erw, 1980grg1.conf..393B}. When restricting to exact solutions of the Einstein-Maxwell equations, this selects the non-accelerating type-D solutions, namely the Kerr-Newman-NUT-(A)dS metrics  (see \cite{Yasui:2011pr,Frolov.17} for comprehensive reviews). While the general Pleba\'nski-Demia\'nski class \cite{SKMHH} admits only a conformal KY tensor, it reduces to a strict KY tensor precisely in this non-accelerating limit \cite{Kubiznak:2007kh}. Thus, our integrability proof applies to any metric in this non-accelerating class, irrespective of its Ricci content (vacuum, $\Lambda$-vacuum, or aligned matter \cite{Iboha.97}), provided the secondary is neutral.

We emphasize that our KY hypothesis to the spacetime is strictly stronger than assuming a Petrov type-D background or the existence of a symmetric KS tensor alone. Spacetimes carrying a non-trivial KS tensor but lacking a KY tensor \cite{AnaBati.16} are excluded from this framework, as they generally lack the additional constants of motion required for linear-in-spin integrability.

As a final remark, the linear-in-spin integrability established in this work serves as the rigorous baseline for exploring higher-order multipolar effects. In a companion paper (Paper III) \cite{Ra.Iso.Dru.IntegO2.26}, we advance this Hamiltonian formalism to the quadratic-in-spin regime. By restricting the background to Einstein spacetimes, we show that the persistence of integrability depends critically on the secondary's internal structure: 
while integrability is maintained for a black-hole secondary (characterized by a Kerr-like spin-induced quadrupole $\kappa = 1$), the symmetry-generated invariants generally break down for generic compact objects $(\kappa \neq 1)$. 
This delineates the precise boundary at which spin-curvature couplings induce non-integrability in extended-body dynamics.
Beyond spin-induced effects, tidal couplings to the background curvature enter at the same quadrupolar order and constitute an independent source of non-integrability, even for a black-hole secondary; this is addressed in a forthcoming work \cite{Ra.NoCarter.26}.

\subsection{Organization of the paper}

The organization of the content is as follows:

\begin{itemize}
    \item Sec.~\ref{sec:one} reviews the MPTD system, focusing on the multipolar formalism of Dixon-Harte (Secs.~\ref{sec:DixHar} and \ref{sec:physint}) and the role of spin supplementary condition (SSC) (Sec.~\ref{sec:SSC})
    We motivate our choice of the TD SSC (Sec.~\ref{sec:TDSSC}) and justify the truncation to linear-in-spin dynamics (Secs.~\ref{sec:MPTDnonlin} and \ref{sec:MPTD}).
    \item Sec.~\ref{sec:Ham} develops our Hamiltonian formulation. 
    We begin with a 14-dimensional phase space $\mcM$ (Sec.~\ref{sec:defPoisson}), which is reduced to a 12-dimensional symplectic leaf $\mcN$ by lifting degeneracies associated with local Lorentz invariance (Sec.~\ref{sec:degPoisson}). We then incorporate the TD SSC via Dirac brackets, yielding the final 10-dimensional physical phase space $\mcP$ (Sec.~\ref{sec:consinv}). The complete formulation is summarized in Sec.~\ref{subsec:blabla} and illustrated in Figure.~\ref{fig:PS}.
    \item Sec.~\ref{sec:KY} focuses on the background geometry, providing an analysis of spacetimes endowed with a KY tensor. After introducing the KY tensor in general relativity (Sec.~\ref{subsec:exsistence-KY}), we derive and collect key geometric identities (Sec.~\ref{subsec:KYids}) that are essential for the subsequent proofs.
    \item Sec.~\ref{sec:PBcalc} presents our main results: the proof of integrability for linear-in-spin dynamics in KY spacetimes. 
    We define the five constants of motion (Sec.~\ref{sec:invMPTD}) and formally state the integrability theorem (Sec.~\ref{subsec:Liouville}). The proof, which relies on the vanishing of various Dirac brackets, is detailed in Secs.~\ref{subsec:orange} and \ref{subsec:bluegrey}, and summarized in Table~\ref{tab:poisson}.
\end{itemize}

Finally, our conclusions and future prospects are discussed in Sec.~\ref{sec:concl}. Several appendices are included to make the paper self-contained and pedagogical: App.~\ref{app:conventions} details our conventions and notations; App.~\ref{app:Hodge} covers Hodge duality; App.~\ref{app:hamsystems} reviews Hamiltonian systems and Poisson geometry (see in particular Table.~\ref{tab:glossary} for an overview); App.~\ref{app:xpstospis} provides proofs regarding phase-space coordinates; and  App.~\ref{app:MMA} supplies explicit coordinate expressions for the Kerr-Newman-de Sitter spacetime used in our companion Mathematica notebook \cite{MMA_IntegO1}. This notebook contains an explicit check of all results presented in the paper, with calculations performed using the \emph{RGTC} package~\cite{Bonanos:2003RGTC} developed by S.~Bonanos \cite{Bonanos:RGTC_webpage}.

\section{Test bodies in general relativity} 
\label{sec:one}

A sometimes overlooked, yet remarkable result of general relativity is that, if a (non-necessarily compact) object is moving in a given background spacetime (with no back reaction) and modelled as a point particle with multipoles, then its equations of motion can be found in closed and covariant form\footnote{In the words of Ehlers and Rudolph \cite{EhRu.77}, ``Even the existence of such a ``reduced multipole description'' could not have been anticipated on general grounds and must be regarded as a remarkable property [...]''.}. Moreover, these equations have been derived in a variety of different and independent ways: multipolar expansions \cite{Tu.59,Di.74,StPu.10}, Lagrangian formalism \cite{St.15}, gravitational self-force \cite{GrWa.08,Po.12,MaPoWa.22}, generalized Killing fields \cite{Ha.12,Ha.15,Ha.20}; see also Chap.~2 in \cite{PhDHal.21} for a survey of such methods. Here, we only summarize the key results from this branch of relativistic mechanics, as presented in \cite{Ha.12,Di.15,Ha.15}. 

\subsection{Brief review of Dixon-Harte formalism} \label{sec:DixHar}

A comprehensive framework for multipolar descriptions of extended bodies was established by Dixon \cite{Di.74,Di.79,Di.15} and completed by Harte \cite{Ha.12,Ha.15}, applicable to any system characterized by a conserved, spatially-compact stress-energy tensor $T^{ab}$. The construction of multipole moments requires specifying a reference worldline (typically timelike), which serves as the origin for the multipolar expansion. Additionally, one must introduce a foliation of the body's worldtube using a one-parameter family of hypersurfaces. In standard practice, each hypersurface comprises the set of geodesics orthogonal to a prescribed timelike vector $t^a$ at each worldline point. While centroid conditions (also called supplementary-spin-conditions, cf. Sec.~\ref{SSC} below) commonly determine the choices of wordline and $t^a$, this prescription is not mandatory. 

The higher the number of multipoles used, the finer the representation of the extended body by the multipolar particle. A whole hierarchy of multipoles enters the description at successive multipolar orders. For example, at monopolar order, the extended body is described by a particle endowed with its four-momentum vector $p^a$ only. At dipolar order, the antisymmetric spin tensor $S^{ab}=S^{[ab]}$ is added on top of $p^a$, for a total of $10=4+6$ degrees of freedom. 

The stress-energy tensor's monopole $I^{ab} = I^{(ab)}$ and dipole $I^{cab} = I^{c(ab)}$ moments take the form
\begin{equation} \label{IspSJ}
    I^{ab} = p^{(a} v^{b)} \quand I^{cab} = S^{c(a} v^{b)},
\end{equation}
where $v^a$ is a tangent vector to the worldline, and these equations serve as definitions for $p^a$ and the angular momentum of the object $S^{ab} = S^{[ab]}$, which are tensor defined along the worldline (See, e.g., Sec.~(IV.2.2) in \cite{Ha.12} for inverse formulae giving $(p_a,S^{ab})$ as explicit hypersurface integrals over the stress-energy tensor). 

Beyond these leading moments, each multipolar order $n\geq2$ adds a tensor of valence $n+2$ (called a $2^n$-pole) to the description of the body, bringing an additional $(n+3)(3n-1)$ independent degrees of freedom due to their algebraic symmetries and orthogonality properties \cite{Di.74}. A complete tower of $2^n$-pole moments $I^{c_1 \cdots c_n ab}$ for any spatially-compact, conserved stress-energy distribution is then obtained, each admitting an explicit integral representation in terms of $T^{ab}$ \cite{Di.74,Di.15,Ha.12,Ha.15,Ha.Ra.26}. 

As is classical in GR, evolution equations (of laws of motion) for the moments can be obtained from stress-energy-momentum conservation $\nabla_{a}T^{ab}=0$. More specifically, the latter holds if and only if the moments satisfy \cite{Di.79,Di.15,Ha.12,Ha.15} \footnote{From the algebraic symmetries of $R_{abcd}$ and the anti-symmetry of $S^{ab}$, one has $ R_{abcd} S^{bc} = -\tfrac{1}{2} R_{adbc} S^{bc}$.}
\begin{subequations} \label{EEgen}
\begin{align}
	\dot{p}_a  &= R_{abcd} S^{bc}v^d + F_a[(I^{c_1\ldots c_nab})_{n\geq2}], 
	\label{EoM}
	\\
	\dot{S}^{ab} &= 2 p^{[a} v^{b]} + N^{ab}[(I^{c_1\ldots c_nab})_{n\geq2}],
	\label{EoP}
\end{align}
\end{subequations}
where the overdot denotes the covariant derivative along the worldline, $\dot{X} := v^a \nabla_a X$. The force $F_a$ and antisymmetric torque $N^{ab} = N^{[ab]}$ arise from the quadrupole and higher-order multipole moments. 

Equations \eqref{EEgen} are the Mathisson-Papapetrou-Tuclzyjew-Dixon (MPTD) equations. Notice the absence of evolution equations for the quadrupole or higher-order moments. As in Newtonian mechanics, the evolution of these higher moments depends on the specifics of the particular body under study (internal fields, equation of state, micro-physics, etc). Explicit expressions for the force $F_a$ and torque $N^{ab}$ are known \cite{Ma.15,Ha.12}, and are linear-in-curvature up to octupole order \cite{Ma.15,Ha.12,Ha.Ra.26}, after which they involve non-linearities in the Riemann tensor \cite{Ha.12,Ha.15}. 

Lastly, we mention that, by construction, the Dixon-Harte framework reviewed here only applies, in principle, to material bodies with a non-vanishing $T_{ab}$ (though see \cite{ScheopnerVines.24} for examples of Dixon moments computed for a Kerr black hole). However, the resulting framework does extend to non-material bodies, and there exist alternative ways to obtain the MPTD equations \eqref{EEgen} for arbitrary (not necessarily material) objects. These include methods pertaining to the framework of Effective Worldline Theory \cite{Ma.15,Po.16,ViKuStHi.16,ScheopnerVines.24}, where the MPTD equations are obtained via an action principle; the gravitational self-force formalism \cite{Po2.10,BaPo.18,PoWa.21,MaPoWa.22} which solves perturbatively the Einstein field equations and derives an equation of motion for the secondary object using matched asymptotics; or elastic body methods such has those presented in \cite{LoomisBrown.17,Brown.20}. Although all yield similar-looking equations of motion, they do rely, fundamentally, on different definitions for the source's multipole moments, and their relationships warrants further exploration. 

\subsection{Physical interpretation and scalar quantities}  
\label{sec:physint}

We consider a fixed background spacetime $(\scE, g_{ab})$ where $\scE$ is a 4-dimensional (4D) manifold covered with four coordinates $x^\alpha$, and $g_{ab}$ is a metric tensor on $\scE$. The actual, physical trajectory of the compact object is put in correspondence with the worldline $\scL\subset\scE$ of the representative point particle endowed with multipole moments, following the Dixon-Harte formalism. 

The physical interpretation of \eqref{EEgen} is clear: the coupling between the spin $S^{ab}$ and the background curvature $R_{abcd}$ drives the evolution of $p^a$, while the misalignment of $p^a$ and $v^a$ (tangent to the worldline) drives the evolution of $S^{ab}$. Despite being referred to as the ``Papapetrou force(s)'', both effects are pure \emph{kinematics}, independent of the body and already exist in classical Newtonian mechanics \cite{Ha.23}. On top of these kinematical effects, \emph{dynamical} effects come from quadrupole and higher-order moments through force and torque $(F_a, N^{ab})$. 

Regardless, three scalar fields along $\scL$ can be defined from the momenta: 
\begin{equation}\label{eq:def-norms}
        \mu^2 := -p_a p^a, \quad 
        S_\circ^2 := \frac{1}{2}S_{ab}S^{ab} \quand
        S_\star^2 := \frac{1}{8}\varepsilon_{abcd}S^{ab}S^{cd},
\end{equation}
where $\varepsilon_{abcd}$ is the Levi-Civita tensor. By assumption, $p^a$ is timelike while $S^{ab}$ is spacelike, resulting in $\mu\geq0$, $S_\circ\geq 0$ and $S_\star \geq 0$. Among these norms, $\mu$ and $S_\circ$ will be referred to as the \textit{mass}, and $S_\circ$ as the \emph{spin norm}. Note that they are intrinsically defined from the particle's multipole, and do not depend on any observer. These quantities are, in general, not conserved under the MPTD system \eqref{EEgen}, and their rate of change depends on the choice of representative worldline and the nature of the force and torque in \eqref{EEgen}. 

Scalar quantities that are unconditionally conserved, however, do exist in the presence of Killing vector fields on spacetime.\footnote{These quantities are at the core of the ``generalized Killing field'' formulation of the framework by Harte \cite{Ha.12,Ha.15}.} 
For any Killing vector $k^a$, define
\begin{equation}\label{defFk}
    \mathcal{F}_k := p_a k^a + \frac{1}{2} \nabla_a k_b S^{ab}.
\end{equation}
Taking a covariant derivative of this equation along the worldline and using \eqref{EEgen} along with the identity $\nabla_a\nabla_bk_c=R_{cbad}k^d$ valid for Killing vectors \cite{Wald}, one finds 
\begin{equation}
    \dot{\mathcal{F}}_k =  F_a k^a + \frac{1}{2} \nabla_a k_b N^{ab}. 
\end{equation}
It is clear that if no force nor torque are exerted on the body, then $\mathcal{F}_k$ is conserved. However, this conservation \emph{still holds for arbitrary force and torque} \cite{Di.74,Ha.12,Ha.15}. In other words, 
\begin{equation} \label{Fkcons}
    \mathcal{L}_kg_{ab}=0 
    \quad \Rightarrow \quad 
    \mathcal{F}_k = p_a k^a + \frac{1}{2} \nabla_a k_b S^{ab} = \text{cst.}
\end{equation}
This result \emph{cannot} be derived from the MPTD equations \eqref{EEgen} alone and comes from more general arguments, see \cite{Ha.12,Ha.15}. This is the generalisation, for a multipolar particle, of the well-known geodesic constant of motion $p_a k^a$ associated to spacetime isometries. The constant of motion \eqref{defFk} also plays a central role in the Hamiltonian formulation of the MPTD equations, as we will see later in Sec.~\ref{sec:Ham}. 

\subsection{Spin supplementary conditions} 
\label{sec:SSC}

Even with a model for the higher order moments, the MPTD system is still under-determined as there are more unknowns (fourteen in $(v^a,p_b,S^{cd})$) than ODEs (ten). This is linked to two observations. First, the tangent vector $v^a$ is arbitrary in \eqref{EEgen} (i.e., not necessarily normalized) as the MPTD equations are actually invariant by re-parametrization of $\scL$. Second, there is freedom in choosing the worldline that represents the body, or equivalently, a procedure to uniquely define the multipole moments. A common way of fixing these issues (and thus having a well-posed ODE system), is to (i) take $v^a$ to be the four-velocity $u^a$ along $\scL$, such that the condition $u^a u_a=-1$ reduces the number of unknowns to 13, and (ii) impose a so-called \emph{spin supplementary condition} (SSC), reducing this number to 10. An SSC takes the form of algebraic constraints of the form 
\begin{equation} \label{SSC}
    f_a S^{ab}=0 \,,
\end{equation}
where $f^a$ is a timelike vector defined along $\scL$. There are only three independent equations encoded in \eqref{SSC} because this equation projected onto $f_b$ gives $0=0$ by antisymmetry of $S^{ab}$. To elucidate the meaning of such SSC, perform a Hodge decomposition (cf. App.~\ref{app:Hodge}) of $S^{ab}$ with respect to the timelike vector $f^a$, effectively splitting $S^{ab}$ into a spin vector $S^a$ and mass dipole vector $D^a$, according to\footnote{Equation \eqref{zip!} assumes that $f_a f^a = -1$. If not, extra factors of $f_a f^a$ must be included.} 
\begin{equation}\label{zip!}
	S^{ab} = \varepsilon^{abcd} f_c S_d + 2 D^{[a} f^{b]} 
    \quad \Longleftrightarrow \quad
	\begin{cases}
		\, S^b = \frac{1}{2} \varepsilon^{abcd} f_a S_{cd}, \\
		D^b = f_a S^{ab}.
	\end{cases}
\end{equation}
The six degrees of freedom contained in the tensor $S^{ab}$ are encoded in the spacelike vectors $S^a$ and $D^a$, which have only three degrees of freedom each, as both verify the constraint of being orthogonal to $f^a$. Clearly, the SSC \eqref{SSC} means, physically, that the mass dipole $D^a$ measured in the spacelike frame of timelike direction $f^a$ vanishes. In that frame, $S^{ab}$ possesses three degrees of freedom and can be entirely described in terms of $S^a$. This decomposition is entirely analogous to the decomposition of the Faraday tensor $F^{ab}$ of electromagnetism into magnetic $B^a$ and electric $E^a$ fields measured by a given observer \cite{GouSR}, with the dictionary $(F^{ab},B^a,E^a)\leftrightarrow (S^{ab},S^a,D^a)$ (see details in App.~\ref{app:Hodge}).

Inserting the decomposition \eqref{zip!} into the definitions of the spin scalars \eqref{eq:def-norms} leads to the following relations
\begin{equation} \label{spinnorms}
    S_\circ^2 = S^a S_a - D^a D_a \quand S_\star^2 = S^aD_a.
\end{equation}
This relation shows that under the SSC \eqref{SSC} ($D^a=0$), the spin scalar $S_\star$ vanishes identically, and the spin norm $S_\circ$ coincides with the norm of the spin \emph{vector} $S^a$. 

A natural question is which frame to choose---i.e., which vector $f^a$ in \eqref{SSC}.
This question goes beyond the scope of the present work, and we refer to the thorough review \cite{WiStLu.19} and references therein for details on SSCs (see also \cite{LuSeKu.14,CoNa.15,TiLuAp.21}). Our choice is explained and motivated below. 

\subsection{The Tulczyjew-Dixon spin supplementary condition} 
\label{sec:TDSSC}

In this article and subsequent works, we shall work with the so-called Tulczyjew-Dixon spin supplementary condition (TD SSC) \cite{Tu.59,Di.64}\footnote{Tulczyjew was the first to acknowledge the fact that the other SSCs did not necessarily lead to unique center-of-mass trajectories in special relativity \cite{Di.15}, and adopted it also in GR in \cite{Tu.59}.}:
\begin{equation}\label{TDSSC}
    C^b := p_a S^{ab} = 0 \,. 
\end{equation}
Combining the TD SSC \eqref{TDSSC} with the quadratic MPTD equations \eqref{EEgen} leads to a number of well-known results, recalled here for convenience. First, as shown first in \cite{EhRu.77} (see also \cite{ObuPue.11} for a covariant proof and \cite{GraHarWal.10} for a generalization with electromagnetic fields), the TD SSC implies the following expression for the tangent vector $v^a$ in terms of $p_a,S^{ab}$ and the background geometry. Assuming (without loss of generality) that $v^a$ satisfies $v^a p_a=-\mu$, one has the exact result \cite{EhRu.77,GraHarWal.10}
\begin{equation} \label{momvelTD}
    v^a = \hat{p}^a-\hat{N}^{ab} \hat{p}_b -\frac{1}{2\mathfrak{m}^2} S^{ab}\left(2F_b-(\hat{p}^c-\hat{N}^{cd} \hat{p}_d) R_{bcde} S^{de} \right),
\end{equation}
where $\hat{p}^a:=p^a/\mu$, $\hat{N}^{ab}=N^{ab}/\mu$ and we introduced the mass-type quantity $\mathfrak{m}$ such that
\begin{equation}\label{eq:def-frakm}
    \mathfrak{m}^2 := \mu^2 + \frac{1}{4} R_{abcd} S^{ab} S^{cd}. 
\end{equation}
Notice that all but the first term on the right-hand side (RHS) of equation \eqref{momvelTD} are orthogonal to $p_a$, making it a sort of 3+1 decomposition of the tangent vector. In particular, one can invert \eqref{momvelTD} as $p^a=\mu v^a + p_\perp^a$, where $p_\perp^a$ is orthogonal to $p_a$ and sometimes referred to as the \emph{hidden momentum} of the object \cite{GraHarWal.10}. 

The momentum-velocity relation \eqref{momvelTD} is a consequence of the general MPTD system \eqref{EEgen} and the TD SSC \eqref{SSC}. One could ask if the converse result holds true, i.e., whether the momentum-velocity relation \eqref{momvelTD} and the MPTD equations together imply $p_aS^{ab}=0$ at all times. This is \emph{almost} the case, for Eqs.~\eqref{EElin} and \eqref{momvelTD} actually imply that $v^c \nabla_c (p_a S^{ab}) = 0$, i.e., that the four-vector $p_aS^{ab}$ is parallel-transported along $\scL$. This feature has no major implication for solving the equations themselves\footnote{From the point of view of a Cauchy problem, it suffices to require $p_aS^{ab}=0$ at some initial time for the TD SSC to hold at all times (by linearity of parallel-transport).}. Yet, it must be handled with care in a Hamiltonian formulation of those same equations. This is discussed in Sec.~\ref{sec:sscnote}. 

Our choice of SSC \eqref{TDSSC} is not random, but motivated by the following reasons:~\footnote{Items (i, ii, iii, iv) result from the literature using more frequently the TD SSC than other, and could potentially be extended to other SSCs if studies were conducted. Items (v, vi, vii, viii), however, single out the TD SSC because to its special properties.} 
\begin{itemize}
    \item[(i)] to our knowledge, it is the only one for which the uniqueness of center-of-mass-like worldline has been given mathematical proofs \cite{Be.67,Mad.69,Scha.I.79,Scha.II.79,Ha.15}. Other SSCs (but not all) tend to correspond to a family of worldlines rather than a unique one \cite{Costa.al.12}.
    \item[(ii)] it leads to a momentum-velocity relation that has been obtained in an \emph{exact form}, to \emph{all multipolar orders} and can also account for the presence of electromagnetic fields \cite{GraHarWal.10}. To our knowledge, this has never been shown for any other SSC.
    \item[(iii)] several works, including \cite{Rudiger.I.81,Rudiger.II.83,ComDru.22,ComDruVin.23}, aimed at finding invariants of motion for the MPTD system. These works proved successful when relying on the TD SSC, but methods do not seem to generalize to other SSCs.
    \item[(iv)] in spacetimes admitting a Killing-Yano tensor, it leads to a covariant notion of angular momentum \cite{DiRu.I.81,DiRu.II.82,ComDru.22}, with well-behaved geodesic and Newtonian limits for spherically symmetric spacetimes \cite{Ra.Iso.PapII.24}. 
    \item[(v)] it is compatible with a \textit{covariant} Hamiltonian formulation. As pointed out in \cite{WiStLu.19}, the reason is that the TD SSC is ``co-moving'', i.e., it relies only on the intrinsic properties of the body ($p_a,S^{ab}$) and not on any background fields.
    \item[(vi)] for such covariant Hamiltonian formulations, the TD SSC uniquely reduces the number of degrees of freedom by the largest amount  possible \cite{WiStLu.19}. In a symplectic setup, this corresponds to 10 phase space dimensions.
    \item[(vii)] when viewed as a constraint on phase space variables, the TD SSC defines a submanifold that is \textit{invariant} under the flow of the free-particle Hamiltonian (cf. \eqref{sec:stabflow}) and of Killing vector invariants (cf. Eq.~\eqref{FkH}). Both properties single out the TD SSC among all SSCs.
    \item[(viii)] the TD SSC seems to be particularly suited to the SO(1,3) Poisson brackets \eqref{PBs} that underpins the MPTD dynamics: the bracket of two TD SSC components generates the mass-type quantity $\mathfrak{m}^2$ (cf. Eq.~\eqref{eq:def-frakm}), the non-vanishing of which entails the very well-posedness of the framework, regardless of Hamiltonian mechanics. 
\end{itemize}

Lastly, let us mention two other results that hold under the TD SSC. We start with exact evolution laws for $\mu$ and $S_\circ$. They read \cite{Gr.al2.10}
\begin{equation} \label{mudotSdot}
    \dot{\mu} = -F_a v^a + \frac{1}{\mu^{2}} N^{ab} p_a \dot{p}_b 
     \quand 
     \dot{S}_\circ = \frac{1}{2S_\circ} N_{ab} S^{ab}.
\end{equation}
These equations show that when no force nor torque is exerted on the body, both $\mu$ and $S_\circ$ are conserved \emph{exactly}. This happens trivially when one neglects the quadrupole and higher order moments (cf. \eqref{mudotSdot}), but also for specific choices of multipole moments.\footnote{In the case of quadratic-in-spin dynamics, the spin-induced quadrupole frequently used in the literature satisfies $N_{ab} S^{ab}=0$ exactly, under the TD SSC. See \cite{Ra.Iso.Dru.IntegO2.26} for details.}

Second, under the TD SSC, one can replace the spin tensor $S^{ab}$ by the equivalent spin \emph{vector} 
\begin{equation} 
\quad S^b_{\tTD} := \frac{1}{2} \varepsilon^{abcd} \hat{p}_a S_{cd} \quad \stackrel{\eqref{TDSSC}}{\Longleftrightarrow} \quad S^{ab} = \varepsilon^{ab}_{\phantom{ab}cd}\hat{p}^c S^d_{\tTD}, 
\end{equation}
such that $S^a_{\tTD}$ is the spin 4-vector measured by an observer of four-velocity $\hat{p}_a$ (recall the general decomposition in \ref{sec:sscnote}). An evolution equation for this vector can be obtained in closed form, see for example equation (38) in \cite{GraHarWal.10}. Yet, in the remaining of this work, we will find it more useful to use the spin tensor $S^{ab}$.

\subsection{Truncated MPTD equations} 
\label{sec:MPTDnonlin}

{The full multipolar MPTD equations \eqref{EEgen}} with the TD SSC \eqref{TDSSC} are seldom used in their entirety, because of their complexity. Their strength relies on the possibility of truncating the multipolar expansion. We mention several of these truncations below.

\subsubsection{Monopolar order}

When truncating at monopolar order, one sets to zero the second multipole moment (the spin $S^{ab}$) and all other moments, implying that the force $F_a$ and torque $N^{ab}$ must vanish as well, cf. \eqref{EEgen}. The SSC \eqref{TDSSC} is then automatically satisfied, and the MPTD equations simply become
\begin{equation}
    \dot{p}_a = 0 \quand \mu v^a = p^a,
\end{equation}
where $\mu$ is conserved [cf. \eqref{mudotSdot}]. These equations imply that $p_a$ is at once tangent to, and parallel-transported along, the worldline: the latter must be a geodesic. Note that the momentum being aligned with the four-velocity is \emph{not} an assumption, but a consequence of the laws of motion just like in Newtonian gravity \cite{Ha.15,HaGa.20}.

\subsubsection{Dipolar order}

At dipolar order, one neglects the quadrupole and higher order moments, keeping $p_a$ and $S^{ab}$. The resulting equations of motion are often called the pole-dipole equations:  
\begin{equation} \label{MPTDdipo}
    {\dot p}_a = R_{abcd} S^{bc} v^d, \quad
    {\dot S}^{ab} = 2p^{[a}v^{b]}, 
\end{equation}
and the momentum-velocity relation \eqref{momvelTD} becomes 
\begin{equation}\label{eq:mo-dipo}
    \mu v^a = {p}^a + \frac{1}{2 \mathfrak{m}^2} S^{ab} R_{bcde} {p}^c S^{de},
\end{equation}
with $\mathfrak{m}$ given in \eqref{eq:def-frakm}. 
Although the dipolar ODEs may appear linear in the spin $S^{ab}$, the whole system \eqref{MPTDdipo}-\eqref{eq:mo-dipo} is nonlinear due to the momentum–velocity relation in~\eqref{eq:mo-dipo}. Remarkably, despite these non-linearities, the mass $\mu$ and the spin norm $S_\circ$ remain exactly conserved at this order, cf.~\eqref{mudotSdot}. 

The non-linear-in-spin, pole--dipole model has been studied repeatedly: for equatorial orbits in black-hole backgrounds \cite{Suzuki:1996gm,Suzuki:1997by,Pio.al.21,SkouLukes.21,Shahzadial.26}, and from the standpoint of its Lagrangian and Hamiltonian formulations \cite{DeriRam.15,WiStLu.19},  its symmetries \cite{BatistaSantos.20}, and through Hamiltonian reduction and a fixed-point analysis \cite{Bizyaev:2025mva}. When reading these results physically, one point is worth bearing in mind. Retaining terms non-linear in the dipole while discarding the higher multipoles mixes contributions of the same order in the spin: schematically, $(\text{dipole})^2$ enters at the order of the quadrupole; $(\text{dipole})^3$ and $(\text{dipole})\times(\text{quadrupole})$ at that of the octupole; and so on. A truncation that keeps the dipolar terms to a given order in spin but drops the multipoles of that same order is therefore not order-consistent. 

The omission is not innocuous, either: the discarded multipoles need not act as a mere correction but can restore structure some quadrupolar models that the truncated model lacks. Examples include: an additional constant of motion as shown in \cite{ComDruVin.23,Ra.Iso.Dru.IntegO2.26}; the (remarkable) proportionality between $p^a$ and $v^a$, which never holds for the pole-dipole particle but holds for some quadrupolar models (see Sec.~I.C.2 in Paper III \cite{Ra.Iso.Dru.IntegO2.26}); or the complete Liouville integrability, as proven in \cite{Ra.Iso.Dru.IntegO2.26}. 
A pole--dipole model carried to $O(\text{spin}^2)$ and one that adds the spin-induced quadrupole at the same order may thus reach opposite conclusions on the very existence of that integral, and hence on whether the orbits are regular or chaotic. In particular, the former can report chaos precisely where the latter, being integrable to that order, predicts none (or a weaker form thereof).
It is in this sense that only a multipole expansion truncated at a fixed, self-consistent order in the spin is physically reliable, and it is that expansion we adopt below. 

\subsection{The dipolar, linear-in-spin system} 
\label{sec:MPTD}

For the aforementioned reasons, we do not retain the non-linear-in-spin sector of the full
pole--dipole MPTD system, and instead linearize all expressions in the spin
tensor $S^{ab}$. This is precisely
the order-consistent truncation discussed above: neglecting the higher multipoles,
as the pole--dipole approximation does, simultaneously discards non-linear-in-spin
contributions of the same order --- those sourced by the force and torque terms on
the right-hand side of Eqs.~(2.2) --- so retaining the dipolar non-linearities
alone would call for a physical justification that is generally not met for
astrophysical compact objects. We make this quantitative in Paper~III, where the
higher multipole moments of compact objects are shown to source non-linear-in-spin
effects comparable to, and at times larger than, any nominal dipolar non-linearity.

We will also take the (so far generic) tangent vector $v^a$ to be the four-velocity $u^a$, such that $u^a u_a = -1$ and the associated parameter is the proper time $\tau$. Applying these assumptions (linearization and 4-velocity) to the MPTD equations and the momentum-velocity relation results in the \emph{linear-in-spin, dipolar MPTD equations}
\begin{subequations}\label{EElin}
	\begin{align}
        \dot{p}_a    & = R_{abcd} S^{bc} u^d, \label{EEp} \\
		\dot{S}^{ab} & = 0, \label{EES} \\
        p^a &= \mu u^a. \label{momvellin}
 	\end{align}
\end{subequations}
This system of equations has been the subject of many studies in the past, and the leading, linear-in-spin effects thoroughly investigated, especially in black hole spacetimes (see the introduction for references). From now on, we will consider this system of equations, and discard any term in the development that involves two or more products of the spin tensor. 

\section{The MPTD equations as a constrained Poisson system} 
\label{sec:Ham}

In this section, we examine the linear-in-spin MPTD system \eqref{EElin} and write it as a Poisson system (a particular type of Hamiltonian systems which comes with degeneracies, see App.~\eqref{Poi} for details about different types of Hamiltonian systems.) First, the Poisson system is defined in Sec.~\ref{subsec:PoissonMPTD} and the resulting equations of motion are derived in Sec.~\ref{sec:EoMPoisson}. Using a change of phase space coordinates in Sec.~\ref{sec:coordPoisson}, the Poisson system is shown to be degenerate (non-symplectic) in Sec.~\ref{sec:degPoisson}, and the degeneracies are lifted accordingly.

\subsection{Definition of the 14-dimensional Poisson system} 
\label{sec:defPoisson}

We wish to turn the system \eqref{EElin} into a \textit{Poisson system} in the general sense, i.e., as defined from the three following ingredients: (i) a \textit{phase space $\mcM$}, defined as a $N$-dimensional manifold endowed with $N$ coordinates $y:=(y^1,\ldots,y^N)\in\RR^N$ ; (ii) a \textit{Poisson structure $\Lambda$}, defined as a skew-symmetric $N\times N$ matrix whose entries, denoted $\Lambda^{ij}(y)$, satisfy the Jacobi identity $\Lambda^{\ell(i}\partial_\ell\Lambda^{jk)}=0$, with $\partial_\ell:=\partial/\partial y^\ell$ ; (iii) a \textit{Hamiltonian $H$}, defined as a scalar field $\mcM\rightarrow \RR$.
The triplet $(\mcM,\Lambda,H)$ is called a \textit{Poisson system}. If the matrix $\Lambda(y)$ has maximal rank for all $y\in\mcM$, one speaks of a \textit{symplectic} structure. If not, it is said to be \textit{degenerate}. 
Regardless of its degeneracy, the Poisson structure $\Lambda$ defines a \textit{Poisson bracket} via:
\begin{equation} \label{PBnonsymp}
\{F,G\} := \sum_{i,j} \,\Lambda^{ij}(y) \,  \frac{\partial F}{\partial y^i} \, \frac{\partial G}{\partial y^j} \,,
\end{equation}
for any $y$-dependent functions $F,G$. While the geometry on $\mcM$ is fixed by $\Lambda$, the physics rely on a choice of Hamiltonian $H$. The latter defines a specific ``law of evolution'' on $\mcM$, in the sense that the evolution of any function $F$ of phase space satisfies \textit{Hamilton's generalized equation}: 
\begin{equation} \label{HamEq}
\frac{\ud F}{\ud \lambdaH} = \{F,H\} \,, 
\end{equation}
where $\lambdaH$ is the evolution parameter uniquely associated to the Hamiltonian $H$. When $F$ in \eqref{HamEq} is replaced by the coordinates $y^i$ covering $\mcM$, solutions to the resulting ODE system describe a preferred set of curves in $\mcM$ that are level curves of $H$. In particular, combining \eqref{HamEq} and \eqref{PBnonsymp} gives the ODEs describing these curves: 
\begin{equation} \label{geny}
    \frac{\ud y^i}{\ud \lambdaH} = \sum_{j=1}^N \,\Lambda^{ij}(y) \, \frac{\partial H}{\partial y^j}.
\end{equation}

\subsubsection{A 14-dimensional formulation of the MPTD equations} \label{subsec:PoissonMPTD}

Let us now show that the MPTD system \eqref{EElin} can be written as a Poisson system like \eqref{geny} in $N=14$ dimensions. 
The phase space $\mcM = \RR^{14}$ is endowed with coordinates 
\begin{equation} \label{phasespaceM}
    y := \left( x^\alpha,p_\alpha,S^{\alpha\beta} \right) 
    \in
    \RR^{4}\times\RR^{4}\times\RR^{6}
\end{equation}
that coincide physically with their covariant spacetime definitions. 
The Poisson structure $\Lambda$ is defined uniquely through a given set of Poisson brackets between the coordinates, since Eq.~\eqref{PBnonsymp} implies $\{y^i,y^j\}=\Lambda^{ij}(y)$. The non-vanishing ones are given by \cite{Souri.70,Kun.72,dAKuvHo.15}
\begin{subequations}\label{PBs}
    \begin{align}
    &\{x^\alpha,p_\beta\} = \delta^\alpha_\beta \,, \\
    &\{p_\alpha,p_\beta\} =-\tfrac{1}{2}R_{\alpha\beta\gamma\delta}S^{\gamma\delta}\,, \\
    &\{p_\alpha,S^{\beta\gamma}\} =  2\Gamma^{[\gamma}_{\delta\alpha} S^{\beta]\delta} \,,  \\ 
    &\{S^{\alpha\beta},S^{\gamma\delta}\} = 2(g^{\alpha[\delta}S^{\gamma]\beta} 
    +g^{\beta[\gamma}S^{\delta]\alpha}) \,, \label{spinsss}
    \end{align}
\end{subequations}
where $\delta^\alpha_\beta$ is the 4D Kronecker symbol. Notice that the RHSs of \eqref{PBs} are independent of $p_\alpha$. To our knowledge, there does not exist a Hamiltonian that, when varied with respect to the brackets \eqref{PBs}, generates the covariant, background-independent, multipolar MPTD equations \eqref{EEgen}.\footnote{We may even go further and claim that there \emph{cannot} be such a Hamiltonian in general. Indeed, 
the Poisson brackets \eqref{PBs} automatically imply the conservation of the spin norm $S_\circ$ (cf. Sec.~\ref{subsec:casimir}) \emph{regardless of the hamiltonian}, whereas the full MPTD system \eqref{EEgen}-\eqref{TDSSC} only implies that $S_\circ$ is conserved if $N^{ab} S_{ab}=0$, cf. \eqref{mudotSdot}. One can certainly imagine situations where this does not hold. In such a case, the brackets \eqref{PBs} cannot lead to the correct equations of motion.} However, when no force and torques are present, it is possible to find such a Hamiltonian. In the non-spinning case, the dynamics are govern by the geodesic equations, which are well-known to be a Hamiltonian system, with Hamiltonian 
\begin{equation} \label{Hgeo}
    H_\text{geo}(x,p)=\frac{1}{2}g^{\alpha\beta}p_\alpha p_\beta,
\end{equation}
and coordinates $(x^\alpha,p_\beta)$ treated as canonical pairs. Numerically, the Hamiltonian is linked to the  dynamical mass $\mu$ of the object through $H=-\mu^2/2$ (cf.~\eqref{Hgeo} and \eqref{eq:def-norms}).

When linear-in-spin effects are added, the MPTD equations are not geodesic, but the mass $\mu$ is still conserved up to non-linear terms in spin. Therefore, a good candidate for a linear-in-spin Hamiltonian is still \eqref{Hgeo}, which we now write as
\begin{equation} \label{Hlinspin}
    H_\text{lin}(x,p,S) = \frac{1}{2}g^{\alpha\beta}p_\alpha p_\beta,
\end{equation}
to emphasize that, even though $H_\text{lin}$ does not depend explicitly on the spin coordinates $S^{\alpha\beta}$, it is defined on the phase space \eqref{phasespaceM}, and spin-dependent terms in the equations of motion can still arise from the non-canonical Poisson brackets \eqref{PBs}, even though $\partial H_{\text{lin}}/\partial S^{\alpha\beta}=0$.

We also mention that the full pole-dipole Hamiltonian (not truncated in spin) was provided in \cite{WiStLu.19}. Slightly adapting to our purposes, it becomes
\begin{equation}
   \label{fullTDdipo}
        H_\text{dipo}(x,p,S) = \frac{1}{2} g^{\alpha\beta} p_\alpha p_\beta + V_\alpha S^{\alpha\beta} p_\beta \,, \quad \text{with} \quad V_\alpha =  \frac{2S^{\beta\gamma}R_{\beta\gamma\alpha}^{\phantom{sbn}\delta}p_\delta}{ R_{\alpha\beta\gamma\delta}S^{\alpha\beta}S^{\gamma\delta}-4g^{\alpha\beta}p_\alpha p_\beta} \,,
\end{equation}
with the non-linear-in-spin correction that encodes all the non-linearities coming from the dipole sector. We do not need them here, but exploit this expression in Paper III \cite{Ra.Iso.Dru.IntegO2.26} where quadratic-in-spin terms are (partially) constructed from this particular Hamiltonian. 

\subsubsection{Hamilton's equations on $\mcM$} 
\label{sec:EoMPoisson}

To verify that \eqref{Hlinspin} does produce the linear-in-spin MPTD equations with the TD SSC, we compute Hamilton's (generalized) equations \eqref{HamEq} with $F\in(x^\alpha,p_\alpha,S^{\alpha\beta})$. Consider for example the equation of motion for $p_\alpha$, which we detail below as a prototype of calculations involved in this work. One has 
\begin{subequations} \label{Hamexample}
    \begin{align}
        \frac{\ud p_\alpha}{\ud\lambdaH} &= \{p_\alpha,H_\text{lin}\} \\
        &= \{p_\alpha,x^\beta\}\frac{\partial H_\text{lin}}{\partial x^\beta} +\{p_\alpha,p_\beta\}\frac{\partial H_\text{lin}}{\partial p_\beta}+\{p_\alpha,S^{\beta\gamma}\}\frac{\partial H_\text{lin}}{\partial S^{\beta\gamma}}\\
        &= -\frac{\partial H_\text{lin}}{\partial x^\alpha} -\frac{1}{2}R_{\alpha\beta\gamma\delta}S^{\gamma\delta}\frac{\partial H_\text{lin}}{\partial p_\beta}+2\Gamma^{[\gamma}_{\delta\alpha} S^{\beta]\delta}\frac{\partial H_\text{lin}}{\partial S^{\beta\gamma}} \\
        &= \Gamma_{\alpha\gamma}^{\beta}p_\beta p^\gamma -\frac{1}{2}R_{\alpha\beta\gamma\delta}S^{\gamma\delta}p^\beta, \label{Hameqp}
    \end{align}
\end{subequations}
where, step-by-step, we have used: Hamilton's equation \eqref{HamEq}, the Leibniz rule, the Poisson brackets \eqref{PBs}, the partial derivatives of \eqref{Hlinspin} and the formula $\partial_\alpha g^{\beta\gamma}=-2\Gamma_{\alpha\delta}^{(\beta}g^{\gamma)\delta}$. Crucially, one does not know \emph{a priori} what the evolution parameter $\lambdaH$ is on the left-hand side (LHS) of this equation. To find it, one must compare the evolution equation to the MPTD equation \eqref{EElin}, which reads, in coordinates
\begin{equation} \label{pdottest}
    u^\beta \nabla_\beta p_\alpha = -\frac{1}{2}R_{\alpha\delta\beta\gamma}S^{\beta\gamma}u^\delta 
    \quad \Longleftrightarrow \quad 
    \frac{\ud p_\alpha}{\ud \hat{\tau}} - p^\beta \Gamma_{\beta\alpha}^\gamma p_\gamma 
    = -\frac{1}{2}R_{\alpha\delta\beta\gamma}S^{\beta\gamma}p^\delta,
\end{equation}
where the second form is obtained by multiplying by the constant $\mu$, using $p_a=\mu u_a$ and the fact that $p^a$ is tangent to the worldline with associated parameter $\hat{\tau}=\tau/\mu$. The equation in \eqref{pdottest} coincides with Hamilton's equation \eqref{Hameqp} if and only if the evolution parameter $\lambdaH$ verifies 
\begin{equation} \label{lamtaumu}
    \lambdaH = \tau/\mu.
\end{equation}
This is the same parameter that is associated to the geodesic Hamiltonian \eqref{Hgeo} \cite{Schm.02,HinFle.08}. Note that, since $\mu$ is a constant of motion at linear order in spin, $\lambdaH=\tau/\mu$ is an affine. 

Using the same recipe as in \eqref{Hamexample} ---Leibniz rule, partial derivatives of \eqref{Hlinspin} and Poisson brackets \eqref{PBs}--- Hamilton's equations for $x^\alpha$ and $S^{\alpha\beta}$ read
\begin{subequations} \label{xSdottest}
    \begin{align}
        \frac{\ud x^\alpha}{\ud\hat{\tau}} &= \{x^\alpha,H_\text{lin}\} =\{x^\alpha,p_\beta\}\frac{\partial H_\text{lin}}{\partial p_\beta} = \frac{\partial H_\text{lin}}{\partial p_\alpha} = p^\alpha, \label{xdot}\\
        \frac{\ud S^{\alpha\beta}}{\ud\hat{\tau}} &= \{S^{\alpha\beta},H_\text{lin}\} = \{S^{\alpha\beta},p_\gamma\}\frac{\partial H_\text{lin}}{\partial p_\gamma} = -\{p_\gamma, S^{\alpha\beta}\} p^\gamma = -2\Gamma_{\delta\gamma}^{[\beta}S^{\alpha]\delta}p^\gamma, \label{Sdot}
    \end{align}
\end{subequations}
Given \eqref{lamtaumu}, we recognize in equation \eqref{xdot} the momentum-velocity relation \eqref{momvellin}, and in equation \eqref{Sdot} the parallel transport of $S^{ab}$, i.e., equation \eqref{EES}. Hamilton's equations \eqref{Hamexample} and \eqref{xSdottest} are thus equivalent to the linear in spin MPTD equations \eqref{EElin}.

\subsubsection{Alternative coordinates on $\mcM$} \label{sec:coordPoisson}

The phase space coordinates $(x^\alpha,p_\alpha,S^{\alpha\beta})$ are useful to prove that the Hamiltonian system \eqref{Hlinspin}-\eqref{PBs} is equivalent to the original system \eqref{EElin}. It is also relevant for interpretation purposes. However, when it comes to reducing the system and to doing Hamiltonian mechanics \textit{per-se}, they are less practical. Let us introduce a new system of coordinates on $\mcM$ which still retains some physical interpretation 
while also admitting drastically simpler Poisson brackets. These coordinates are well-known in the literature, we only give a review and complete with additional details.

First, we require some geometrical construction within the spacetime $(\scE,g_{ab})$. Consider an orthonormal tetrad field\footnote{In the notation $(e_A)^\alpha$, $A$ labels the four different vectors, and $\alpha$ the four components of the $A$-th vector.} $\{(e_A)^a\}_{A\in\{0,\ldots,3\}}$, whose components in the natural basis are $(e_A)^\alpha$. Let $\omega_{aBC}:= g_{bc}(e_B)^b \nabla_a (e_C)^c$ be the connection 1-forms associated to the tetrad (we follow Sec.~3.4b of \cite{Wald} for orthonormal tetrad results and notations). Since $\omega_{aBC}=-\omega_{aCB}$, there are six independent such 1-forms. Their components in the natural basis, $\omega_{\alpha BC}$, are the connection coefficients and those in the tetrad, $\omega_{ABC}$, are the so-called Ricci rotation coefficients. As before, all these objects are defined throughout spacetime, and become simple functions of the phase space coordinates $x^\alpha$ in the phase space picture. New coordinates on $\mcM$ are then obtained by (i) keeping the coordinates $x^\alpha$, (ii) replacing the linear momentum variables $p_\alpha$ by $\pi_\alpha$, a combination of linear and angular momentum, and (iii) replacing $S^{\alpha\beta}$ by $S^{AB}$, the tetrad components of the spin tensor. Explicitly, we define these new coordinates $(x^\alpha,\pi_\alpha,S^{AB}$ by setting
\begin{subequations} \label{pStopiS}
    \begin{align}
    p_\alpha &= \pi_\alpha +\frac{1}{2} \omega_{\alpha BC} S^{BC}, \label{ptopi}\\
    S^{\alpha\beta} &= S^{AB} (e_A)^\alpha (e_B)^\beta .\label{StoS}
    \end{align}
\end{subequations}
Solving these equations for $\pi_\alpha$ and $S^{AB}$ shows that this coordinate change is indeed invertible. The Poisson brackets of the new coordinates can then be computed from the rule \eqref{PBnonsymp} and the old brackets \eqref{PBs}. This calculation does not appear in the literature, to our knowledge, so we have detailed it in App.~\ref{app:xpstospis}. It is lengthy but straightforward, using not so well-known formulae involving the connection 1-forms (in particular Eq.~(3.4.20) of \cite{Wald}). The non-vanishing Poisson brackets are found to be:
\begin{subequations} \label{very}
    \begin{align}
        \{x^\alpha,\pi_\beta\} &= \delta^\alpha_\beta \,, \\
        \{S^{AB},S^{CD}\} &=2\eta^{A[D}S^{C]B} + 2\eta^{B[C}S^{D]A}  \label{very2}\,.
    \end{align}
\end{subequations}
Note that, even though the six $S^{AB}$ do not verify the canonical Poisson brackets, the four pairs $(x^\alpha,\pi_\alpha)$ do, and have vanishing Poisson brackets with $S^{AB}$. The spin variables $S^{AB}$ have Poisson brackets akin to the commutators of the so(1,3) algebra (see, e.g., Chap.~7 in \cite{GouSR}). This is reminiscent of the definition of $S^{AB}$ as the components in an orthonormal tetrad field, the latter forming a set left invariant by the Lorentz group SO(1,3). This property is even more clear when we use, instead of the six independent $S^{AB}$, the $3+3$ variables $(S^1,S^2,S^3,D^1,D^2,D^3)$ defined by 
\begin{equation} \label{defSD}
S^I=\tfrac{1}{2}\varepsilon^{I}_{\phantom{I}JK}S^{JK} \quand D^I = S^{0I} ,
\end{equation}
where $\varepsilon_{IJK}$ is the 3D Levi-Civita symbol, equal to $1$ (resp. $-1$) when $IJK$ is an even (resp. odd) permutation of $123$, and $0$ otherwise. Although they are but notations in phase space, $S^I$ and $D^I$ have a clear physical meaning: they are the tetrad components of the Euclidean spin and mass dipole vectors as defined in Eq.~\eqref{zip!} (i.e., with $(e_0)^a$ replacing $f^a$ there). They are \emph{not} the tetrad components of $S_\tTD^a,D_\tTD^a$. The brackets \eqref{very2}, with these notations, take their simplest form
\begin{subequations} \label{PBsSD}
    \begin{align}
    \{x^\alpha,\pi_\beta\} &= \delta^\alpha_\beta ,\label{bktzpi}\\
    \{S^I,S^J\} &= \varepsilon^{IJ}_{\phantom{IJ}K} S^K , \label{bktSS} \\
    \{D^I,S^J\} &= \varepsilon^{IJ}_{\phantom{IJ}K} D^K, \label{bktDS}\\
    \{D^I,D^J\} &= - \varepsilon^{IJ}_{\phantom{IJ}K} S^K .\label{bktDD} 
    \end{align}
\end{subequations}
The brackets \eqref{PBsSD} are identical to the commutators of the generators for the Lie algebra so(1,3) \cite{Gourgoulhon}. We also stress that brackets \eqref{PBsSD} are equivalent to \eqref{very} (through Eq.~\eqref{defSD}), it is a renaming more than a coordinate transformation. 

Although it will not be needed in this work, one may express the original Hamiltonian \eqref{Hlinspin} in terms of the new coordinates $(x^\alpha,\pi_\beta,S^{I},D^J)$ on $\mcM$, by inserting Eqs.~\eqref{pStopiS} into Eq.~\eqref{Hlinspin}. One obtains
\begin{equation} \label{HbeginSD}
H =\frac{1}{2}g^{\alpha\beta}\pi_\alpha\pi_\beta + \frac{1}{2} g^{\alpha\beta}\pi_\alpha\omega_{\beta CD}S^{CD}. 
\end{equation}
Even though \eqref{HbeginSD} calls for a rather natural identification into a ``non-spinning'' term $\tfrac{1}{2}g^{\alpha\beta} \pi_\alpha \pi_\beta$ corrected by a linear-in-spin term, the latter should be interpreted with care, as there are spin-related quantities hidden in $\pi_\alpha$, cf. Eq.~\eqref{pStopiS}. We will avoid any such interpretation, and rather treat the whole Hamiltonian as one describing an object with both linear and angular momentum (orbital and spin) intertwined. Note however, that this Hamiltonian \eqref{HbeginSD} is very practical to work with, and used often in the literature \cite{WitzHJ.19,WitzAA.22,Pio.25,Witzany:2026eqc}. 

\subsection{12-dimensional symplectic system $\mcN$} 
\label{sec:degPoisson}

We now focus on the Poisson brackets \eqref{PBsSD} between the new coordinates $y^{i} := (x^\alpha,\pi_\alpha,S^I,D^I)$. Following the notations of App.~\ref{sec:Poi}, we are in the case $N=14$ and we construct the Poisson matrix $\Lambda^{ij}(y)=\{y^i,y^j\}$, where $(y^1,\ldots,y^{14})$ denote the coordinates $y^{i}$, in that order. 

\subsubsection{Degeneracies of the Poisson structure}

Directly reading through Eqs.~\eqref{PBsSD}, we find that the matrix $\Lambda$ is block diagonal, and given by $\Lambda = \text{diag}(\mathbb{J}_8,\mathfrak{S})$, 
where
\begin{equation}
\mathbb{J}_8=\left( \begin{array}{cc}
  0 & \mathbb{I}_4 \\
-\mathbb{I}_4 &0 
\end{array} \right) \,, \quad 
\mathfrak{S}=\left( \begin{array}{cc}
  \mathcal{S} & \mathcal{D} \\
\mathcal{D} & -\mathcal{S} 
\end{array} \right) \,.
\end{equation}
In the above equations, $\mathbb{I}_4$ it the $4\times 4$ identity matrix, so that $\mathbb{J}_8$ is the canonical $8\times 8$ Poisson matrix (it would be the Poisson matrix of four pairs of canonical coordinates in an 8-dimensional symplectic system), while $\mathfrak{S}$ is a $6\times 6$ antisymmetric matrix constructed from two SO(3) matrices associated to $S^I$ and $D^I$, namely
\begin{equation}
\mathcal{S}=\left( \begin{array}{ccc}
  0 & S^3 & -S^2 \\
  -S^3 & 0 & S^1 \\
  S^2 & -S^1 & 0 \\
\end{array} \right)
\quand
\mathcal{D}=\left( \begin{array}{ccc}
  0 & D^3 & -D^2 \\
  -D^3 & 0 & D^1 \\
  D^2 & -D^1 & 0 \\
\end{array} \right) .
\end{equation}

The matrix formulation of the Poisson structure $\Lambda$ permits an easy calculation of its rank at each point of $\mcM$. Direct inspection reveals that $\text{rank}(\mathbb{J}_8)=8$ and $\text{det}(\mathfrak{S})=4$. Since $\Lambda=\text{diag}(\mathbb{J}_8,\mathfrak{S})$, we obtain
\begin{equation}
\text{rank}(\Lambda)=12 < 14 = \text{dim}(\mcM) \,.
\end{equation}
Therefore, as explained in App.~\ref{sec:Poi}, the Poisson structure generated by the brackets of the coordinates \eqref{PBsSD} is \textit{degenerate}. Consequently, there exists exactly $\text{dim}(\mcM)-\text{rank}(\Lambda)=14-12=2$ \emph{Casimir invariants} for this structure. We note that, since the rank (and the vanishing of the determinant) are invariant under  diffeomorphisms, these features (degeneracy and existence of two Casimirs) also hold for the initial brackets \eqref{PBs}.

\subsubsection{Casimir invariants} 
\label{subsec:casimir}

The brackets \eqref{PBsSD} admit two independent Casimir invariants, which can be explicitly computed\footnote{For example, using the Killing form of the $\mathfrak{so}(p,q)$ algebra \cite{Bau.09}, or computing of the null-space of $\Lambda$, of which the gradients of Casimirs form a basis, cf. Eq.~\eqref{PBnonsymp},.}. 
Following tradition \cite{WiStLu.19}, we shall take
\begin{equation} \label{Casimirnew}
    \mathcal{C}_{\circ} :=\vec{S}\cdot\vec{S} - \vec{D}\cdot\vec{D} \quad \text{and} \quad \mathcal{C}_{\star} := \vec{S}\cdot\vec{D} \,,
\end{equation}
where we have used the Euclidean notations $\vec{S}=(S^1,S^2,S^3)$ and $\vec{D}=(D^1,D^2,D^3)$. We can verify directly with the brackets \eqref{PBsSD} and the Leibniz rule that they are Casimirs, i.e., that any functions $F := F(x,\pi,S,D)$ defined on the phase space $\mcM$, 
\begin{equation}
    \forall F \in \mathcal{C}^{\infty}(\mcM)\,, \quad
    \{F,\mathcal{C}_{\circ}\}=0 \quand \{F,\mathcal{C}_{\star}\}=0\,.
\end{equation}
From a physical point of view, these Casimirs are easily interpreted: re-writing Eqs.~\eqref{Casimirnew} in terms of the variables $S^{AB}$, we obtain 
\begin{equation} \label{Casimir}
\mathcal{C}_{\circ} =\frac{1}{2}\eta_{AB}\eta_{CD}S^{AC}S^{BD} \quad \text{and} \quad \mathcal{C}_{\star} = \frac{1}{8}\varepsilon_{ABCD}S^{AB}S^{CD} \,.
\end{equation}
Notice that \eqref{Casimir} coincide with the covariant definition \eqref{eq:def-norms} expanded in an orthonormal tetrad. Therefore, the Casimirs are numerically equal to the (squares of the) covariant spin norms:
\begin{equation}\label{eq:Casmir_vs_Snorm}
    \mathcal{C}_{\circ} (S,D) = S_\circ^2 
    \quand 
    \mathcal{C}_{\star} (S, D) = S_\star^2 \,,
\end{equation}
justifying the notation of the Casimirs. This result should be understood with care. The Poisson structure defined by the brackets \eqref{PBsSD} (or equivalently \eqref{PBs}) automatically implies the conservation of both spin norms $S_\circ^2$ and $S_\star^2$, as these are Casimirs $(\mathcal{C}_{\circ},\mathcal{C}_{\star})$ of the structure. As pointed out in  \cite{WiStLu.19}, these spin scalars are not constant along the worldline $\scL$ for the general MPTD system, but they are conserved under the dipolar MPTD system with the TD SSC (regardless of linearization in spin). Indeed, $p_a S^{ab}=0$ readily implies $S_\star=0$ from Eq.~\eqref{spinnorms} and $S_\circ=\text{cst}$ from Eq.~\eqref{mudotSdot}.

\subsubsection{Symplectic leaves} 
\label{sec:sympleaves}

As reviewed in App.~\ref{sec:Poi}, the Poisson brackets \eqref{PBsSD} define a  degenerate structure on the 14-dimensional manifold $\mcM$. Yet $\mcM$ is foliated  by non-degenerate submanifolds --- the so-called \emph{symplectic leaves} $\mcN$ --- defined as the level sets of the two Casimirs. Setting both Casimir invariants \eqref{Casimirnew} to fixed values $S^2_\circ$ and $S^2_\star$, the symplectic leaf $\mcN$ is the 12-dimensional submanifold defined by the two independent algebraic equations~\eqref{eq:Casmir_vs_Snorm}
where these relations are algebraic constraints between the coordinates $(S, D)$ on $\mcM$. The manifold $\mcN$ is symplectic and can, at least locally, be endowed  with canonical coordinates, discussed at length in \cite{Ra.AJP.26}. On $\mcN$, the values $S_\circ, S_\star$ are pure parameters rather than functions of the phase space coordinates as they were on $\mcM$. This distinction matters  when classifying conserved quantities: \emph{Casimirs} are conserved for any Hamiltonian on a degenerate Poisson manifold, whereas \emph{first integrals} are Hamiltonian-dependent invariants of a symplectic system. The difference is  especially relevant when counting independent integrals of motion to assess integrability.

Let us now discuss the Poisson structure on a given leaf $\mcN$. Since $\mcN$ is  defined by holding both Casimirs \eqref{Casimirnew} constant, we are dealing with  a constrained Hamiltonian system. By the standard theory of constrained systems (reviewed in App.~\ref{sec:Poi}), the Poisson structure $\Lambda^\mcN$ on $\mcN$ is the restriction of $\Lambda$ to $\mcN$, and in general  $\{~,~\} \neq \{~,~\}^\mcN$. We will encounter this explicitly in  Eq.~\eqref{Diracall} of Sec.~\ref{sec:consinv}. When the constraints are Casimir  invariants, however, a classical result \cite{Derigl.22} shows that $\Lambda^\mcN$  and $\Lambda$ are essentially identical, which follows directly from the Casimirs  commuting with everything on $\mcM$. Therefore, the 14D Hamiltonian system  $(\mcM,\Lambda,H)$, restricted to a leaf $\mcN$, becomes the 12D system $(\mcN,\Lambda,H_\mcN)$, where $\mcN$ is coordinatised by any invertible subset of 12 coordinates from $(x^\alpha,\pi_\alpha,S^I,D^I)$, and the original brackets \eqref{PBsSD} still govern the evolution on $\mcN$,
\begin{equation}
     \forall (F, G) \in \mathcal{C}^{\infty}(\mcN)\,, \quad
     \{F,G\}^\mcN=\{F,G\}.
\end{equation}
In practice, it is more convenient to work with 12 canonical coordinates on $\mcN$ directly (cf. ~\cite{Ra.Iso.PapII.24,Ra.AJP.26}), rather than selecting an arbitrary subset from those on $\mcM$. 
Hereafter we use the same symbols $\{\,,\}$ and $H$ on both $\mcM$ and $\mcN$; the number of phase-space variables (14 or 12) resolves any ambiguity.~\footnote{Any occurrence of the two  non-chosen variables in $H_\mcN$ (or anywhere else) can be expressed in terms of the 12 chosen ones by inverting the two Casimir relations \eqref{Casimirnew}.}

\subsection{Reduction to the 10D physical phase space $\mcP$ via the TD SSC}
\label{sec:consinv}

We have reduced the original 14D phase space $\mcM$ to a 12D symplectic leaf $\mcN$, where the Poisson structure is non-degenerate. However, the linearized MPTD\,+\,TD SSC system \eqref{EElin} contains only 10 independent unknowns, not 12. The remaining reduction comes from enforcing the TD SSC as a set of algebraic constraints on $\mcN$, yielding a 10D physical phase space that we call $\mcP$. With $H$ and its evolution parameter already fixed, we now explain how to properly impose the TD SSC at the Hamiltonian level using the so-called Dirac brackets, and how $\mcP$ emerges geometrically.\footnote{All our results on constrained Hamiltonian systems follow Sec.~I.5 of \cite{Arn} (see also \cite{Derigl.22}).} A summary of the full reduction $\mcM\rightarrow\mcN\rightarrow\mcP$ is given in Sec.~\ref{subsec:blabla}. The exposition here is kept general, and we refer to our work \cite{Ra.Iso.PapII.24} for a concrete application of the framework to linear-in-spin dynamics in the Schwarzschild spacetime.

\subsubsection{Constraint surface and the physical phase space $\mcP$}
\label{sec:sscnote}

On $\mcN$, the canonical variables are \emph{independent}; in particular, they do not satisfy the TD SSC. Projecting Eq.~\eqref{TDSSC} onto the tetrad, the TD SSC reads
\begin{equation} \label{SSCappplied}
C^B(x,\pi,S,D) := \pi_A S^{AB} + \frac{1}{2}\omega_{ACD}S^{CD}S^{AB} = 0 \,,
\end{equation}
where $\pi_A:=(e_A)^\alpha\pi_\alpha$ is the tetrad-projected momentum, a function of $(x^\alpha,\pi_\alpha)$. Equation~\eqref{SSCappplied} defines a submanifold of $\mcN$, which we denote $\mcT$, on which the canonical variables are no longer independent. 

\begin{figure}[t!]
    \begin{center}
    	\includegraphics[width=0.8\linewidth]{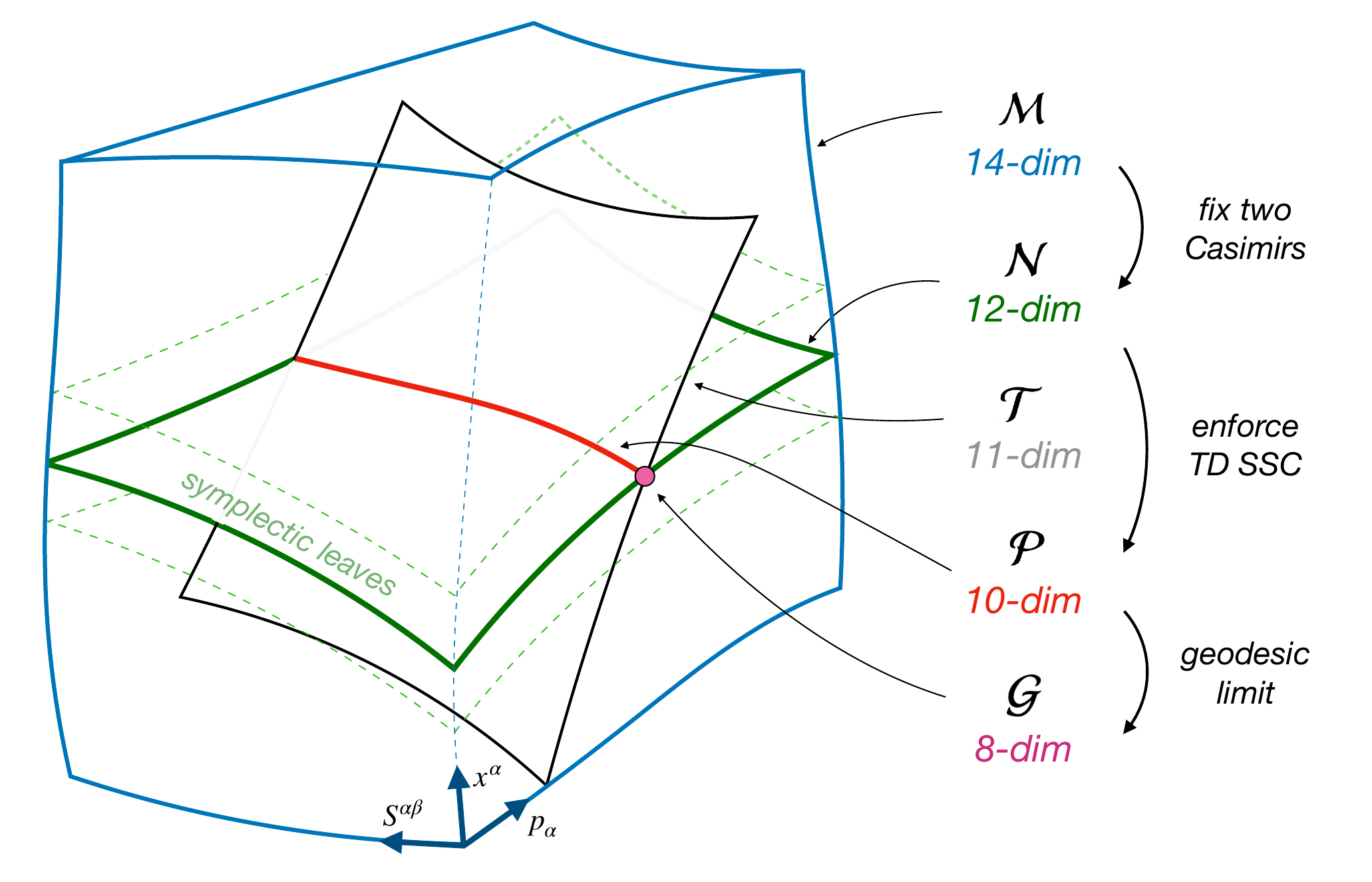}
        \caption{Different submanifolds are required to lift the degeneracies associated with the Casimir invariants ($\mcM\rightarrow\mcN$) and to correctly implement the TD SSC ($\mcN\rightarrow\mcT\rightarrow\mcP$). The well-known geodesic Hamiltonian \cite{Schm.02,HinFle.08} on the 8D phase space $(x^\alpha,p_\alpha)\in\mathcal{G}$ is recovered by restricting to the leaves $S_\circ=0$ and setting all spin variables to zero.}
        \label{fig:PS}
    \end{center}
\end{figure}

Let us now count the dimension of $\mcT$. Expanding \eqref{SSCappplied} in the Euclidean notation introduced earlier yields
\begin{subequations} \label{expandSSC}
    \begin{align}
       C^0 &= -\vec{\pi}\cdot\vec{D} = 0 \,, \label{fi} \\
       \vec{C} &= \pi_0 \vec{D} - \vec{\pi} \times \vec{S} = 0 \label{fj}\,.
    \end{align}
\end{subequations}
Equation~\eqref{fi} is a consequence of Eq.~\eqref{fj}, so only three of the four constraints are independent and $\text{dim}(\mcT) = 12-3 = 11$. Moreover, Eqs.~\eqref{fj} readily imply
\begin{equation} \label{sstar0}
    S_\star^2 = \vec{S}\cdot\vec{D} = 0,
\end{equation}
consistently with the general statement made below \eqref{spinnorms} that any SSC enforces $S_\star=0$. Therefore, only the symplectic leaves $\mcN$ with $\mcC_\star=0$ actually intersect $\mcT$. We work at this intersection, 
\begin{equation} \label{defP}
\mcP := \mcT \cap \mcN, \qquad \text{with} \quad (\mathcal{C}_{\circ},\mathcal{C}_{\star}) = (S_\circ,0),
\end{equation}
where the Poisson structure is non-degenerate (thanks to $\mcN$) and the TD SSC is automatically enforced (thanks to $\mcT$). On this submanifold, \emph{any} two of the four constraints \eqref{expandSSC} suffice to define $\mcP$, and consequently
\begin{equation} \label{dimP}
\text{dim}(\mcP) = 10,
\end{equation}
matching the number of fundamental degrees of freedom of the linearized MPTD\,+\,TD SSC system. We call $\mcP$ the \emph{physical} phase space, as all physically relevant solutions of \eqref{EElin} lie within it. The reduction $\mcM\rightarrow\mcN\rightarrow\mcP$ is summarized geometrically in Fig.~\ref{fig:PS}. 

\subsubsection{Stability of $\mcP$ under the Hamiltonian flow}
\label{sec:stabflow}

For Hamiltonian mechanics on $\mcP$ to make sense, $\mcP$ must be invariant under the flow of $H$, i.e., any solution to Hamilton's equations originating in $\mcP$ must remain in $\mcP$ at all times. This stability condition is expressed as
\begin{equation} \label{PstableflowH}
    \text{$\mcP$ is stable under the flow of $H$} 
    \quad \Longleftrightarrow \quad 
    \{C^A,H\} = O(C^A),
\end{equation}
with $O(C^A)$ denoting terms vanishing on $\mcP$. We verify \eqref{PstableflowH} for our Hamiltonian \eqref{Hlinspin} by direct calculation. Using the Leibniz rule and the brackets \eqref{very}, one finds\footnote{This calculation is most easily performed using the intermediate brackets $\{x^\alpha,\pi_A\}=(e_A)^\alpha$ and $\{\pi_A,\pi_B\}=2\eta^{CD}\pi_D\omega_{C[AB]}$, which follow from Eqs.~\eqref{pStopiS} and \eqref{PBs}, together with the commutation relations for the tetrad vectors; see Eq.~(3.4.23) of \cite{Wald}.}
\begin{equation} \label{VflowH}
    \frac{\ud C^A}{\ud\hat{\tau}} = \{C^A,H\} = \Omega^A_{\phantom{A}B} \, C^B = O(C^A) \,,
\end{equation}
with $\Omega^A_{\phantom{A}B}:=\eta^{AD}\pi^C\omega_{CBD}$. This confirms \eqref{PstableflowH}: any trajectory of $H$ originating on $\mcP$ remains on $\mcP$. That $\mcT$ (and hence $\mcP$) be invariant under the flow of $H$ is, in fact, expected: the linearized MPTD system itself was derived under the TD SSC. From a purely computational viewpoint, one could stop here, evolve the dynamics under $H$ with initial conditions in $\mcP$, and obtain physically meaningful trajectories. However, to establish integrability in the Liouville-Arnold sense \cite{Arn,Liou1855}, the first integrals of motion must be in involution with respect to the symplectic structure on $\mcP$, not that on the larger $\mcN$. This requires knowing what the brackets on $\mcP$ are, to which we now turn.

\subsubsection{Dirac bracket on $\mcP$} 
\label{sec:ssV}

As established above, $\mcP$ is described by any two independent constraints out of the four in \eqref{expandSSC}. Let us denote these two by $C^\mcA$ and $C^\mcB$, where $\mcA,\mcB$ are two \emph{fixed} indices in $\{0,\ldots,3\}$ (as opposed to abstract indices $A,B,\ldots$, which can be summed over). Coordinates on $\mcP$ can be chosen by taking any 10 of the 12 coordinates on $\mcN$ (as we did when going from $\mcM$ to $\mcN$).

According to the classical theory of constrained Hamiltonian systems \cite{Derigl.22}, the projected Poisson structure $\Lambda_\mcP$ on $\mcP$ is well-defined provided that the brackets $\{C^\mcA,H\}$ and $\{C^\mcB,H\}$ both vanish on $\mcP$. As established in \eqref{VflowH}, this is indeed the case. However, well-definedness alone does not guarantee non-degeneracy: $\Lambda_\mcP$ is \emph{symplectic} if and only if the $2\times 2$ matrix $M^{\mcA\mcB}:=\{C^\mcA,C^\mcB\}$ is invertible on $\mcP$.\footnote{When there are more than two constraints, the analogous condition is that the matrix of pairwise brackets be invertible on the constraint surface \cite{Bro.22}.} Using the $\mcN$-brackets, we find
\begin{equation} \label{MAB}
 M^{\mcA\mcB} = \{ C^{\mcA}, C^{\mcB} \} = \eta^{CD}\pi_{C}\pi_{D}\, S^{\mcA \mcB} - 2 \pi^{[\mcA} C^{\mcB]} \,,
\end{equation}
which reduces, on $\mcP$, to $-\mu^2 S^{\mcA\mcB}$ up to higher-order-in-spin terms. Since $\mu$ is the particle's mass and $S^{\mcA\mcB}$ a non-vanishing tetrad component of the spin tensor, $M^{\mcA\mcB}$ is generically non-zero on $\mcP$. Therefore, $\Lambda_\mcP$ is symplectic, and the induced bracket on $\mcP$, hereafter denoted $\{,\}^\mcP$, is given by the Dirac formula \cite{Arn,Bro.22}\footnote{The formula \eqref{Diracall} was first introduced by Dirac \cite{Dirac.50} in the context of Lagrangian quantum mechanics. In our symplectic approach (no Lagrangian, no Legendre transform), this formula is a consequence of how Poisson structures are pulled back/pushed forward onto submanifolds, cf.~\cite{Zambon.11,Burs.13,Derigl.22}.}
\begin{equation} \label{Diracall}
\{F,G\}^\mcP := \{F,G\}|_\mcP - \{F,C^\mcA\}\, (M^{-1})_{\mcA\mcB}\, \{C^\mcB,G\}|_\mcP \,, 
\end{equation}
for any two functions $F,G$ defined on $\mcN$. On the right-hand side, the $\mcN$-brackets are computed first (e.g., using the 12 canonical coordinates on $\mcN$), and then the symbol $|_\mcP$ means ``project onto $\mcP$'', i.e., simplify the result using the constraints $C^A=0$. Notice that, by construction, setting $G=C^\mcC$ for any $\mcC\in\{\mcA,\mcB\}$ in \eqref{Diracall} yields
\begin{equation} \label{CisCasimirofDB}
    \forall F\in\mathcal{C}^{\infty}(\mcN)\,,\quad \{F,C^\mcC\}^\mcP = 0,
\end{equation}
as follows from $(M^{-1})_{\mcA\mcB}\,M^{\mcB\mcC} = \delta_\mcA^\mcC$. Equation~\eqref{CisCasimirofDB} simply expresses that the two chosen constraints are Casimirs of the Dirac bracket: this is how the TD SSC is enforced at the level of the Poisson structure.

Since we deal with only two constraints, the antisymmetric $2\times 2$ matrix $M$ has a single independent entry $\{C^\mcA,C^\mcB\}$ and can be written as $M = \{C^\mcA,C^\mcB\}\,\mathbb{J}_2$, where $\mathbb{J}_2$ is the canonical $2\times 2$ symplectic matrix $\left(\begin{smallmatrix} 0 & 1 \\ -1 & 0 \end{smallmatrix}\right)$. Formula~\eqref{Diracall} then simplifies to
\begin{equation} \label{Diracfinal}
    \{ F, G \}^\mcP = \{ F, G \}|_\mcP
    + 
    \frac{\{ F, C^\mcA \} \{ G, C^\mcB \} - \{ F, C^\mcB \} \{ G, C^\mcA \}}{\mu^2\, S^{\mcA\mcB}}\bigg|_\mcP \,,
\end{equation}
where $\mu^2 = -2H_\mcP$, $C^\mcA, C^\mcB$, and $S^{\mcA\mcB}$ are all evaluated as functions on $\mcP$. For practical examples of this formula and the overall framework, see our further work \cite{Ra.Iso.PapII.24} focused on the Schwarzschild spacetime.

\subsubsection{A practical bracket shortcut on $\mcP$}
\label{sec:shortcut}

We now record a crucial consequence of \eqref{Diracall} that will drastically simplify the forthcoming analysis. Consider a function $I\in\mathcal{C}^\infty(\mcN)$ such that $\{I,C^A\}=O(C^A)$, i.e., whose $\mcN$-bracket with the constraints vanishes on $\mcP$. Then the second term in \eqref{Diracall} is itself $O(C^A)$ and vanishes under the projection $|_\mcP$. Consequently,
\begin{equation} \label{KeyResult}
   \{I,C^A\} = O(C^A) \quad \Longrightarrow \quad \forall F\in\mathcal{C}^\infty(\mcN)\,,\quad \{F,I\}^\mcP = \{F,I\}|_\mcP.
\end{equation}
In words: the $\mcP$-bracket of any function $F$ with such an $I$ reduces to the much simpler $\mcN$-bracket evaluated on the constraint surface. A canonical example is the Hamiltonian itself: by Eq.~\eqref{VflowH}, $H$ satisfies the hypothesis of \eqref{KeyResult}, and therefore $\{F,H\}^\mcP = \{F,H\}|_\mcP$ for any $F$. Hamilton's equations on $\mcP$ can thus be computed using the $\mcN$-brackets and simplifying the result with the constraints, without ever invoking the second term in \eqref{Diracall}. Our integrability result in Sec.~\ref{sec:PBcalc} will rely on this shortcut, as several of the relevant first integrals satisfy the hypothesis of \eqref{KeyResult}.

\subsubsection{On the importance of using the correct brackets}
\label{sec:importance}

The shortcut \eqref{KeyResult} should not, however, mislead the reader into thinking that the $\mcN$- and $\mcP$-brackets are interchangeable. They are not, and substituting one for the other generally leads to inconsistent equations. To illustrate, consider the $\mcN$-bracket
\begin{equation} \label{CaCb}
  \{ C^{\alpha}, C^{\beta} \} 
  = 
  - \left( \mu^2 + \tfrac{1}{4} R_{\gamma\delta\rho\sigma} S^{\gamma\delta} S^{\rho\sigma} \right) S^{\alpha \beta} + O(C^{\alpha}),
\end{equation}
where $O(C^\alpha)$ denotes a term proportional to $C^\alpha$ irrelevant for this illustration. If one were to ``apply the SSC'' by naively setting $C^\alpha = 0$ in \eqref{CaCb}, one would conclude
\begin{equation} 
  \left( \mu^2 + \tfrac{1}{4} R_{\gamma\delta\rho\sigma} S^{\gamma\delta} S^{\rho\sigma} \right) S^{\alpha \beta} \overset{?}{=} 0,
\end{equation}
which is incorrect: the LHS is the product of $\mathfrak{m}^2$ (recall~\eqref{momvelTD}) and the spin tensor, neither of which vanishes in general. The resolution is that ``taking an $\mcN$-bracket" and ``enforcing the SSC" do \emph{not} commute. Only the $\mcP$-brackets, by construction, are compatible with the SSC. 

\subsection{Summary of the reduction $\mcM\rightarrow\mcN\rightarrow\mcP$} 
\label{subsec:blabla}

We can now summarize our final Hamiltonian formulation of the linearized MPTD\,+\,TD SSC equations, valid for any background metric. A detailed and illustrative application of this formalism to the Schwarzschild spacetime can be found in \cite{Ra.Iso.PapII.24}.

\begin{itemize}
    \item Endow $\mcP \simeq \RR^{10}$ with 10 coordinates chosen from the 12 canonical ones on $\mcN$. These coordinates all admit a natural physical interpretation in spacetime;
    \item Build the Hamiltonian $H_\mcP$ by expressing \eqref{HbeginSD} in terms of these 10 coordinates, resulting in a covariant expression (arbitrary spacetime coordinates and tetrad field);
    \item Choose two constraints $C^\mcA, C^\mcB$ from the TD SSC \eqref{expandSSC} and compute the $\mcP$-brackets via the Dirac formula \eqref{Diracfinal}, or via the shortcut \eqref{KeyResult} whenever its hypothesis is met.
\end{itemize}

The result is a well-defined, covariant, non-degenerate Hamiltonian system on the 10D phase space $\mcP$. Any physical solution to the MPTD\,+\,TD SSC system \eqref{EElin} corresponds to a trajectory in $\mcP$, with the TD SSC automatically enforced at each point. These trajectories obey Hamilton's equations
\begin{equation}
\frac{\ud F}{\ud \hat{\tau}} = \{F,H_\mcP\}^\mcP \,,
\end{equation}
where $F:\mcP\rightarrow\RR$ is any phase space function and $\hat{\tau}$ is the proper time per unit mass $\mu$. The formulation is covariant and automatically guarantees the conservation of the spin norm $S_\circ$ (as a Casimir on $\mcP$) and of the particle's mass, fixed by the value of the Hamiltonian via $\mu^2 = -2 H_\mcP$.

\section{Killing-Yano tensors and symmetries} 
\label{sec:KY}

In this section, we review general material on Killing-Yano tensors~\cite{Yano.52} in general relativity, keeping the presentation self-contained and tailored to the needs of subsequent sections. The literature on the subject is vast and scattered across several decades and communities~\cite{Carter.68,Ki.69,KiPhD.69,Coll.71,HughSomm.73,Coll.74,Hauser.I.75,Hauser.II.75,Coll.76,Steph.78,Rudiger.I.81,DiRu.I.81,DiRu.II.82,Rudiger.II.83,Wolf.98,JeziLuka.06,CookDray.09,AndBacTho.15,Frolov.17,Ha.20,LinSar.21,ComDru.22,ComDruVin.23,Ha.23}, and we refer to Paper III~\cite{Ra.Iso.Dru.IntegO2.26} for a more extensive discussion.

\subsection{Existence of a Killing-Yano tensor} \label{subsec:exsistence-KY}

As reviewed in the introduction, the existence of extra constants of motion, in addition to those arising from spacetime isometries, is well-known to be possible under the existence of Killing-St\"ackel (KS) tensors for geodesics and Killing-Yano (KY) tensors at linear order in spin. Since the latter implies the former, we make a simple and single assumption: that the spacetime under study admits a KY tensor $f_{ab}$. It is defined by the following two conditions \cite{Yano.52}
\begin{equation} \label{deff}
   f_{(ab)}=0 \quand \nabla_{(a}f_{b)c} = 0.
\end{equation}
The first condition means that $f_{ab}$ is antisymmetric and the second is its defining differential equation, called the Killing-Yano equation. We will further assume that $f_{ab}$ is \emph{non-degenerate}, in the sense of that
\begin{equation}\label{KYrank4}
    \varepsilon_{abcd}f^{ab}f^{cd} \neq 0.
\end{equation}
This condition means that the antisymmetric matrix $f_{\alpha\beta}$ is of rank 4 \cite{DiRu.I.81,DiRu.II.82}. If \eqref{KYrank4} does not hold, then $f_{ab}$ has rank 2, and the resulting constants of motion built from it will be degenerate with those from the spacetime isometries, which does not lead to integrability even for geodesics (see also the discussion in \cite{KamMar.86}).

The existence of a KY tensor in a given spacetime has major consequences on its geometry, in particular its symmetries and curvature. In particular, it implies that:
\begin{itemize}
    \item there exists of a Killing-St\"ackel (KS) tensor $K_{ab} := f_{ac}f^{c}_{\ph b}$,
    \item there exist two commuting Killing vectors $\xi_a := \tfrac{1}{3}\nabla^b f^\star_{ab}$ and $\eta^a := K^{ab} \xi_b$,
    \item the Weyl tensor $C_{abcd}$ is of Petrov type D (only the Weyl scalar $\Psi_2$ is nonzero),
    \item the Ricci tensor $R_{ab}$ is a linear combination of the KS tensor $K_{ab}$ and the metric $g_{ab}$. 
\end{itemize}
The first two results are the primary ingredient behind the integrability for the MPTD equations at linear order in spin proven in this paper (as well as quadratic order in spin, cf. \cite{Ra.Iso.Dru.IntegO2.26}). The last two results constrain the Riemann curvature tensor of a spacetime admitting a KY tensor to have only four independent degrees of freedom instead of twenty\footnote{In the sense of the Newman-Penrose formalism \cite{SKMHH,Chandra}, these are encoded in the scalars $(\Psi_2,\Phi_{11},R)\in\CC\times\RR^2$, see \cite{Ra.Iso.Dru.IntegO2.26} for details.}. We will \emph{not} need these results on spacetime curvature in this work limited to linear-in-spin integrability. However, they are fundamental for the quadratic-in-spin result of Paper III \cite{Ra.Iso.Dru.IntegO2.26}. 

The implications of the existence of a KY tensor are scattered in the relevant literature, and sometimes partially stated or with unnecessary assumptions. To compensate for this, and for the ease of reading, we will prove them in the following paragraphs, establishing important (and sometimes new) identities along the way. We also refer to Sec.~III. of Paper III \cite{Ra.Iso.Dru.IntegO2.26} for more identities, and alternative proofs using the NP formalism. 

\subsection{Derived Killing objects} \label{subsec:KYids}

We start with several definitions of the aforementioned Killing objects, and establish identities proving their Killing nature. We start with six important results, from which all subsequent ones will be derived. 

Most importantly, we will heavily rely on the Hodge dual of the KY tensor (or more generally of any bivector). It is defined as follows
\begin{equation}
    f_{ab}^\star := \frac{1}{2} \varepsilon^{ab}_{\phantom{ab}cd}f^{cd},
\end{equation}
where $\varepsilon_{abcd}$ is the totally antisymmetric tensor Levi-Civita tensor associated with the metric. More information on Hodge duality is gathered in App.~\ref{app:Hodge}.

\subsubsection{A first identity}

First, we state an important identity obtained by combining $f^{\star\star}_{ab}=-f_{ab}$ (cf. App.~\ref{app:Hodge}) and equation \eqref{eps2}, as well as the assumption \eqref{KYrank4}. It reads
\begin{equation}\label{ffstar}
    f_{ac}^\star \,f^{c}_{\phantom{c}b}=-\mathfrak{Z}\, g_{ab},\quad \text{where} \quad \mathfrak{Z}:=\tfrac{1}{4} f_{ab}^\star f^{ab}.
\end{equation}
Notice that with this definition, the non-degeneracy assumption \eqref{KYrank4} reads $\mathfrak{Z}\neq 0$. By analogy with the spin tensor \eqref{spinnorms}, the scalar $\mathfrak{Z}$ is one of the two ``norms'' of the KY tensor, the other one being $f_{ab}f^{ab}$. 

\subsubsection{Killing-St\"ackel tensor}

Since $f_{ab}$ is anti-symmetric (cf. \eqref{deff}), the valence-$(0,2)$ tensor $f_{ac}f^{c}_{\ph b}$ is symmetric in $ab$. In addition, owing to the the KY equation \eqref{deff}, one readily has
\begin{equation} \label{KSdef}
    K_{ab}:=f_{ac}f^{c}_{\ph b} \quad \Rightarrow \quad \nabla_{(a}K_{bc)}=0,
\end{equation}
which means that $K_{ab}$ is a Killing-St\"ackel (KS) tensor. In the Kerr spacetime, this tensor was discovered by Walker and Penrose \cite{WalkPen.70}, and is responsible for the existence of the Carter constant discovered by Carter \cite{Carter.68}, ensuring that geodesics in Kerr are integrable.

\subsubsection{First Killing vector}

Another property emerging from the antisymmetry of $f_{ab}$ and equation \eqref{deff} is that the tensor $\nabla_{a}f_{bc}$ is totally anti-symmetric, i.e., a 3-form. Consequently, by Hodge duality, there exists a vector $\xi^a$ such that (see App.~\ref{app:Hodge} for details)
\begin{equation} \label{defxi}
    \nabla_a f_{bc}=\varepsilon_{abcd}\xi^d \quad \Leftrightarrow \quad \xi_a =\frac{1}{3}\nabla^b f^\star_{ab},
\end{equation}
where the second equality follows by contracting the first with $\varepsilon^{abcd}$ and using equation \eqref{eps2}. The vector $\xi^a$ so-defined is a Killing vector, as will be show below. 
Note that taking the Hodge dual of equation \eqref{defxi} readily gives 
\begin{equation} \label{dfstar}
    \nabla_a f_{bc}^\star=-2g_{a[b}\xi_{c]}, \quad \Rightarrow \quad \nabla_{(a} f_{b)c}^\star= \xi_{(a}g_{b)c}-g_{ab}\xi_{c},
\end{equation}
where the second equation implies that $f_{ab}^\star$ is a so-called \emph{conformal} KY tensor, as classically defined in the literature \cite{Frolov.17,Ha.20}. We will not need any of its properties, however. 

\subsubsection{Integrability conditions}

One can derive important identities on $f_{ab}$ known as \emph{integrability conditions}. These are called so because they must hold if a spacetime is to possess a KY tensor, allowing one to make claims about the (non-)existence of a KY tensor without solving the actual KY equation \eqref{deff}. There exist many integrability conditions in the literature. We only deal with those that are useful to us later on.

Taking a covariant derivative of \eqref{deff} and combining different index combinations of the resulting equation with the Ricci identity, one obtains the following integrability conditions\footnote{For a proof of the first identity, see, e.g., equations (159) to (166) of \cite{ComDru.22} or references on page 357 of \cite{DiRu.I.81}. See also \cite{LinSar.21} and references therein.}
\begin{subequations}
    \begin{align}
    \nabla_a \nabla_b f_{cd} &= - 3 R_{a[bc}^{\phantom{abc}e} f_{d]e}, \label{intcond} \\
    R_{abe[c} f_{d]}^{\ph e} &= - R_{cde[a} f_{b]}^{\ph e}. \label{intcondHarte}
    \end{align}
\end{subequations} 
The first identity can be seen as a generalization of the similar formula for Killing vectors, known as the Kostant formula (cf. the second equality in \eqref{KillKost}. The method used to get that formula is the same that can be applied to \eqref{deff} to obtain \eqref{intcond}. The second identity \eqref{intcondHarte} can be found in \cite{Ha.23}. It seems that an equivalent formula cannot be found for KS tensors (i.e., $\nabla_a\nabla_bK_{cd}$ does not appear to be writeable solely in terms of $R_{abcd}$ and $K_{ab}$.)

From the integrability condition \eqref{intcond} alone we obtain two important corollaries. On the one hand, contracting with $\varepsilon^{bcdf}$ and combining with \eqref{eps1} and \eqref{defxi}, we find 
\begin{equation} \label{Dxi}
     \nabla_a \xi_b = -\frac{1}{4}R_{ab}^{\ph\ph cd} f_{cd}^\star+\frac{1}{2}R_a^{\ph c} f^\star_{cb}.
\end{equation}
On the other hand, taking the trace of equation \eqref{intcond} with respect to the indices $ac$, using the Ricci identity and the total anti-symmetry of $\nabla_{b}f_{cd}$, we establish that
\begin{equation} \label{Rfsym}
    R_{(a}^{\ph\; c} f_{b)c}=0.
\end{equation}
The consequences of equations \eqref{Dxi} and \eqref{Rfsym} is two-fold. First, contract \eqref{Rfsym} with $f^{\star \,a}_d f^{\star \,b}_e $ and use \eqref{ffstar} to find
\begin{equation} \label{Rfstarsym}
    R_{(a}^{\ph\; c} f^\star_{b)c}=0.
\end{equation}
This is yet another integrability condition for the existence of a KY tensor. Combining \eqref{Rfstarsym} with \eqref{Dxi} shows that $\xi^a$ satisfies
\begin{equation}
    \nabla_{(a}\xi_{b)}=0,
\end{equation}
and thus $\xi^a$ is a Killing vector. Importantly, this is true regardless of the curvature, and holds as long as there exists a KY tensor. 

\subsubsection{Second Killing vector}

A second Killing vector is obtained by contracting the KS tensor \eqref{KSdef} with the first Killing vector, 
\begin{equation} \label{defeta}
    \eta^a:=K^{ab} \xi_b.
\end{equation}
Differentiating this definition and using \eqref{KSdef}, \eqref{defxi} and \eqref{intcond} allows us to express $\nabla_{a}\eta_b $ as the sum of three terms: $\nabla_{a}\eta_b = \varepsilon_{abde} \xi^e f^{dc} \xi_c - \frac{1}{4} K_b^{\ph c} R_{acde} f^{de}_{\star} - \frac{\mathfrak{Z}}{2} R_a^{\ph d} f_{db}.$ The second term in this expression can be re-expressed by contracting the second integrability condition in \eqref{intcond} with $f_\star^{ab}f_f^{\ph c}$. This implies
\begin{equation} \label{Deta}
     \nabla_a \eta_b = \varepsilon_{abde} \xi^e f^{dc} \xi_c + \frac{1}{4} f_a^{\ph f} f_b^{\ph c} R_{fcde} f^{de}_{\star} + \frac{\mathfrak{Z}}{2} R_a^{\ph c} f_{bc},
\end{equation}
in which all three terms on the RHS are clearly anti-symmetric in $ab$ (recall \eqref{Rfsym}). Therefore, we have shown that
\begin{equation}
    \nabla_{(a}\eta_{b)}=0,
\end{equation}
and thus $\eta^a$ is a Killing vector, regardless of curvature.

The degenerate case $\mathfrak{Z}=0$ (which includes Schwarzschild) falls outside our framework and must be treated separately. When $f_{ab}$ is degenerate, the second Killing vector $\eta^a = K^{ab}\xi_b$ reduces to $\eta^a \propto \xi^a$ (and vanishes when $\xi^a$ lies in the null space of $K_{ab}$); consequently, the invariant $\mathfrak{X}$ ceases to be independent of~$\Xi$, and only four commuting first integrals $(H,\Xi,K,Q)$ remain:integrability is no longer guaranteed by the KY structure alone. Specific spacetimes with $\mathfrak{Z}=0$ may nonetheless remain integrable through \emph{additional} isometries---in Schwarzschild, the $SO(3)$ symmetry restores the missing invariant---but this owes to extra symmetry rather than to the degeneracy itself. A full discussion is given in Sec.~V.A.4 of Paper III~\cite{Ra.Iso.Dru.IntegO2.26}.

\subsubsection{Commutation of the two Killing vectors} 
\label{subsec:comu}

With the two formulae \eqref{Dxi} and \eqref{Deta} for the covariant derivatives of $\xi^a$ and $\eta^a$, we can show that the latter \emph{commute} in the sense of differential geometry. Define for any two vector fields $X^a,Y^a$ their \emph{commutator} \cite{Wald}
\begin{equation} \label{defcomu}
[X,Y]^a := X^b \nabla_b Y^a - Y^b \nabla_b X^a,
\end{equation}
which is also a vector field. Then, applying this to the two Killing vectors \eqref{defxi}-\eqref{defeta}, inserting equations \eqref{Dxi} and \eqref{Deta} into \eqref{defcomu}, and gathering the Riemann terms and Ricci terms together respectively, we find that 
\begin{equation} \label{comuKill}
    [\xi,\eta]^a = 0.
\end{equation}

\subsubsection{A last identity}

Lastly, we mention an identity that was established in \cite{Rudiger.II.83} (Eq.(4.6) there), or also in \cite{ComDru.22} but in vacuum only. Regardless, we can improve both results and construct a similar formula that holds in any spacetime admitting a KY tensor (non-necessarily vacuum) and reads:
\begin{equation} \label{RstarK}
  R^{\star}_{cabd} K^{cd} = 4 \nabla_a \nabla_b \mathfrak{Z}- g_{ab}\, \Delta \mathfrak{Z},
\end{equation}
where $\Delta:=g^{ab}\nabla_a\nabla_b$. The proof can be found in \cite{Rudiger.II.83}, or using the tools developed in our subsequent work, Paper III \cite{Ra.Iso.Dru.IntegO2.26}.

\subsubsection{Physical interpretation: covariant angular momentum} \label{CovAngMom}

Although the KY tensor is often introduced as a purely mathematical device, it admits several physical interpretations that illuminate its role in relativistic dynamics. Indeed, define the following quantity
\begin{equation} \label{SDL}
\mathcal{L}^b := p_a f^{ab} .
\end{equation}
Then $\mathcal{L}^a$ can be interpreted as a covariant notion of angular momentum, as first pointed out by Dietz and R\"udiger \cite{DiRu.I.81,DiRu.II.82}. This vector is, by construction, orthogonal to $p^a$, and if $p^a$ is tangent to a geodesic, then $p^a\nabla_a \mathcal{L}^b =0$ by the Leibniz rule and the KY equation \eqref{deff}. This fundamental property was used by Marck and collaborators to construct parallel-transported frames in Kerr \cite{Marck.83} and Killing-Yano spacetimes \cite{KamMar.86}. This framework is prevalent in the study of linear-in-spin effects in Kerr dynamics and EMRI waveform modeling \cite{MaPoWa.22,DruHug.I.22,DruHug.II.22,MaPoWa.22,Pio.25,SkouWitz.25}. 

In addition, in the Schwarzschild spacetime (that is, in the exterior of any spherically symmetric source), the spatial components of $\mathcal{L}^a$ admit the Newtonian limit $\vec{r}\times\vec{p}$ at spatial infinity, much like for spherically symmetric potentials in Newtonian gravitation. This analogy is explored in \cite{Ra.Iso.PapII.24}, where explicit formulae are given. In this context, the KY can be thought of as an operator acting on the linear momentum that generates the angular momentum. The latter always lives the subspace orthogonal to $p_a$, just like $\vec{r}\times\vec{p}$ is always perpendicular to $\vec{p}$ in Newtonian mechanics. 

\section{Constants of motion, Poisson brackets and Integrability} 
\label{sec:PBcalc}

In this section, we start by defining scalar quantities from the spacetime geometry and the particle's momenta, that will end up being constants of motion for the linear MPTD system under the TD SSC. This is done in Sec.~\ref{sec:invMPTD}. Then Sec.~\ref{subsec:Liouville} describes the ingredients needed to prove that these constants of motion are first integrals of the Hamiltonian system we built in Sec.~\ref{sec:Ham}, and that the latter is \emph{integrable}. This constitutes the main result of this paper. Lastly, Secs.~ \ref{subsec:orange} and \ref{subsec:bluegrey} contain detailed calculations of the vanishing Poisson brackets involved in the integrability results, as encoded into table \ref{tab:poisson}. 

\subsection{Constants of motion for the MPTD system} \label{sec:invMPTD}

We begin by reviewing the literature on invariants of motion for the MPTD system itself and adapting their definitions to our purposes. Since our Hamiltonian generates the same equations, we expect these invariants to become the ``first integrals'' of our system. 

\subsubsection{Invariants from Killing vectors} \label{sec:kinvSSC}

In the context of general relativity, most spacetime metrics of interest possesses Killing vectors, i.e., a vector field $k^a$ that Lie-drags the metric $\mathcal{L}_k(g_{ab})=0$. This definition implies two important properties satisfied by $k^a$, the Killing equation and the Kostant formula \cite{Wald}:
\begin{equation} \label{KillKost}
    \nabla_{(a} k_{b)}=0 \quand \nabla_a\nabla_b k_c = R_{cbad}k^d.
\end{equation}

It is well-known for the MPTD equations \eqref{EEgen} that, if $k^a$ is such a Killing vector, then the quantity \cite{Di.74,Ha.12}
\begin{equation} \label{Killinginv}
\mathcal{F}_k:=p_a k^a + \frac{1}{2} S^{ab}\nabla_a k_b 
\end{equation}
remains constant along the worldline of the particle. In the Hamiltonian formulation, this result translates naturally. Let $k^\alpha$ be the natural components of some Killing field $k^a$. In our setup, $k^\alpha$ are four functions of the phase space coordinates $x^\alpha$. Then combining the Hamiltonian \eqref{Hlinspin} and the Poisson brackets \eqref{PBs} on $\mcM$ leads to (without any spin-truncation)
\begin{equation} \label{HC}
\{ \mathcal{F}_k, H \} = p^{\alpha} p^{\beta} \nabla_{\alpha} k_{\beta} + \frac{1}{2}
(\nabla_{\alpha} \nabla_{\beta}
k_{\gamma}-R_{\gamma \beta \alpha \delta} k^{\delta} ) p^{\alpha} S^{\beta \gamma} = 0 \,,
\end{equation}
where the first and second terms vanish independently thanks to \eqref{KillKost}. 

Killing invariants like \eqref{Killinginv} have an additional interesting property. Set $C^\alpha:=p_\beta S^{\beta\alpha}$, so that the TD SSC amounts to the vanishing of the vector $C^\alpha$. Then, using again the brackets \eqref{PBs} and the identities \eqref{KillKost}, we find (without any spin-truncation)
\begin{equation} \label{FkH}
\{ C^{\alpha}, \mathcal{F}_k \} = C^{\beta} \nabla_{\beta} k^{\alpha}  \quad \Longrightarrow \quad \{ C^{\alpha}, \mathcal{F}_k \}|_\mcP =0,
\end{equation}
i.e., the Killing invariants $\mathcal{F}_k$ commute with the constraints induced by the TD SSC. Inserting Eqs.~\eqref{HC} and \eqref{FkH} into the brackets \eqref{Diracfinal} on $\mcP$, we find that for any Killing vector field $k^\alpha$ of the background spacetime,
\begin{equation} \label{phiD}
\{\mathcal{F}_k, H\}^\mcP = 0 
\,,
\end{equation}
i.e., $\mathcal{F}_k$ is an invariant for the constrained system on $\mcP$. A similar calculation shows that two Killing invariants in involution on $\mcM$ remain so for the bracket on $\mcP$; that $\Xi$ and $\fX$ are themselves in involution on $\mcM$ --- because their underlying Killing vectors commute --- is established in Sec.~\ref{subsec:bluegrey}. We stress that these results hold for Killing invariants (or any combination thereof), and not, in general, for other invariants that may not be associated to Killing vector fields. 

\subsubsection{Invariants from Killing-Yano symmetry}

In addition to vector fields, Killing tensors of higher valence may exist and lead to invariants of motion \cite{Frolov.17}. Of particular interest are the (symmetric) Killing-St\"ackel tensors $K_{ab}$ and the (antisymmetric) Killing-Yano tensors $f_{ab}$, defined in equations \eqref{deff} and \eqref{KSdef}, respectively. For our purposes, we take inspiration from the pioneering work of R\"udiger on Killing invariants for the MPTD system \cite{Rudiger.I.81,Rudiger.II.83,ComDru.22}. In these works, the authors look for scalar quantities built from $p_a,S^{ab}$ and the geometry that are conserved under the dipolar MPTD system under the TD SSC. As a result, they find that in spacetimes with a KY tensor, two such invariants exist. The first one is
\begin{equation} \label{KYK}
K := f^\star_{ab} S^{ab}.
\end{equation}
It was introduced for the first time by R\"udiger in \cite{Rudiger.I.81,Rudiger.II.83}. In contrast to Killing vector invariants \eqref{Killinginv}, which are \emph{exactly} conserved for the general dipolar MPTD system (without any SSC), the R\"udiger invariant $K$ is sometimes called a ``quasi-invariant'' \cite{ComDru.22}, since its derivative along $\scL$ is not zero, but of quadratic order in spin. No such distinction will be made here, however, as we work consistently to linear order in spin.

The other invariant, which we call the (generalized) Carter constant, was also defined by R\"udiger in \cite{Rudiger.I.81,Rudiger.II.83}, and is\footnote{Our definition of $L_{abc}$ is equivalent to that given in references \cite{Rudiger.I.81,ComDru.22}.}
\begin{equation} \label{KYQ}
Q := K_{ab} p^a p^b + L_{abc} S^{ab} p^c, \quad \text{with} \quad L_{abc} := \frac{2}{3}\nabla_{[a}K_{b]c} + \frac{4}{3} \varepsilon_{abcd}\nabla^d \mathfrak{Z},
\end{equation}
where $K_{ab}$ and $\mathfrak{Z}$ are built from the Killing-Yano tensor, cf. Eqs.~\eqref{KSdef} and \eqref{ffstar}. Like $K$, the Carter constant $Q$ is only conserved at linear order in spin. However, both $K$ and $Q$ admit generalizations to quadratic order in spin (at least for a particular quadrupole model), see Paper III \cite{Ra.Iso.Dru.IntegO2.26} for details.  

In the non-spinning (geodesic) case $S^{ab}=0$, one has $K=0$ and $Q=K^{ab}p_a p_b$, which only requires the existence of a KS tensor, not a KY tensor \cite{DiRu.I.81,DiRu.II.82}. In the Kerr spacetime, $K^{ab}p_a p_b$ is then the well-known Carter constant \cite{Carter.68} that ensures the integrability of geodesics.

\subsubsection{Physical interpretation of the constants of motion}

Some physical insight about the constants of motion defined above can be gained by combining their definitions with the decomposition \eqref{zip!} of the spin tensor (with $f^a=\hat{p}^a$ there). It is then possible to re-write equations \eqref{KYK} and \eqref{KYQ} solely in terms of the spin vector $S_\tTD^a$ and the covariant angular momentum $\mathcal{L}^a$ defined previously in \eqref{SDL}. One finds
\begin{subequations} \label{newKQ}    
    \begin{align} 
        K &= - \frac{2}{\mu} \mathcal{L}_a S^a_{\tTD}, \label{newKSSC}\\
        Q &= \mathcal{L}^a \mathcal{L}_a - 2 \mu f_{a b} S_\tTD^a \xi^b + \xi_a p^a f^\star_{bc}S^{bc}+ O(C^a), \label{newQSSC}
    \end{align}
\end{subequations}
where the fact that $\mathcal{L}^a$ and $S^a_\tTD$ are orthogonal to $p_a$ was used, and $O(C^a)$ describes a term proportional to $C^b=p_a S^{ab}$, which vanishes under the TD SSC 
\eqref{TDSSC}. 

With expression \eqref{newKSSC} at hand, we see that the R\"udiger constant is the component of the spin vector parallel to the covariant angular momentum $\mathcal{L}^a$ defined in \eqref{SDL}. This is the key to solving the linear-in-spin dynamics of spinning objects in spacetimes with a KY tensor, building on parallel-transported frames \cite{Marck.83,VdM.20}.

From Eq.~\eqref{newKQ}, we also see that $Q$ is the sum of the squared norm of $\mathcal{L}^a$ (the Carter constant for the spinless case) and two corrections. The first one is $\propto f_{a b} S_\tTD^a \xi^b$: the component of the KY tensor in the 2-space spanned by the spin vector and the first Killing vector: it does not vanish in general. For the second correction, equations \eqref{Killinginv} and \eqref{KYK} imply that it is equal to the product $\mathcal{F}_\xi\cdot K$ to leading order in spin. Since both are constants of motion, it is common in the literature to entirely drop this contribution that is separately conserved. 

We will work with the original definitions \eqref{KYK}-\eqref{KYQ} throughout; the equivalent forms \eqref{newKQ} are recorded only for readers willing to make contact with the literature, where this decomposition is commonly used. 

\subsection{The Liouville-Arnold theorem} \label{subsec:Liouville}

We are now ready to state precisely the integrability results that will be proven subsequently. We have a 10D hamiltonian system defined by the Hamiltonian \eqref{Hlinspin} and the Poisson brackets \eqref{Diracfinal}. We also have 5 scalar functions on that phase space, namely the Hamiltonian $H$ itself, the two Killing invariants $\Xi:=\mathcal{F}_\xi$ and $\fX:=\mathcal{F}_\eta$, the R\"udiger constant $K$ and the (generalized) Carter constant $Q$. For convenience, we gather below their covariant definitions as functions on the original 14D phase space $\mcM$, i.e., as functions of $(x^\alpha,p_\beta,S^{\gamma\delta})$ :
\begin{subequations}\label{defHXiKQ}
    \begin{align} 
        H &= \frac{1}{2} g^{\alpha\beta} p_\alpha p_\beta, \quad 
        \Xi = p_\alpha \xi^\alpha + \frac{1}{2} S^{\alpha\beta} \nabla_\alpha \xi_\beta, \quad 
        \fX = p_\alpha \eta^\alpha + \frac{1}{2} S^{\alpha\beta} \nabla_\alpha \eta_\beta, \label{defHXifX}\\
        K &= f_{\alpha\beta}^\star S^{\alpha\beta} \quand 
        Q = K^{\alpha\beta} p_\alpha p_\beta + \frac{2}{3}\left(  \nabla_{[\alpha} K_{\beta]}^{\ph \gamma} + 2\varepsilon_{\alpha\beta}^{\ph\ph \gamma\delta} \nabla_\delta \mathfrak{Z} \right) S^{\alpha\beta}p_\gamma, \label{defKQ}
    \end{align}
\end{subequations}
where the Killing objects $(\xi^a,\eta^a,K_{ab},\mathfrak{Z})$ are all given in terms of the KY tensor $f_{ab}$ in equations \eqref{defxi}, \eqref{defeta}, \eqref{KSdef} and \eqref{ffstar} respectively. 

According to the classical Liouville-Arnold theory \cite{Arn}, the Hamiltonian system will be \emph{integrable} if the 5 constants of motion $(H,\Xi,\fX,K,Q)$ are linearly independent and Poisson-commuting (their pairwise Poisson brackets vanish). Linear independence is a consequence of that of the non-spinning case\footnote{Indeed, in the non-spinning case, we know that $(H,\Xi,\fX,Q)$ are independent since the system describes geodesics and is integrable \cite{PKVK.07,Frolov.17}. For the spinning case, these geodesic constants get a linear-in-spin correction that does not change their linear independence, while the fifth constant $K$ vanishes in the non-spinning limit, and is thus necessarily linearly independent from $(H,\Xi,\fX,Q)$.} and we thus focus on the Poisson-commuting part. 

The five functions $(H,\Xi,\fX,K,Q)$ are defined on the reduced phase space $\mcP$, so their mutual involution must be assessed with the Dirac bracket \eqref{Diracfinal} rather than the bare $\mcM$-bracket. Beyond the restricted bracket $\{F,G\}|_\mcP$ (first term in \eqref{Diracfinal}), the Dirac bracket carries a correction bilinear in the constraint brackets $\{F,C^\alpha\}$ and $\{C^\beta,G\}$ (second term in \eqref{Diracfinal}). Establishing integrability thus requires not only the ten brackets among the constants, but also the five brackets of each constant with the constraint $C^\alpha$ feeding this correction --- fifteen in total, one for each pair of distinct elements of $(H,\Xi,\fX,K,Q,C^\alpha)$. We collect them in Table~\ref{tab:poisson}, grouped by the way each is shown to vanish.

\begin{table}[h]
  \centering
  \renewcommand{\arraystretch}{1.35}
  \setlength{\tabcolsep}{6pt}
  \begin{tabular}{c|ccccc|c}
    $\{\,\cdot\,,\,\cdot\,\}|_\mcP$ & $H$ & $\Xi$ & $\fX$ & $K$ & $Q$ & $C^\alpha$ \\ \hline
    $H$   & $\quad 0\quad $    & \PBdone{\ref{sec:kinvSSC}} & \PBdone{\ref{sec:kinvSSC}} & \PBtodo{\ref{sec:PBKH}}  & \PBtodo{\ref{sec:PBQH}}  & \PBdone{\ref{sec:NotePBKill}} \\
    $\Xi$ & \PBna  & $0$    & \PBind{\ref{PBXifX}} & \PBind{\ref{sec:NotePBKill}} & \PBind{\ref{sec:NotePBKill}} & \PBdone{\eqref{FkH}} \\
    $\fX$ & \PBna  & \PBna  & $0$    & \PBind{\ref{sec:NotePBKill}} & \PBind{\ref{sec:NotePBKill}} & \PBdone{\eqref{FkH}} \\
    $K$   & \PBna  & \PBna  & \PBna  & $0$    & \PBtodo{\ref{sec:PBKQ}}  & \PBnz{~\ref{sec:NotePBKC}} \\
    $Q$   & \PBna  & \PBna  & \PBna  & \PBna  & $0$    & \PBtodo{\ref{sec:PBQC}} \\
  \end{tabular}
  \caption{Pairwise brackets $\{\,\cdot\,,\,\cdot\,\}|_\mcP$ on $\mcP$ among the constants $H,\Xi,\fX,K,Q$ ($5\times5$ block) and with the constraint $C^\alpha$ (last column). Since $\{F,C^\alpha\}|_\mcP=0$ for $F\in(H,\Xi,\fX,Q)$, the Dirac correction never contributes and $\{\,\}^\mcP=\{\,\}|_\mcP$ on every pair, as argued in \ref{sec:shortcut}. The table gives references for the calculation of each bracket, and the color indicates the status of the proof at this stage in the paper : \colorbox{green!12}{done directly (Sec.~\ref{sec:invMPTD})}, \colorbox{blue!10}{done indirectly (Sec.~\ref{subsec:bluegrey})}, \colorbox{orange!18}{done below, (Sec.~\ref{subsec:orange})}, \colorbox{gray!18}{nonzero, but does not contribute (Sec.~\ref{subsec:bluegrey})}. By antisymmetry of Poisson brackets, the lower triangle (\textendash) is redundant, and the diagonal vanishes identically.}
  \label{tab:poisson}
\end{table}

Table~\ref{tab:poisson} reduces the problem considerably. Its last column vanishes for every constant but one: $\{\Xi,C^\alpha\}|_\mcP$ and $\{\fX,C^\alpha\}|_\mcP$ vanish from Eq.~\eqref{phiD}, $\{H,C^\alpha\}|_\mcP=0$ expresses the stability of the constraint surface under the Hamiltonian flow (Sec.~\ref{sec:stabflow}, Eq.~\eqref{VflowH}), and $\{Q,C^\alpha\}|_\mcP=0$ is shown in Sec.~\ref{subsec:orange}; the sole inconsequential exception is $\{K,C^\alpha\}|_\mcP\neq0$, addressed in Sec.~\ref{subsec:bluegrey}. Now, the Dirac correction to a pair $(F,G)$ is bilinear in $\{F,C^\alpha\}$ and $\{C^\beta,G\}$ (recall \eqref{Diracall}) so it vanishes as soon as \emph{one} of the two constants commutes with the constraints. Since $K$ is the only constant with a nonzero constraint bracket and never pairs with itself, every correction term vanishes on $\mcP$, and the Dirac bracket reduces to the restricted $\mcM$-bracket on all ten pairs,
\begin{equation} \label{eq:DiracReduces}
    \{F,G\}^\mcP = \{F,G\}\big|_\mcP, \qquad F,G \in (H,\Xi,\fX,K,Q).
\end{equation}
Integrability therefore hinges solely on the ten brackets of the $5\times5$ block in table \ref{tab:poisson}. Those among the Killing vector invariants are already known to vanish (see \ref{subsec:bluegrey} below); the remaining three, involving $K$ or $Q$, are computed directly in Sec.~\ref{subsec:orange}, along with $\{Q,C^\alpha\}$.

\subsection{Poisson brackets involving $K$ and $Q$} \label{subsec:orange}

In this subsection, we compute the four brackets depicted in orange in Table~\ref{tab:poisson}. These vanish for different and non-trivial reasons. As for all calculations, the proof is covariant, and relies on the identities involving the KY tensor established in Sec.~\ref{sec:KY}. 

We start with general, covariant expressions for the brackets: using the definitions \eqref{defHXiKQ} the brackets \eqref{PBs}, and repeated applications of the Leibniz rule, leads to 
\begin{subequations} \label{allPBs}
    \begin{align} 
    \{ K,H \} & = \nabla_a f_{bc}^{\star} \,p^a S^{bc} + O (C^a, 2), \label{PBKH} \\
    \{ Q,H \} &= \nabla_{a}K_{bc} p^ap^bp^c -(R_{abce}K^e_{\ph d}+\nabla_d L_{abc})\,S^{ab} p^c p^d + O(C^a, 2), \label{PBQH}\\
     \{ K,Q \} &= 2 \left( K_c^{\ph d} \nabla_d f^{\star}_{ab} - 2 f^{\star d}_a L_{dbc} \right) S^{ab} p^c + O(C^a,2), \label{PBKQ} \\
    \{ Q,C^d \} & = (\nabla_c K_{ab} - 2 L_{cab}) p^a p^b S^{cd} + O(C^a,2). 
    \label{PBQC}
    \end{align}
\end{subequations}
where we remind the reader that $O(C^a,2)$ indicates terms that are either quadratic-in-spin or proportional to the SSC, and thus vanish on $\mcP=0$ (where $C^a=0$). Our goal is to show that all RHSs of equations \eqref{allPBs} are also $O(C^a,2)$. We treat them one by one in the following four subsections. We use Latin indices without risk of confusion, since all intermediary results are valid for the tensor fields involved and not just their components. 

\subsubsection{The Poisson bracket $\{K,H\}$} \label{sec:PBKH}

For the bracket \eqref{PBKH}, we first write the spin tensor in terms of the spin vector using, \eqref{zip!} in the form $S^{bc}=\varepsilon^{bc}_{\ph\ph de}\hat{p}^dS^e+O(C^a)$. Second, we use equation \eqref{dfstar} for the Hodge dual of the KY tensor. Simplifying the result, we obtain
\begin{equation}
    \{ K,H \} = 2 \varepsilon_{abcd}\xi^a p^b\hat{p}^c S^d + O(C^a, 2),
\end{equation}
where the first term vanishes from $\hat{p}^b=p^b/\mu$ and the anti-symmetry of $\varepsilon_{abcd}$. Therefore, we have $\{ K,H \}=O(2)$ on $\mcP$ and since $\{H,C^\alpha\}=O(2)$ on $\mcP$, we obtain 
\begin{equation}
    \{ K,H \}^{\mcP} = 0 + O(2).
\end{equation}

In particular, $K$ is a first integral for $H$. Equivalently, we have recovered the fact that $K$ is constant under the linearized MPTD system under the TD SSC. This holds for any spacetime endowed with a non-degenerate KY tensor. Remarkably, it also holds for spacetime with a degenerate KY tensor ($\mathfrak{K}=0$, cf. \eqref{deff}). Indeed, the the argument above only relies on \eqref{dfstar} only, which is true for any type of KY tensor, not just non-degenerate ones.

\subsubsection{The Poisson bracket $\{Q,H\}$} \label{sec:PBQH}

For the bracket \eqref{PBQH}, the non-spinning term $\nabla_{a}K_{bc}p^ap^bp^c$ vanishes by virtue of $K_{ab}$ being a KS tensor, cf. equation \eqref{KSdef}. For the linear-in-spin term, we will show that it vanishes too, by proving that the geometric tensor in parenthesis, namely 
\begin{equation} \label{Fabcd}
   F_{abcd} := R_{abde}K^e_{\ph c}+\nabla_d L_{abc}=R_{abde}K^e_{\ph c} + \frac{2}{3}\nabla_d \nabla_{[a}K_{b]c} +\frac{4}{3} \varepsilon_{a b c f} \nabla_d \nabla^f \mathfrak{Z},
\end{equation}
is anti-symmetric in $cd$. Since $F_{abcd}$ is clearly anti-symmetric in $ab$, it suffices to show that it also has the exchange symmetry $F_{abcd}=F_{cdab}$. The proof of this statement is a matter of definitions combined with the following identities:
\begin{align}
    \nabla_d \nabla_a K_{c b} - \nabla_b \nabla_c K_{d a} & = 2 ( R_{f d a (c } K^f_{\ph b )} - R_{f b c (d} K^f_{\ph a )} ), \label{id1}\\
    2 \nabla_{[a } \nabla_{ b]} K_{c d} & = - 2 R_{abe(c } K^e_{\ph d)}, \label{id2}\\
    4 \nabla_a \nabla_b \mathfrak{Z} &= R_{cabd}^{\star} K^{cd} +g_{ab} \Delta \mathfrak{Z}  , \label{id3}\\
    \varepsilon_{c a b}^{\phantom{a b c} f} R_{g d f h}^{\star} & = - (g_{b h} R_{g d a c} + g_{a h} R_{g d c b} + g_{c h} R_{g d b a}), \label{id4}
\end{align}
all of which are valid in any spacetime admitting a KY tensor, without any extra condition on the curvature. In particular: \eqref{id1} can be found on page 5 of \cite{CookDray.09}, \eqref{id2} is the Ricci identity for a symmetric tensor (cf. equation (3.2.12) in \cite{Wald}), \eqref{id3} follows from the KY existence conditions and was given in \eqref{RstarK}, and \eqref{id4} combines the definition of the Hodge dual and identity \eqref{eps3}.

%
Identities \eqref{id1}--\eqref{id2} express the second term on the right-hand side of~\eqref{Fabcd} purely in terms of products of $R_{abcd}$ and $K_{ab}$, while identities \eqref{id3}--\eqref{id4} do the same for the third term.
Once both terms are in this form, the algebraic symmetries of the Riemann tensor imply that $F_{abcd}$ is symmetric under $ab\leftrightarrow cd$, whence $F_{abcd}=F_{cdab}$.
Consequently, equation \eqref{PBQH} can be written as $\{ Q,H \} = -F_{ab[cd]}S^{ab}p^cp^d + O(C^a,2)$ which clearly vanishes from symmetries. Again, from $\{ Q,H \}=O(2)$ on $\mcP$ and since $\{H,C^\alpha\}=O(2)$ on $\mcP$, we obtain
\begin{equation}
    \{ Q,H \}^\mathcal{P} = 0 + O(2).
\end{equation} 

\subsubsection{The Poisson bracket $\{K,Q\}$} \label{sec:PBKQ}

For the bracket \eqref{PBKQ}, we first focus on the geometric tensor in parenthesis. Inserting the definition \eqref{KYQ} of $L_{abc}$, expanding the anti-symmetry therein and using $\nabla_d K_{ac} = - 2 \nabla_{(a} K_{c)d}$ (which follows from \eqref{KSdef}), we find
\begin{equation} \label{interKF1}
    K_c^{\ph d} \nabla_d f^{\star}_{ab} + 2 f^{\star d}_b L_{dac} = K_c^{\ph  d} \nabla_d f^{\star}_{a b} - \frac{4}{3} f^{\star d}_b \nabla_{(a } K_{ c) d} - \frac{2}{3} f^{\star d}_b \nabla_a K_{c d} + \frac{8}{3} f^{\star d}_b \varepsilon_{d a c f} \nabla^f \mathfrak{Z}.
\end{equation}
Then, we use the Leibniz rule on the second and third terms. The resulting expressions can be simplified using  $f^{\star d}_bK_{cd}=\mathfrak{Z}f_{cb}$ (a consequence of \eqref{ffstar}) and  \eqref{dfstar}. Equation \eqref{interKF1} then becomes
\begin{align}\label{interKF2}
    K_c^{\ph d} \nabla_d f^{\star}_{ab} + 2 f^{\star d}_b L_{dac} = K_c^{\ph  d} &\nabla_d f^{\star}_{a b} - \frac{4}{3} \nabla_{(a} (\mathfrak{Z} f_{c)b}) -\frac{2}{3} \nabla_a (\mathfrak{Z} f_{cb}) + \frac{8}{3} f^{\star d}_b \varepsilon_{dacf} \nabla^f \mathfrak{Z} \nonumber\\
    &\phantom{=} - \frac{4}{3} K^d_{\ph c} g_{a[b} \xi_{ d]} - \frac{4}{3} K^d_{\ph  a} g_{c [b } \xi_{ d]} - \frac{4}{3} K_c^d g_{a [b } \xi_{ d]}.
\end{align}
Now we can contract this equation with $S^{ab}p^c$ to re-construct the right-hand side of \eqref{PBKQ}. This kills many terms with a metric, since any trace of $S^{ab}p^c$ either vanishes or is $O(C^a)$. Other terms are treated with the identity $f^{\star d}_b \varepsilon_{dacf} = g_{ba} f_{cf} + g_{bf} f_{ac}+ g_{bc}f_{fa}$ (which follows from the definition for the Hodge dual combined with \eqref{eps3}), as well as $\nabla_a \mathfrak{Z}=f_{ba}\xi^b$ (a consequence of \eqref{ffstar} and \eqref{defxi}). In the end, one obtains
\begin{align}\label{interKF3}
    \{K,Q\}=\frac{4}{3} (f_{ab} f_{cd} + 2 f_{bc} f_{ad} + \mathfrak{Z}\, \varepsilon_{abcd}) S^{ab} p^c \xi^d + O(C^a,2).
\end{align}
The fact that the quantity in parentheses vanishes is not trivial, but follows from the following identity, already found and proved in \cite{ComDru.22}:
\begin{equation} \label{fffeps}
  f_{ab} f_{cd} +2f_{c[a} f_{b]d}  =  - \mathfrak{Z}\, \varepsilon_{abcd},
\end{equation}
which holds because the left-hand side is totally antisymmetric, and thus proportional to $\varepsilon_{abcd}$ in four dimensional spacetimes. The scalar of proportionality (here $-\mathfrak{Z}$) is obtained by contracting both sides with $\varepsilon^{abcd}$ and using equations \eqref{eps0} and \eqref{ffstar}. Ultimately, the combination of \eqref{fffeps} and \eqref{interKF3} gives the final result:
\begin{equation}
    \{K,Q\}^\mcP= O(2).
\end{equation}

The R\"udiger and Carter constants thus commute under the $\mcP$-brackets. Crucially, this is not true for the $\mcM$-bracket, as is mentioned for a Kerr background in \cite{ComDru.22}\footnote{However, the fact that $\{K,Q\}^\mcM\neq0$ does \emph{not} imply that the system is non-integrable, as stated in \cite{ComDru.22}. Indeed, the $\{\}^\mcM$-brackets are \emph{not} the ones adapted to describe the MPTD equations, as we discussed in Sec.~\eqref{sec:consinv}.}. 

\subsubsection{The Poisson bracket $\{Q,C^\alpha\}$} \label{sec:PBQC}

Lastly, we turn to the bracket \eqref{PBQC}. Inserting the definition \eqref{KYQ} of $L_{cab}$ in it, we see that the contribution from the Levi-Civita tensor vanishes when contracted with $p^a p^b$. Expanding the anti-symmetry, we are left with
\begin{equation}
    \{ Q, C^d \} = \frac{1}{3}\bigl( \nabla_{c} K_{ab}+2 \nabla_{a} K_{cb}\bigr) p^a p^b S^{cd} + O(C^a,2),
\end{equation}
but the quantity in parenthesis vanishes thanks to the symmetry brought by $p^ap^b$, since $K_{ab}$ satisfies 
$\nabla_c K_{a b} = - 2 \nabla_{(a} K_{b) c}$ by the KS equation \eqref{KSdef}. Thus, we have shown that 
\begin{equation}
    \{ Q,C^d \} = O(C^a,2). 
\end{equation}

Interestingly, the presence of $S^{cd}$ in this bracket does nothing, the geometric tensor vanishes only under the symmetry $(ab)$ brought by $p^ap^b$, i.e., we just used the general result $\nabla_c K_{ab}-2L_{c(ab)}=0$.

\subsection{Comments on other Poisson brackets} 
\label{subsec:bluegrey}

In this subsection, we comment on the vanishing (or lack thereof) of the remaining Poisson brackets in table \ref{tab:poisson}, those involving the Killing vector invariants $(\Xi,\fX)$, (in blue there) and the special bracket $\{K,C^\alpha\}$. 

\subsubsection{Poisson-commutation of Killing vector invariants} \label{PBXifX}

In \ref{subsec:comu}, we established that the two Killing vectors generated by the KY tensor commute, in the sense of \eqref{comuKill}. Here, we show that this commutation at the level of the spacetime manifold elegantly implies the commutation between their invariants (cf. \eqref{Killinginv}) at the level of the Poisson brackets. To this end, we use the definitions of $\Xi,\fX$ (cf. \eqref{defHXiKQ}), the bi-linearity of Poisson brackets to obtain
\begin{equation} \label{eq:comu1}
    \{\Xi,\fX\} = \{ p_a \xi^a, p_b \eta^b \} + \frac{1}{2} \{ S^{a b} \nabla_a \xi_b, p_c \eta^c \} + \frac{1}{2} \{ p_a \xi^a, S^{bc} \nabla_b \eta_c \} + \frac{1}{4} \{ S^{a b} \nabla_a \xi_b, S^{c d} \nabla_c \eta_d \}.
\end{equation}

Then, with help from the brackets themselves \eqref{PBs}, the Leibniz rule, and some index re-arrangement, this becomes

\begin{equation} \label{eq:comu2}
    \{\Xi,\fX\} = p_a [\eta, \xi]^a + \frac{1}{2} S^{a b} R_{a b c d} \xi^c \eta^d - S^{b c} \nabla_a \xi_b \nabla_c \eta^a,
\end{equation}
where we have introduced the commutator, defined in \eqref{defcomu}. The last term in \eqref{eq:comu2} can be turned into a more usable form as follows: 
\begin{subequations}\label{eq:comu3}
\begin{align} 
  \nabla_a \xi_b \nabla_c \eta^a & = \nabla_b \xi^a \nabla_a \eta_c, \\
  & = \nabla_b (\xi^a \nabla_a \eta_c) - \xi^a \nabla_b (\nabla_a \eta_c), \\
  & = \nabla_b [\xi, \eta]_c + \nabla_b (\eta^a \nabla_a \xi_c) - \xi^a \nabla_b (\nabla_a \eta_c),  \\
  & = \nabla_b [\xi, \eta]_c + \nabla_b \eta^a \nabla_a \xi_c + \eta^a \nabla_b (\nabla_a \xi_c) - \xi^a \nabla_b (\nabla_a \eta_c),
\end{align}
\end{subequations}
where we have successively used: the Killing equation in the form $\nabla_c \eta^a=-\nabla^a \eta_c$, the Leibniz rule on $\nabla_b$, the definition \eqref{defcomu} of the commutator, and the Leibniz rule on $\nabla_b$ again. Now, contracting \eqref{eq:comu3} with $S^{bc}$, gathering terms of interest on the LHS, and using Riemann-spin trick formula
$S^{b c} R_{c a b d} = - \frac{1}{2} S^{b c} R_{b c a d}$, we can write

\begin{equation} \label{eq:comu4}
  S^{b c} \nabla_a \xi_b \nabla_c \eta^a = \frac{1}{2} S^{a b} \nabla_a [\xi, \eta]_b - \frac{1}{2} S^{a b} R_{a b c d} \eta^c \xi^d.
\end{equation}

Lastly, upon inserting equation \eqref{eq:comu4} into \eqref{eq:comu2} and simplifying, we get the following fundamental result
\begin{equation}
  \{ \Xi, \fX \} = p_a [\xi, \eta]^a - \frac{1}{2} S^{a b}
  \nabla_a [\xi, \eta]_b. 
\end{equation}

This formula is well-known in the non-spinning (geodesic) case, where the link between Killing vectors and their invariants are linked by morphisms and Lie algebras \cite{Frolov.17}. Here we have shown, covariantly, that this link persists for the full Dixon invariant, including the spinning degrees of freedom. In particular, one can see that if the Killing vectors commute (as is the case for us since we showed \eqref{comuKill}), then their associated constants of motion commute too, under the brackets \eqref{PBs}. 

\subsubsection{Comment on the brackets involving a Killing invariant} \label{sec:NotePBKill}

In the background, spacetime, let $x^{\alpha} = (\mathfrak{t}, x^i)$ denote a coordinate system adapted to the Killing vector field $k^a=(\partial_\mathfrak{t})^a$, such that the metric coefficients do not depend explicitly on $\mathfrak{t}$, i.e., $\partial_\mathfrak{t} g_{\alpha \beta} = 0$. For example, in the Kerr spacetime covered with Boyer-Lindquist coordinates $(t,r,\theta,\phi)$, this $\mathfrak{t}$ could either be $t$ or $\phi$. In this setting, one can show the following identity:
\begin{equation} \label{equalpit}
  \mathcal{F}_k = \pi_{\mathfrak{t}},
\end{equation}
where the LHS is the Killing invariant $\mathcal{F}_k$ associated to the Killing field $k^a=(\partial_\mathfrak{t})^a$ defined  in \eqref{defFk} and the RHS is the (new) momentum variable $\pi_\mathfrak{t}$ introduced in \eqref{ptopi} (part of the quasi-symplectic coordinates). The equality between the two is proved in App.~\ref{app:KillEqual}. 

In the Kerr spacetime covered with Boyer-Lindquist coordinates $(t,r,\theta,\phi)$, the above equation reads $E=-\pi_t,L=\pi_\phi$, where $E$ is the particle's energy and $L$ is the (axial component of) the angular momentum, as measured at spatial infinity.

With \eqref{equalpit}, let us now consider any function $F$ in the set $(H,\Xi,K,Q,C^\alpha)$, which are all function of the 10 quasi-symplectic phase space coordinates $(x^\alpha,\pi_\alpha,S^{AB})$ on $\mathcal{N}$. Crucially, the dependence of $F$ on the background geometry is only through covariant expressions involving the metric, the Riemann tensor, the KY tensor and their covariant derivatives. In particular, none of these functions depend on the coordinate $\mathfrak{t}$, when spacetime coordinates are adapted to the $(\partial_\mathfrak{t})^a$-isometry, i.e., that $\mathfrak{t}$ is part of the $x^\alpha$. 

What equation \eqref{equalpit} shows is that the conjugated momentum to $\mathfrak{t}$ is nothing but the momentum $\pi_\mathfrak{t}$. Thus, since $\Xi=\pi_\mathfrak{t}$ in these coordinates, one simply has
\begin{equation}
    \{\mathcal{F}_k,F\}= \{\pi_\mathfrak{t},\mathfrak{t}\} \cdot \frac{\partial F}{\partial \mathfrak{t}} = -1\cdot0=0,
\end{equation}
where we have used the Leibniz rule, the fact that $(\mathfrak{t},\pi_\mathfrak{t})$ is a conjugated pair, and that $F$ does not depend on $\mathfrak{t}$ since the latter does not appear in any function $F$. This immediately leads to all Poisson brackets involving at least one Killing vector invariant $(\Xi,\fX)$ invariant to vanish.

\subsubsection{Comment on the bracket $\{K,C^\alpha\}$} \label{sec:NotePBKC}

The bracket $\{K,C^\alpha\}$ does not generally vanish. Indeed, we easily find, using \eqref{PBs} and the Leibniz rule:
\begin{equation} \label{PBKC}
    \{K,C^\gamma\} = 2 f_{\alpha \beta}^{\star} p^{\alpha} S^{\beta
  \gamma} + O (2, C^{\alpha}),
\end{equation}
and this does not generally vanish, even on $\mcP$ where $C^\alpha=0$. However, this is not a problem for two reasons. 

First, the Poisson-commutation of all but one of the first integrals with $C^\alpha$ is sufficient to ensure integrability. Indeed, the expression of the $\mcP$-bracket \eqref{Diracall}, the second term on the right hand side always involves a product of the form $\{X,C^\alpha\}\{Y,C^\beta\}$, where $X,Y$ are two distinct elements from the five first integrals $(H,\Xi,\mathfrak{X},Q,K)$. Hence, since four of them satisfy $\{\cdot,C^\alpha\}|_\mcP=0$ (namely $(H,\Xi,\mathfrak{X},Q)$), all those second terms will vanish.

Second, it is possible to define an alternative constant of motion $\tilde{K}$ such that (i) $\tilde{K}=K$ on $\mcP$, and (ii) $\{\tilde{K},C^\alpha\}|_\mcP=0$. We show how to do this below. Suppose that we define $\tilde{K}=K+V_\alpha C^\alpha$, with $V_\alpha$ any quantity function of the $\mcM$ phase-space variables $(x^\alpha,p_\beta,S^{\alpha\beta})$. Since $C^\alpha=0$ on $\mcP$ by definition, item (i) above is automatically satisfied. Computing $\{\tilde{K},C^\alpha\}$, we find using Eqs.~\eqref{CaCb} and \eqref{PBKC}
\begin{equation}
    \{\tilde{K},C^\gamma\} = (2 f_{\alpha \beta}^{\star} p^{\alpha} - \mu^2 V_{\beta}) S^{\beta \gamma} +
O (2, C^{\beta}).
\end{equation}
Therefore, if we chose $V_\beta=2\mu^{-2} p^{\alpha}f_{\alpha \beta}^{\star}$, we readily obtain that $\{\tilde{K},C^\gamma\}|_\mcP = 0$, as claimed. 

Most interestingly, using the definition of $\tilde{K}$ and the decomposition \eqref{zip!} for the spin tensor, we notice that the $O(C^\alpha)$ correction we just added to $K$ to construct $\tilde{K}$ precisely cancels the $O(C^\alpha)$ term in the decomposition of the spin tensor applied to $K$. In other words, we find the following property
\begin{equation}
    \tilde{K}: (x,p,S)\in \mcM \mapsto- 2 f_{\alpha \beta} \hat{p}^{\alpha} S^{\beta},\quad \text{such that} \quad \{\tilde{K},C^\alpha\}|_\mcP=0.
\end{equation}
This is precisely the spin-parallel component of the covariant angular momentum \cite{DiRu.I.81,DiRu.II.82} already discussed around equation \eqref{newKQ}. This could have been our definition of the R\"udiger constant from the beginning, as done in other works  \cite{Pio.25,WitzHJ.19,WitzAA.22}. In any case, this difference is only relevant in our work with Dirac brackets, since, at the level of the equations of motion, or on the physical phase space $\mcP$, one simply has $K=\tilde{K}$. 

\section{Conclusions and prospects} 
\label{sec:concl}

\subsection{Summary and analytical implications} 
\label{sec:sumconc}

In this paper we have proved Liouville--Arnold integrability for the linear-in-spin MPTD dynamics under the TD SSC, in any four-dimensional spacetime admitting a non-degenerate KY tensor, with no restriction from background field equations (Secs.~\ref{sec:Ham}--\ref{sec:PBcalc}; see in particular Table~\ref{tab:poisson} and FIG.~\ref{fig:tableconstant} for a summary).
Our work identifies the KY symmetry as the sole geometric source of this integrability. It demonstrates that the regular motion of spinning bodies does not rely on the restrictive assumptions of vacuum field equations or asymptotic flatness. While the Kerr metric remains a prominent astrophysical example, there is nothing unique about it regarding linear-in-spin integrability; our coordinate-free proof applies universally across the entire class of spacetimes with a KY symmetry, irrespective of their specific Ricci content.
This result aligns with findings from another framework for spinning particle motion \cite{Gibbons:1993ap,Kubiznak:2011ay,Cariglia:2014ysa}, further highlighting the universal link between KY symmetry and integrability.

This rigorous integrability has immediate analytical and numerical implications. The existence of a complete set of involutive first integrals guarantees that the equations of motion can, in principle, be solved by quadrature, meaning the exact orbital trajectories can be determined analytically without relying on numerical integration.
We have explicitly realized this in a companion article (Paper II) \cite{Ra.Iso.PapII.24}, where the reduction of the physical phase space in a Schwarzschild background yields exact analytical solutions in terms of elliptic functions. This approach is consistent with recent analytical treatments of the Kerr spacetime using specific tetrad frames \cite{WiPio.23, SkouWitz.25, Piovano:2025aro}, and our coordinate-free proof ensures that such methods can be systematically extended to any geometry admitting a non-degenerate KY tensor~\cite{Abhinove:2026prep} (see also~\cite{Witzany:2026eqc} for other attempts in this direction).
Furthermore, the non-degenerate Hamiltonian structure established here provides the exact geometric foundation required for developing structure-preserving numerical schemes, such as symplectic integrators \cite{Wu.al.21, Wang.al.21}. By ensuring the exact conservation of the phase-space geometry and the associated constants of motion over long-term evolutions, these symplectic tools will be invaluable for distinguishing regular orbits from chaotic ones in higher-order multipolar dynamics (see, e.g., \cite{ZeLuWi.20}).

\subsection{Future prospects in gravitational-wave modeling} 
\label{sec:concpro}

The rigorous integrability established here has immediate consequences for modelling compact binaries with asymmetric masses. The global existence of action-angle variables is now guaranteed, and their construction has been explored in both Schwarzschild~\cite{Ra.Iso.PapII.24} and Kerr~\cite{WitzHJ.19,Witzany:2024ttz,Pio.25} backgrounds. These coordinates provide the natural framework for analysing conservative and transient orbital resonances~\cite{FlHi.12,Isoyama:2013yor,vandeMeent:2013sza,BrGeHiPRL.15,BrGeHi.15,Isoyama:2021jjd,Lynch.al.24}: unlike geodesic motion, spinning trajectories exhibit three-fold resonances involving a spin-precession frequency in addition to the radial and polar components, whose imprints on the gravitational-wave phase may be important for EMRI/IMRI searches with LISA~\cite{Mukherjee:2019jhd,VdM.20}.

Looking further ahead, the covariant Hamiltonian structure developed here provides a starting point for several directions in gravitational self-force theory (see, e.g., the reviews~\cite{Barack:2009ux,BaPo.18}). These include formulating a first law of binary mechanics for spinning extended bodies~\cite{Bl.al.13,An.al.20,RaLe.22,BlaFlaSpin.23}, deriving exact flux-balance laws for the non-isometric constants, i.e., the R\"udiger and generalized Carter constants~\cite{Grant:2024ivt,Skoupy:2024jsi,Mathews:2025nyb,Drummond:2026haw,Skoupy:2026ewu,Cui:2026qsk}, and ultimately integrating these tools with waveform generation frameworks such as FastEMRIWaveforms~\cite{Chua:2020stf,Katz:2021yft,Speri:2023jte,Chapman-Bird:2025xtd} and WaSABI~\cite{Honet:2025lmk}.

Finally, the linear-in-spin integrability established here serves as the baseline for exploring higher-order multipolar effects. In a companion paper (Paper III) \cite{Ra.Iso.Dru.IntegO2.26}, we extend this Hamiltonian framework to the quadratic-in-spin regime. Restricting the background to Einstein spacetimes, we show that the persistence of integrability depends on the secondary's internal structure. While integrability is maintained for a black-hole secondary characterized by a spin-induced quadrupole ($\kappa = 1$), the invariants generated by the KY tensor generally break down for generic compact objects ($\kappa \neq 1$). This breakdown explicitly demonstrates how spin-curvature couplings at the quadrupole order can induce chaos in extended-body dynamics.

\begin{acknowledgments}
The present work has been greatly informed by continuous interactions with numerous colleagues at all stages throughout the past years.
We thank K.~Van Aelst, A.~Albouy, F.~Blanco, A.~Cardenas-Avendano, C.~Dittmer, S.~Dolan, A.~Druart, L.~Drummond, J.~Fejoz, G.~Lukes-Gerakopoulos, R.~Gonzo, \'E.~Gourgoulhon, A.~Grant, A.~Harte, S.~Hughes, A.~Le Tiec, J.~Mathews, G.~A.~Piovano, X.~M.~Porter, A.~Pound, M.~Rahman, V.~ Skoup\'{y}, A.~N.~Seenivasan, M.~Shahzadi, S.~Tanay and V.~Witzany for useful comments and feedback. 
P.R. is indebted to D.~Brown for his guidance through the Dirac-Bergmann algorithm at earlier stages of this work.
S.I. also thank R.~Fujita, T.~Kakei, H.~Nakano, N.~Sago and T.~Tanaka for interesting discussions on the TD spin and the Marck formulation of a Hamiltonian system, and A.~J.~K. Chua for his continuous encouragement.  
Finally, we gratefully acknowledge the Institute for Mathematical Sciences at the National University of Singapore for hosting the workshop `Mathematical Methods for the General Relativistic Two-body Problem' (August 2025), during which the final part of this work was completed, and for their warm hospitality.
P.R. acknowledges funding from the France 2030 program, as part of the 3NC project. S.I. is supported by the Ministry of Education, Singapore, under the Academic Research Fund Tier 1 A-8001492-00-00 (FY2023).
\end{acknowledgments}

\bibliography{ListeRef.bib,ListeRef_Sis.bib}

\appendix

\section{Conventions and notations}
\label{app:conventions}

Throughout the paper, we use geometric units in which $G=c=1$ with the metric signature $(-,+,+,+)$. Abstract indices are $a,b,c,\dots$; coordinate components use Greek indices $\alpha,\beta,\dots$; and orthonormal tetrad indices are $A,B,\dots$. All vary between $0$ and $3$. 
Symmetrization and anti-symmetrization are $T_{(ab)} := \tfrac{1}{2}\,(T_{ab} + T_{ba})$ and $T_{[ab]} := \tfrac{1}{2}\,(T_{ab} - T_{ba})$, respectively. Our curvature conventions follow Wald \cite{Wald}:
$\nabla_{a}\nabla_b \omega_c = R_{abc}^{\phantom{abc}d}\omega_d$ 
for any 1-form $\omega_a$.
We use the shorthand ``Killing-Yano spacetime'' when referring to a spacetime endowed with a (rank-4) Killing-Yano tensor (such as defined in \eqref{deff}--\eqref{KYrank4}). Calligraphic indices $\mcA,\mathcal{B},\mathcal{C},\ldots$ are used to denote a fixed index (not summed over). An arrow is sometimes used to denote Euclidean 3-vectors, using 
interchangeably $\vec{v}=(v^1,v^2,v^3)=v^I$, as well as usual notations $\cdot$ for the scalar product and $\times$ for the cross product.
In the phase space, pairs of canonical coordinates are always of the form $(q,\pi_q)$, with $\pi_q$ being the conjugated momentum of some degree of freedom $q$. Poisson brackets are denoted with the brackets $\{\cdot,\cdot\}$. 
A table for the frequently used symbols, description and references is given in Table.~\ref{Table}.

\begin{table}[!ht]
    \caption{List of frequently used symbols.}
    \vspace{0.2cm}
	\begin{tabular}{ccc}
		\toprule
        \textbf{Notation}         & \textbf{Description}                  & \textbf{Relevant reference}  \\
        \midrule  
		\textbf{Spacetime}      &                                     &     \\
		$\scE$                & Lorentzian manifold                   &           \\
        $x^\alpha$              & spacetime coordinates                 &          \\
        $(\partial_\alpha)^a$     & natural vector basis                       &           \\
        $(\ud x^\alpha)_a$     & natural 1-form basis                       &           \\
		$g_{ab}$                & metric tensor on $\scE$                     &           \\
		$R_{abcd}$              & Riemann curvature tensor               &           \\
        $\nabla_a$              & covariant derivative                    &           \\
		$\varepsilon_{abcd}$    & Levi-Civita tensor                  &         \\
        $(e_A)^a$               & orthonormal tetrad                        &  \eqref{sec:coordPoisson}         \\
        $\omega_{aBC}$          & connection 1-forms                    &  \eqref{sec:coordPoisson}         \\
        \midrule     
  
		\textbf{Symmetries}         &                                      &         \\
		$k^a$                   & generic Killing vector                       &  \eqref{KillKost}         \\
        $f_{ab}$                & Killing-Yano (KY) tensor                    &  \eqref{deff}  \\
        $K^{ab}$                & Killing-St\"ackel (KS) tensor                &  \eqref{KSdef}         \\
        $\xi^a,\eta^a$          & KY-related Killing vectors                       &  \eqref{defxi}, \eqref{defeta}          \\
        $\Xi,\mathfrak{X}$      & Killing vector invariants                  &  \eqref{defHXifX}         \\
		$K,Q$               & R\"udiger and Carter constant                       &  \eqref{defKQ}         \\
		\midrule     
  
		\textbf{Particle}         &                                      &         \\
        $\scL$                  & particle's worldline                 &           \\
        $(p_a,S^{ab})$               & linear and angular momenta              &   \eqref{IspSJ}      \\
		$\mu,S_\circ,S_\star$         & dynamical mass, spin norms       &  \eqref{eq:def-norms}        \\
		$\tau,u^a$                & proper time, four-velocity           &  \eqref{momvelTD}         \\
        $S^a_\tTD,C^a$     & spin and mass dipole wrt $p^a$     &  \eqref{zip!} \, $[f^a=\hat{p}^a]$         \\
        $S^a,D^a$               & spin and mass dipole wrt $(e_0)^a$        &  \eqref{zip!} \, $[f^a=(e_0)^a]$         \\
        \midrule
        
		\textbf{Phase space}           &                                &         \\
		$\mcM$                    & 14D phase-space manifold                &  \eqref{phasespaceM}          \\
        $\mcN$                  & 12D symplectic leaves                 &  \ref{sec:sympleaves}         \\
        $\mcP$                  & 10D physical phase space               &  \eqref{defP}         \\
        $(\pi_\alpha,S^{AB})$ & alternative coordinates & \eqref{pStopiS} \\ 
        $(S^I,D^I)$             & alternative coordinates for SO(1,3)       & \eqref{defSD} \\ 
        $\mathcal{C}_\circ,\mathcal{C}_\star$ & Casimir invariants       &  \eqref{Casimirnew}         \\
        $H,H_\mcP$                     & Hamiltonian on $\mcM$ (or $\mcN$), on $\mcP$                          &  \eqref{Hlinspin}, \eqref{HbeginSD}         \\
        $\Lambda,\{\,,\}$                 & Poisson structure, brackets on $\mcM$ or $\mcN$                    &  \eqref{PBs}, \eqref{PBsSD}         \\
        $\Lambda_\mcP,\{\,,\}^\mcP$         & Poisson structure, brackets on $\mcP$                    &  \eqref{Diracall}         \\
        ${C}^A,{C}^B$     & Scalar constraints from TD SSC                   &  \eqref{SSCappplied}         \\
        \bottomrule
	\end{tabular}
    \label{Table}
\end{table}

\section{Levi-Civita Contractions and Hodge Duality} \label{app:Hodge}

\subsection{Contractions of Levi-Civita tensors}

From \cite{Wald} (cf. Eq.~(B.2.13)), the following contractions of the Levi-Civita tensor hold:  
\begin{subequations}
	 \begin{align}
    	 \varepsilon^{abcd}\varepsilon_{abcd} &= -24 ,  \label{eps0}\\
    	 \varepsilon^{abch}\varepsilon_{abcd} &= - 6 \delta^h_d , \label{eps1}\\
    	 \varepsilon^{abgh}\varepsilon_{abcd} &= - 4 \delta^{[g}_c \delta^{h]}_d , \label{eps2}\\
    	 \varepsilon^{afgh}\varepsilon_{abcd} &= - 6 \delta^{[f}_b \delta^g_c \delta^{h]}_d \label{eps3} , \\
    	 \varepsilon^{efgh}\varepsilon_{abcd} &= -24 \delta^{[e}_a \delta^f_b \delta^g_c \delta^{h]}_d .
    \end{align}
\end{subequations}

\subsection{Hodge duality for 2-forms and 3-forms}

Let $A_{ab}$ be an anti-symmetric tensor. We define its Hodge dual by
\begin{equation} \label{defHodge}
    A_{ab}^\star := \frac{1}{2} \varepsilon_{ab}^{\ph\ph cd}A_{cd}.
\end{equation}
Because of the total anti-symmetry of $\varepsilon_{abcd}$, we could have multiplied the tensor $A_{cd}$ in \eqref{defHodge} over any other pair of indices of the Levi-Civita tensor, resulting, at worse, by a sign difference. 
Applying the Hodge dual one more time to \eqref{defHodge} and using equation \eqref{eps2}, we obtain
\begin{equation}
    A_{ab}^{\star\star} = - A_{ab}.
\end{equation}

Consider now a 3-form, i.e., a totally anti-symmetric tensor $B_{abc}$, and contract it with the Levi-Civita tensor over its three indices to construct a vector $V^a$. Then, contracting with the Levi-Civita tensor once more and using \eqref{eps1}, we find that 
\begin{equation} \label{Hodgedual13}
    V^a = \varepsilon^{abcd} B_{bcd} \quad \Longleftrightarrow \quad B_{abc} = \frac{1}{6} \varepsilon_{abcd} V^d.
\end{equation}
We say that $V^a$ and $B_{abc}$ are Hodge duals to one another, in the sense of the Hodge star. 
Notice that by construction, $B_{abc}V^a=0$. 

\subsection{Hodge decomposition of a 2-form}

Given any 2-form $F_{ab}$ and a unit vector $t^a$ ($|t_at^a|=1$), it is always possible to put $F_{ab}$ in correspondence with a pair of vectors $E^a$ and $B^a$ that are orthogonal to $t^a$, as follows (see Sec.~(14.5) in \cite{GouSR} for details)
\begin{equation}\label{HodgeDecomGen}
	F^{ab} = \varepsilon^{abcd} t_c B_d + 2 E^{[a} t^{b]} 
    \quad \Longleftrightarrow \quad
	\begin{cases}
		\, B_b = t^a F^\star_{ab}, \\
		E_b = t^a F_{ab}.
	\end{cases}
\end{equation}

Although the notations $(F^{ab},t_{a})$ here denote arbitrary tensors, they are chosen to reflect the well-known electric-magnetic decomposition of the Faraday tensor $F^{ab}$ into the electric $E^a$ and magnetic $B^a$ vectors measured in a frame of timelike direction $t^a$. The decomposition \eqref{HodgeDecomGen} is called a Hodge decomposition. They differ by the choice of vector $t^a$ along which the decomposition is made. In our work, we apply it to the spin tensor $S^{ab}$ in the direction of the four-momentum, resulting in equation \eqref{zip!}. 

\section{Hamiltonian systems and generalizations} 
\label{app:hamsystems}

In this appendix, we provide a summary of several notions related to Hamiltonian systems. To be clear, by \textit{Hamiltonian system}, we mean a general framework in which a scalar function $H$ defined on some phase space manifold $\mcM$ gives rise, through some geometrical structure $\Lambda$, to a set of first order ODEs. Our aim is to be concise and self-contained, and to give an explicit definition of every notion that is, usually, defined implicitly. There exists many different formulations of Hamiltonian systems, with different levels of abstraction and generality. These differences mainly have to do with the dimensionality of the phase space and the geometric structure thereon. We shall cover the ones necessary for our purposes. 

Even though all these notions can be given a coordinate-free exposition (with tensor algebra and differential geometry on the phase space manifold), we will instead choose to always work with coordinates since 1) in applications one will practically never use the coordinate-free application and 2) we do not want to mix things up with the covariant formulation of objects in GR. We refer to the recent introductory exposition \cite{Derigl.22} which is at once detailed and pedagogical. Otherwise, classical textbooks such as \cite{Arn} and \cite{Vaisman.12} are recommended.

\subsection{Symplectic systems (non-degenerate)} \label{sec:symp}

The main distinction between different notions of Hamiltonian systems has to do with the geometrical structure on the phase space, which generates a vector field (trajectories) from the Hamiltonian function. This structure is a closed bilinear form, which, when non-degenerate, is called ``symplectic''. Therefore, in what follows, symplectic is a place-holder for ``non-degenerate''. The degenerate case will be dealt with later in Sec.~\ref{sec:Poi}.

\subsubsection{Canonical}

Perhaps without knowing it, one is always introduced to \textit{symplectic} systems during one's first course on analytical mechanics. There, one learns that a $2n$-dimensional Hamiltonian system is a set of ordinary differential equations (ODEs) for some unknowns quantities $(q,p)=(q^i,p_i)_{i\in\{1,\ldots,n\}}$. There are always an even number of such unknowns (here $2n$ for some $n\in\NN$), and the ODEs that they satisfy are called \textit{Hamilton's equations}:
\begin{equation} \label{Hameq}
\frac{\ud q^i}{\ud \lambda} = \frac{\partial H}{\partial p_i} \quad \text{and} \quad \frac{\ud p_i}{\ud \lambda} = -\frac{\partial H}{\partial q^i} \,,
\end{equation}
where $H:\RR^{2n}\rightarrow \RR$ is a scalar function of the $(q,p)$ called the \textit{Hamiltonian}, and $\lambda$ is the \textit{parameter associated to $H$}, usually (but not necessarily) related to some kind of ``time''.\footnote{We only cover the case where the Hamiltonian $H$ is \textit{autonomous}, i.e., that $H$ does not depend explicitly on the time parameter $\lambda$. We will always work with autonomous systems in this series of work.} What makes this formulation \textit{canonical} is that the $2n$ variables $(q_i,p_i)$ always come in $n$ pairs, referred to as \textit{conjugated pairs}. These variables span a $2n$-dimensional \textit{phase space $\mathcal{P}\subset\RR^{2n}$}, on which each point corresponds to a (possible) configuration of the underlying physical system. On top of this classical picture, one will occasionally be introduced to the notion of \textit{Poisson bracket}. This is defined as an operation between two smooth functions $F: \mathcal{P} \rightarrow \RR$ and $G: \mathcal{P} \rightarrow \RR$, defined by 
\begin{equation} \label{PBCano}
\{F,G\}= \sum_{j=1}^{n} \, \left( \, \frac{\partial F}{\partial q_j}  \frac{\partial G}{\partial p^j} - \frac{\partial G}{\partial q_j}  \frac{\partial F}{\partial p^j} \,  \right) \,.
\end{equation}
The Poisson bracket is useful because it can be used to re-write the evolution of any function $F:\mcP\rightarrow\RR$ along a solution to Hamilton's equations \eqref{Hameq} as follows. Using the chain rule, as well as Eqs.~\eqref{Hameq} and \eqref{PBCano}, one readily finds
\begin{equation} \label{FH}
\frac{\ud F}{\ud \lambda} 
= \{F,H\} \,.
\end{equation}
Therefore, the evolution of any function $F:\mcP\rightarrow\RR$ along a phase space trajectory is given by its Poisson bracket with $H$. From the definition \eqref{PBCano}, the Poisson brackets of two coordinates, say $q^i,p_j$, are also easy to find. One obtains the well-known formula
\begin{equation} \label{canocoord}
\{q^i,p_j\}=\delta^i_j \,, 
\end{equation}
where $\delta^i_j$ is the Kronecker symbol. Equation \eqref{canocoord} is called the \textit{canonical Poisson brackets}, and is a characterization of canonical coordinates. In other words, Eq.~\eqref{canocoord} is equivalent to the word ``canonical'' in the phrase ``canonical Hamiltonian system'', or to the word ``conjugated'' in ``conjugated pairs of variables'', all these notions being tautological. There are infinitely many distinct systems of canonical coordinates, all linked to one another through \textit{canonical transformations}. The reason canonical variables on $\mcP$ are so useful in Hamiltonian mechanics is because 1) the general form of Hamilton's equation \eqref{Hameq}, 2) the definition \eqref{FH} of the Poisson bracket of two arbitrary functions, and 3) the fundamental brackets between canonical coordinates \eqref{canocoord}, are all invariant under canonical transformations. In other words, if $(q,p)\in\RR^{2n}\mapsto(Q,P)\in\RR^{2n}$ is a canonical transformation, then one can simply substitute $(q,p)$ by $(Q,P)$ in Eqs.~\eqref{Hameq}, \eqref{FH} and \eqref{canocoord}, where it is understood that the Hamiltonian $H$ and the functions $F,G$ are expressed in terms of $(Q,P)$.

\subsubsection{Non-canonical}

In spite of their usefulness, a canonical formulation is but a very special way of doing Hamiltonian mechanics. Sometimes, as is the case in the present article, such a formulation is not available in the first place. What happens if we use some coordinates that do not satisfy the canonical brackets \eqref{canocoord}? This, in essence, is answered by the broader field of \textit{symplectic geometry}: a more abstract generalization of Hamiltonian mechanics, of which we give a simplified overview below.

Let us start from some canonical coordinates $(q,p)\in\RR^{2n}$ covering a phase space $\mcP$, and a non-canonical change of coordinates, i.e., a diffeomorphism $\Phi:(q,p)\in\RR^{2n}\mapsto y\in\RR^{2n}$. Note that, since the new coordinates $y=(y^1,\ldots,y^{2n})$ are not canonical, there is no reason to identify them as pairs, like the $(q,p)$. Let us pick two, say $(y^i,y^j)$. Seeing these as two functions of the $(q,p)$, we can compute their Poisson bracket using the canonical formula Eq.~\eqref{PBCano}, and obtain the result in terms of $(q,p)$. According to the characterization \eqref{canocoord} of canonical coordinates, we expect $\{y^i,y^j\}\neq \delta^{ij}$ since $y$ are \textit{not} canonical. Nevertheless, using the inverse map $\Phi^{-1}$, we can express any occurrence of $(q,p)$ in the resulting formula in terms of the $y$. We thus obtain the Poisson bracket $\{y^i,y^j\}$ expressed solely in terms of $y$. Doing this for each pair determines a set of Poisson brackets for the $y$ coordinates, given by $4n^2$ formulae of the form:
\begin{equation} \label{Poissonmat}
\{y^i,y^j\} = \Lambda^{ij}(y) \,,
\end{equation}
where the explicit form of the right-hand side will strongly depend on the chosen coordinates. The $y$-dependent matrix $\Lambda$ whose entries are $\Lambda^{ij}(y)=\{y^i,y^j\}$ is called the \textit{Poisson matrix} for the coordinates $y$. Equation \eqref{Poissonmat} generalizes the canonical formula \eqref{canocoord} to non-canonical coordinates.

The Poisson matrix \eqref{Poissonmat} of the coordinates $y$ on $\mcP$ is necessary and sufficient to do anything one would have done with canonical coordinates. For example, the Poisson bracket of any two functions $F(y),G(y)$ on is simply given by 
\begin{equation} \label{PBnoncano}
\{F,G\} = \sum_{i,j} \, \Lambda^{ij}(y) \,  \frac{\partial F}{\partial y^i} \, \frac{\partial G}{\partial y^j} \,.
\end{equation}
where only the Leibniz rule and \eqref{Poissonmat} was used. Equation \eqref{PBnoncano} generalizes the canonical formula \eqref{PBCano} to non-canonical coordinates $y$. Similarly, the evolution of one coordinate $y^i$ under some Hamiltonian $H(y)$ is obtained directly from \eqref{PBnoncano} with $F=y^i$ and $G=H$, leading to
\begin{equation} \label{tLgh}
\frac{\ud y^i}{\ud \lambda} = \sum_{j}\, \Lambda^{ij}(y) \, \frac{\partial H}{\partial y^j} \,,
\end{equation}
again, generalizing Hamilton's canonical equations \eqref{Hameq} to the non-canonical case. Note that Eqs.~\eqref{PBnoncano} and \eqref{tLgh} can be simply written as the product of gradients of scalar functions and the matrix $\Lambda$.

The non-canonical formulation shows that, given some phase space $\mcM$ and a Hamiltonian $H$, the key quantity that encode all information about the geometry (and thus the dynamics of the system) are the Poisson brackets between some coordinates $y$ on $\mcM$ (non-necessarily canonical). Naturally, one could ask whether, given some manifold $\mcM$ endowed with some coordinates $z$ on it, a set of brackets is actually sufficient in order to define a Hamiltonian system. The answer lies in a yet more general formulation, that of Poisson systems, which, in general, are degenerate. 

\subsection{Degenerate systems (non-symplectic)} \label{sec:Poi}

\subsubsection{Generalities}
 
The Poisson system formulation of a Hamiltonian system relies on the following ingredients. Let it be given (i) a manifold $\mcM$ of dimension $N\in\NN$ endowed with some coordinates $y=(y^1,\ldots,y^N)$, (ii) a scalar function $H(y)$ on $\mcM$ and (iii) a set of Poisson brackets between the $y$. In what we have presented so far, items (i) and (ii) are well-defined, but (iii) is not: the definition \eqref{PBnoncano} of Poisson bracket relied on the assumption that the system was non-degenerate. In the present context, the Poisson bracket is defined as an operation between two functions $F(y),G(y)$ on $\mcM$, defined by
\begin{equation} \label{PBnonsympapp}
\{F,G\} = \sum_{i,j} \,\Lambda^{ij}(y) \,  \frac{\partial F}{\partial y^i} \, \frac{\partial G}{\partial y^j} \,.
\end{equation}
where the coefficients $\Lambda_{ij}(y)$ satisfy the following two properties:\footnote{These two properties are equivalent to the other definitions involving bi-vectors or Poisson brackets \cite{Derigl.22}.}
\begin{subequations} \label{condi}
    \begin{align}
        \text{Skew-symmetry:} \quad & \Lambda^{ij}=-\Lambda^{ji} \,, \\
        \text{Jacobi identity:} \quad & \Lambda^{i\ell} \partial_\ell \Lambda^{jk} + \Lambda^{j\ell} \partial_\ell \Lambda^{ki} + \Lambda^{k\ell} \partial_\ell \Lambda^{ij} = 0 \,, \label{Jaco}
    \end{align}
\end{subequations}
where a sum from $\ell=1$ to $\ell=N$ is implied and $\partial_\ell = \partial/\partial y^\ell$. In other words, defining a Poisson system is a very general process: it is done by the choice of a $N\times N$ antisymmetric matrix whose coefficients $\Lambda^{ij}(y)$ verify the condition \eqref{Jaco}. Again, one refers to $\Lambda(y)$ as the Poisson matrix of the system. Replacing $(F,G)$ in Eq.~\eqref{condi} by two coordinates $(y^i,y^j)$, we find the meaning of its coefficients: 
\begin{equation} \label{Poi}
\Lambda^{ij}(y) = \{y^i,y^j\} \,.
\end{equation}

What then makes a Poisson system \textit{different} from a symplectic system in non-canonical coordinates? This question is natural since, after all, Eqs.~\eqref{PBnonsympapp} and \eqref{Poi} for Poisson systems look very much like Eqs.~\eqref{PBnoncano} and ~\eqref{Poissonmat} for non-canonical systems. 

There is essentially one difference, hidden in the Poisson matrix $\Lambda$. In a symplectic system, $\Lambda$ is necessarily non-singular, i.e., it has maximal rank at every point in phase space. This is because the $y$ coordinates are always a diffeomorphism away from canonical coordinates $(q,p)$, in which $\Lambda=\mathbb{J}_{2n}$. Since the determinant is invariant under changes of coordinates, we have $\text{det}(\Lambda)=\text{det}(\mathbb{J}_{2n})=1\neq0$, and thus $\text{rank}(\Lambda)=2n$. In a Poisson system, however, the Poisson matrix $\Lambda$ can be degenerate. For example, this is automatically the case if $\mcM$ is odd-dimensional, because $\text{rank}(\Lambda)$ is always even by the skew-symmetry of $\Lambda$. Yet, even when $\mcM$ is even-dimensional, the rank of $\Lambda$ can be non-maximal, i.e., non-equal to $N$. This is the case for the brackets \eqref{PBsSD}, for example, and other related brackets like those presented in \cite{Ra.AJP.26}.

If $\Lambda$ is degenerate, then $\text{rank}(\Lambda)=p$ for some $0<p\leq N$. From the rank theorem, this means that their exists $k=N-p$ independent vectors spanning the null space of $\Lambda$. It can be shown that these vectors are the gradient of some scalar fields on $\mcM$, called \textit{Casimirs}.\footnote{They are also called \emph{Casimir elements}, \emph{quadratic invariants} or \emph{Casimir operators} in the field theory of Lie algebra and quantum mechanics.} If $C(y)$ is such a Casimir, then since its gradient $\nabla C$ belongs to the null space of $\Lambda$, if follows from Eq.~\eqref{PBnonsympapp} that for any $F:\mcM\rightarrow\RR$, $\{F,C\}=0$. In other words, Casimirs are special functions of the $y^i$ whose Poisson brackets with \textit{anything} vanish on $\mcM$. There are exactly $k=n-p$ independent such Casimirs. A remarkable result of Poisson geometry is that $\mcM$ is foliated by submanifolds of constant Casimirs. These submanifolds, of dimension $p$, are called \textit{symplectic leaves} and they are symplectic, i.e., non-degenerate.

\subsubsection{Dirac brackets}
\label{sec:dirac}

The symplectic leaves of the previous section are submanifolds onto which the Poisson structure restricts effortlessly, precisely because they are level sets of Casimirs. In practice, though, one frequently wishes to confine the dynamics to a submanifold $\mcP\subset\mcM$ that is \emph{not} of this kind: a surface cut out by a family of constraints
\begin{equation} \label{constraints}
    C^\alpha(y) = 0\,, \qquad \alpha = 1,\ldots,k\,,
\end{equation}
whose Poisson brackets with the other phase-space functions do not, in general, vanish. This is exactly the situation met in the main text, where $\mcM$ is the $14$-dimensional phase space of the dipolar system and the $C^\alpha := p_\beta S^{\beta\alpha}$ are the four functions whose vanishing enforces the TD SSC [cf. Sec.~\ref{sec:consinv}].

Restricting the ambient bracket \eqref{PBnonsympapp} to $\mcP$ by brute force --- evaluating $\{F,G\}$ and then imposing $C^\alpha=0$ --- does not yield a sensible bracket on $\mcP$, for two intertwined reasons: the outcome depends on how $F$ and $G$ are continued away from the surface, and the flow generated by the ambient bracket need not be tangent to $\mcP$, so that the constraints fail to be preserved along the motion.

The obstruction is encoded in the antisymmetric \emph{constraint matrix}
\begin{equation} \label{constraintmat}
    \Delta^{\alpha\beta}(y) := \{C^\alpha,C^\beta\}\,.
\end{equation}
Two cases must be distinguished. When $\Delta^{\alpha\beta}$ vanishes on $\mcP$, the constraints are called \emph{first-class}: they generate gauge transformations and require a separate (quotient) treatment. When $\Delta^{\alpha\beta}$ is invertible on $\mcP$ --- which forces $k$ to be even --- the constraints are \emph{second-class}, and the relevant construction is the one introduced by Dirac. Writing $\Delta_{\alpha\beta}$ for the inverse matrix, $\Delta_{\alpha\gamma}\Delta^{\gamma\beta}=\delta^\beta_\alpha$, the \emph{Dirac bracket} of two functions $F,G$ on $\mcM$ is defined as
\begin{equation} \label{Diracapp}
    \{F,G\}^{\mathrm D} := \{F,G\} - \{F,C^\alpha\}\,\Delta_{\alpha\beta}\,\{C^\beta,G\}\,,
\end{equation}
with an implicit sum over $\alpha,\beta=1,\ldots,k$, every bracket on the right-hand side being the ambient one \eqref{PBnonsympapp}. The TD SSC constraints \eqref{SSCappplied} are second-class, so this is the only case we require.

Like the ambient bracket, the Dirac bracket \eqref{Diracapp} is skew-symmetric and satisfies the Jacobi identity \eqref{condi}; it is therefore itself a legitimate Poisson bracket on $\mcM$. Its defining virtue appears upon bracketing any function $F$ with a constraint: using \eqref{constraintmat} and the definition of the inverse,
\begin{equation} \label{DiracCas}
    \{F,C^\gamma\}^{\mathrm D}
    = \{F,C^\gamma\} - \{F,C^\alpha\}\,\Delta_{\alpha\beta}\,\Delta^{\beta\gamma}
    = \{F,C^\gamma\} - \{F,C^\alpha\}\,\delta^\gamma_\alpha = 0\,.
\end{equation}
That is, the constraints $C^\alpha$ are \emph{Casimirs of the Dirac bracket}: with respect to $\{\cdot,\cdot\}^{\mathrm D}$, they Poisson-commute with every function on $\mcM$. In the language of the previous section, the Dirac construction deforms the Poisson structure just enough to promote the constraint surface $\mcP$ to one of its symplectic leaves.

This single property \eqref{DiracCas} dissolves both earlier difficulties at once. Since $\{H,C^\alpha\}^{\mathrm D}=0$, the Hamiltonian flow computed with the Dirac bracket is tangent to $\mcP$, so the constraints are dynamically preserved. And since the $C^\alpha$ are Casimirs, the value of $\{F,G\}^{\mathrm D}$ on $\mcP$ is insensitive to the way $F$ and $G$ are extended off the surface. The Dirac bracket thus descends to a well-defined, non-degenerate Poisson bracket on the reduced phase space $\mcP$ --- its intrinsic symplectic structure. It is this restriction that we denote $\{\cdot,\cdot\}^\mcP$ in the main text.

A practical feature of \eqref{Diracapp}, used repeatedly in our calculations, is that the correction term factorizes through the constraint brackets of \emph{each} argument separately. Consequently, if one of the two functions Poisson-commutes with every constraint on $\mcP$, the Dirac bracket collapses to the ambient one,
\begin{equation} \label{Diracsimp}
    \{F,C^\alpha\}\big|_\mcP = 0 \ \ (\forall\,\alpha)
    \quad\Longrightarrow\quad
    \{F,G\}^\mcP = \{F,G\}\big|_\mcP \quad (\forall\,G)\,.
\end{equation}
a fact that considerably shortens the involution analysis of Sec.~\ref{sec:PBcalc}.

\subsection{Integrable systems and action-angle variables}
\label{sec:integ}

According to Eq.~\eqref{FH}, a phase-space function is conserved along every trajectory if and only if it Poisson-commutes with the Hamiltonian; such a function is called a \emph{first integral} (equivalently, a constant or invariant of motion). On a $2n$-dimensional symplectic phase space, the system is said to be \emph{(Liouville) integrable} if it admits $n$ first integrals $(I_1,\ldots,I_n)$ that are (i) functionally independent almost everywhere and (ii) in pairwise involution, $\{I_i,I_j\}=0$. The Hamiltonian is always one of them. When these conditions hold, Liouville's theorem guarantees that the equations of motion can be integrated by quadratures.

The geometric content of the theorem, due to Arnold \cite{Arn}, is sharper. Wherever the common level sets $\{I_i=\mathrm{const}\}$ are compact and connected, they are diffeomorphic to $n$-dimensional tori $\mathbb{T}^n$ --- the \emph{invariant tori} --- and each trajectory winds around its torus without ever leaving it. In a neighbourhood of such tori there exist a privileged set of canonical coordinates $(\theta^i,J_j)$, the \emph{action-angle variables}: the $n$ \emph{actions} $J_i$ are first integrals labelling the torus, defined by the loop integrals
\begin{equation} \label{action}
    J_i = \frac{1}{2\pi}\oint_{\gamma_i} p_j\,\ud q^j
\end{equation}
over the $n$ independent cycles $\gamma_i$ of the torus, while the $n$ \emph{angles} $\theta^i$ are $2\pi$-periodic coordinates around those cycles, conjugate to the actions: $\{\theta^i,J_j\}=\delta^i_j$.

It is essential that action-angle variables form a \emph{complete} canonical set. Their defining virtue is that, in these coordinates, the Hamiltonian depends on the actions alone,
\begin{equation} \label{HofJ}
    H = H(J_1,\ldots,J_n) \,,
\end{equation}
so that every angle is cyclic, i.e., $\partial_{\theta^i}H=0$. Property \eqref{HofJ} can only hold for the full set $(\theta^i,J_i)_{i=1,\ldots n}$: were one to keep only a subset of action-angle pairs alongside other, generic coordinates, those generic coordinates would necessarily reappear in $H$ and spoil \eqref{HofJ}. There is consequently no such thing as a \emph{partial} action-angle system --- on a given torus neighborhood, either all $2n$ coordinates are action-angle variables or none are. 

When \eqref{HofJ} holds, the angles advance linearly,
\begin{equation} \label{omega}
    \frac{\ud\theta^i}{\ud\lambda} = \frac{\partial H}{\partial J_i} =: \Omega^i(J)\,,
    \qquad\Longrightarrow\qquad
    \theta^i(\lambda) = \theta^i_0 + \Omega^i\,\lambda\,,
\end{equation}
where the $\omega^i$, constant on each torus, are the \emph{Hamiltonian frequencies} (also called fundamental frequencies). The motion is then quasi-periodic with respect to the Hamiltonian time $\lambda$, and densely fills the torus, \emph{unless} the frequencies are commensurate: a \emph{resonance} occurs when
\begin{equation} \label{reso}
    k_i\,\Omega^i = 0 \quad\text{for some}\quad k=(k_1,\ldots,k_n)\in\mathbb{Z}^n\setminus\{0\}\,,
\end{equation}
in which case the trajectory degenerates onto a lower-dimensional sub-torus. 

Finally, the very existence of action-angle variables hinges on the compactness of the level sets, that is, on the motion being \emph{bounded}. Along unbounded directions --- escaping or plunging trajectories, or the non-compact $t$ direction of relativistic orbits --- the invariant manifolds are cylinders $\mathbb{T}^k\times\RR^{\,n-k}$ rather than tori, and one retains genuine angle variables only for the librating degrees of freedom while keeping linear coordinates along the non-compact ones. For bound geodesics in Kerr spacetime, this construction can be seen in \cite{Schm.02} and \cite{HinFle.08}, and in a more general context in \cite{Fio.07}.


\begin{table}[t]
  \centering
  \renewcommand{\arraystretch}{1.3}
  \footnotesize
  \begin{tabular}{ccc}
    \hline
    \textbf{Term}          & \textbf{Meaning} & \textbf{Ref.} \\ \hline
    Poisson bracket        & bilinear, antisymmetric, Jacobi operation $\{F,G\}=\Lambda^{ij}\partial_iF\,\partial_jG$ & Eq.~\eqref{PBnonsymp} \\
    Poisson matrix         & $\Lambda^{ij}(y)=\{y^i,y^j\}$; encodes the geometric structure & Eq.~\eqref{Poi} \\
    Poisson system         & manifold $+\,H\,+$ Poisson bracket, with $\Lambda$ possibly degenerate & Sec.~\ref{sec:Poi} \\
    Symplectic system      & Poisson system with non-degenerate (full-rank) $\Lambda$ & Sec.~\ref{sec:symp} \\
    Degenerate system      & Poisson system with $\Lambda$ not of full rank ($\det\Lambda=0$) & Sec.~\ref{sec:Poi} \\
    Canonical system       & symplectic system in coordinates where $\Lambda=\mathbb{J}_{2n}$, i.e. $\{q^i,p_j\}=\delta^i_j$ & Eq.~\eqref{canocoord} \\
    Non-canonical system   & symplectic system in coordinates where $\Lambda\neq\mathbb{J}_{2n}$ & Eq.~\eqref{Poissonmat} \\
    Canonical variables    & an even number of coordinates obeying $\{q^i,p_j\}=\delta^i_j$ & Eq.~\eqref{canocoord} \\
    Casimir                & function $C$ with $\{F,C\}=0$ for all $F$; spans the null space of $\Lambda$ & Sec.~\ref{sec:Poi} \\
    Symplectic leaf        & level set of the Casimirs; symplectic; dimension $=\mathrm{rank}\,\Lambda$ & Sec.~\ref{sec:Poi} \\
    Dirac bracket          & induced bracket on a second-class constraint surface, $\{\,\}^{\mathrm D}$ & Eq.~\eqref{Diracapp} \\
    First integral         & phase space function conserved along the flow, $\{I,H\}=0$ & Eq.~\eqref{FH} \\
    Integrable system      & $2n$-dim symplectic system with $n$ independent first integrals in involution & Sec.~\ref{sec:integ} \\
    Invariant torus        & compact connected level set of the $n$ integrals, $\cong\mathbb{T}^n$ & Sec.~\ref{sec:integ} \\
    Cyclic variable        & such that $H$ is independent of its conj. coord. (part of a cano set) & Eq.~\eqref{HofJ} \\
    Action-angle variables & complete canonical set $(\theta^i,J_i)$ with $H=H(J)$; angles cyclic & Eq.~\eqref{HofJ} \\
    Hamiltonian frequency  & $\Omega^i=\partial H/\partial J_i$; rate of advance of the angle $\theta^i$ & Eq.~\eqref{omega} \\
    Resonance              & a phase space (locus of) point(s) where $k_i\Omega^i=0$ for some $k\in(\mathbb{Z}^\star)^n$ & Eq.~\eqref{reso} \\
    \hline
  \end{tabular}
  \caption{Glossary of the notions used throughout this appendix, with the equation or
  section where each is defined.}
  \label{tab:glossary}
\end{table}

\section{Identities for the quasi-symplectic coordinates $(x^\alpha,\pi_\alpha,S^I,D^I)$} 
\label{app:xpstospis}

\subsection{Derivation of the Poisson brackets for $(x^\alpha,\pi_\alpha,S^I,D^I)$}

In this appendix, we prove the claim made in Sec.~\ref{sec:coordPoisson} that the brackets \eqref{PBs} for the coordinates $(x^\alpha,p_\alpha,S^{\alpha\beta})$ turn into the brackets \eqref{PBsSD} for the coordinates $(x^\alpha,\pi_\alpha,S^{AB})$, under the change of coordinates defined by
\begin{subequations} \label{defapp}
    \begin{align}
    p_\alpha &= \pi_\alpha +\frac{1}{2} \omega_{\alpha BC} S^{BC}, \label{ptopiapp}\\
    S^{\alpha\beta} &= S^{AB} (e_A)^\alpha (e_B)^\beta,\label{StoSapp}
    \end{align}
\end{subequations}
where the connection coefficients $\omega_{\alpha B C}$ and the tetrad components $(e_A)^\alpha)$ are just functions of $x^\alpha$ by construction. 

First, we note that $S^{AB}$ depends only on $x^\alpha,S^{\alpha\beta}$ via Eq.~\eqref{StoSapp}. From this result alone we get $\{x^\alpha,S^{AB}\}=0$ using Eq.~\eqref{StoSapp}, $\{x^\alpha,\pi_\beta\}=\delta^\alpha_\beta$ from Eq.~\eqref{ptopiapp}, and, using the Leibniz rule, $\{S^{AB}, S^{CD}\}$ is simply the bracket obtained by projecting \eqref{spinsss} onto the orthonormal tetrad, thus giving Minkowski symbols $\eta^{AB}$ in place of the metric coefficients $g^{\alpha\beta}$. 

Next, the bracket $\{\pi_\alpha,S^{AB}\}$ is shown to vanish using the definitions \eqref{defapp}, the Leibniz rule, and the definition of $\omega_{\alpha}$ as covariant derivatives of the tetrad components. 

Lastly, to find the bracket $\{\pi_\alpha,\pi_\beta\}$, we isolate $\pi_\alpha$ from Eq.~\eqref{ptopiapp} and insert the resulting expression in the brackets. The Leibniz rule and the brackets already established give
\begin{equation} \label{Wald3}
\{\pi_\alpha,\pi_\beta\} = -\tfrac{1}{2}R_{\alpha\beta AB}S^{AB}+\tfrac{1}{2}S^{AB}(\nabla_\alpha\omega_{\beta AB}-\nabla_\beta \omega_{\alpha AB})+ \omega_{\alpha DB} \omega_{\beta CA} S^{AB} \eta^{CD} \,,
\end{equation}
where the algebraic antisymmetry $\omega_{\alpha BC}=-\omega_{\alpha CB}$ was also used. Factoring $S^{AB}$ in the right-hand side of Eq.~\eqref{Wald3} then reveals a vanishing expression obtained from a classical result related to computing curvature with connection coefficients, namely Eq.~(3.4.20) of Wald's textbook \cite{Wald}. This concludes the proof of the brackets \eqref{PBsSD}.

\subsection{Jacobi identity}\label{app:jaco}

Showing that the Poisson brackets \eqref{PBs} satisfy the Jacobi identity is straightforward but lengthy. However, the calculation can be shortened as follows. First, we use the coordinates $(y^1,\ldots,y^14)=(x^\alpha,\pi_\alpha,S^I,D^I)$, which are in a 1-to-1 correspondence with $(x^\alpha,\pi_\alpha,S^{\alpha\beta})$, and, second, the equivalence between the Jacobi identity. Whether the Jacobi identity holds is, indeed, invariant under isomorphisms. Second, we can use the Jacobi identity in the equivalent form $\sum_{\text{cycle}}\{y^i,\{y^j,y^k\}\}=0$, cf. \cite{Derigl.22}. Since any bracket involving one $x^\alpha$ or one $\pi_\alpha$ is either 0 or 1, the Jacobi identity will be verified for any triplet that includes at least one instance of $x^\alpha$ or $\pi_\alpha$. For the other triplets, only four combinations are possible, thanks to the anti-symmetry of Poisson brackets. For example,
\begin{align}
     \{ S^I\!, \{ S^J\!, S^K \} \} +\{ S^K\!, \{ S^I\!, S^J \} \}  + \{ S^J\!, \{ S^K\!, S^I
  \} \} \cr
  \qquad = (\varepsilon^{I J L} \varepsilon^{K L M} \!+ \varepsilon^{J K L}
  \varepsilon^{I L M} \!+ \varepsilon^{K I L} \varepsilon^{J L M}) S^M \,, 
\end{align}
where we used the brackets $\eqref{PBsSD}$ and the fact that $\varepsilon^{IJK}\in\{-1,0,1\}$. Using the identity $\varepsilon^{I J K} \varepsilon^{L M K} = \delta^{I L} \delta^{J M} -\delta^{I M} \delta^{J L}$, the right-hand side of this equation is readily seen to vanish. All three other possibilities (schematically $\{S,\{S,D\}\}$, $\{S,\{D,D\}\}$ and $\{D,\{D,D\}\}$), follow exactly the same pattern and vanish for the same reason.

\subsection{Link between the momenta $\pi_\alpha$ and Killing vector invariants $\mathcal{F}_k$ } \label{app:KillEqual}

For a spacetime admitting a Killing field $k^a=(\partial_\mathfrak{t})^a$ and a set of adapted coordinates $(\mathfrak{t},x^i)$, one has the equality $\mathcal{F}_k=\pi_\mathfrak{t}$, where $\mathcal{F}_k$ is the invariant associated to a Killing field \eqref{defFk} and $\pi_\mathfrak{t}$ is the new momentum coordinate \eqref{ptopi}. 
Given those definitions, we want to prove that
\begin{equation}
  \mathcal{F}_k := p_a (\partial_\mathfrak{t})^a + \frac{1}{2} S^{a b} \nabla_a (\partial_\mathfrak{t})_b = p_\mathfrak{t} -
  \frac{1}{2} \omega_{\mathfrak{t} A B} S^{A B}.
\end{equation}

The contributions from $p_a$ are clearly equal. The non-trivial part lies in the term involving the spin tensor $S^{ab}$. Since $\omega_{\mathfrak{t}AB}$ is antisymmetric in the indices $A, B$, and $\nabla_a k_b=(\nabla\partial_\mathfrak{t})_{ab}$ is antisymmetric in $a,b$ as well, it suffices to establish
\begin{equation} \label{nabla_omega}
  \nabla_A (\partial_\mathfrak{t})_B = - \omega_{\mathfrak{t} A B},
\end{equation}
where the left-hand side refers to the tetrad components of the 2-form $(\nabla \partial_\mathfrak{t})_{ab}$, and the right-hand side involves the Ricci rotation coefficients $\omega_{\alpha A B} \equiv (e_A)^{\beta} \nabla_{\alpha} (e_B)_{\beta}$. By definition,
\begin{align}
  \omega_{\alpha A B} &= (e_A)^{\beta} \partial_{\alpha} (e_B)_{\beta} - (e_A)^{\beta} \Gamma^{\gamma}_{\alpha \beta} (e_B)_{\gamma}, \\
  \Gamma^{\gamma}_{\alpha \beta} &= \frac{1}{2} g^{\gamma \delta} \left( \partial_{\alpha} g_{\delta \beta} + 2 \partial_{[\beta} g_{\delta] \alpha} \right).
\end{align}
Taking the $\alpha=\mathfrak{t}$ component of these equations and using the fact that $\partial_\mathfrak{t} (e_B)_{\beta} = 0$, as well as $(\partial_\mathfrak{t})_{\beta} = g_{\mathfrak{t} \beta}$, we obtain:
\begin{align} \label{idsKt}
  \omega_{\mathfrak{t} A B} &= (e_A)^{\beta} \nabla^{\gamma} k_{\beta} (e_B)_{\gamma} = - (\nabla k)_{A B}, \\
  \Gamma^{\gamma}_{\mathfrak{t} \beta} &= - g^{\gamma \delta} \partial_{[\delta} k_{\beta]} = - g^{\gamma \delta} \nabla_{\delta} k_{\beta} = - \nabla^{\gamma} k_{\beta},
\end{align}
where the Killing equation was used in the last step to write $\nabla_{\delta} k_{\beta} = 2 \partial_{[\delta} k_{\beta]}$. Equations \eqref{idsKt} readily imply \eqref{nabla_omega}, and \eqref{equalpit} thus follows.

\section{Coordinate expressions in the Kerr-Newman-de Sitter spacetime} 
\label{app:MMA}

Our results apply to any spacetime endowed with a Killing-Yano (KY) tensor. As mentioned in the present paper and reviewed in Paper III \cite{Ra.Iso.Dru.IntegO2.26}, the curvature of any such spacetime is restricted: out of the possible 20 independent components of Riemann tensor, only 4 can be non-zero when a non-degenerate KY exists. These four independent components can be encoded into the complex Weyl scalar $\Psi_2$, the real Ricci scalar $R=R_{ab}g^{ab}$ and another real component of the Ricci tensor $\phi_2$ (see Sec. III.D. in Paper III \cite{Ra.Iso.Dru.IntegO2.26} for details). 

One prototypical example of such a spacetime, endowed with a KY tensor, is the Kerr-Newman-de Sitter (KNdS) spacetime. It is parametrized by four physical parameters: mass $M$, spin $a$, electric charge $q$ and cosmological constant $\Lambda$. Each physical parameter is related to one of the aforementioned curvature components, in the sense that 
\begin{equation} \label{justif}
    \text{Re}[\Psi_2] \propto M,\quad \text{Im}[\Psi_2] \propto a, \quad R \propto \Lambda \quand \phi_2 \propto q,
\end{equation}

In this appendix we provide expressions for the components of several geometric tensors involved in our analysis, in the natural basis associated to Boyer-Lindquist coordinates $(t,r,\theta,\phi)$ covering this 4-parameter family of KNdS spacetimes. They, and other expressions, are used in the Mathematica companion notebook \cite{MMA_IntegO1}, wherein explicit coordinate calculations of all this paper's results are explicitly checked.

\subsection{Symbols and metric}

We start with the following useful symbols used throughout this appendix and the Mathematica companion notebook: 
\begin{subequations}
    \begin{align}
    \varpi&:= a^2+r^2, \quad
    \Sigma := r^2+a^2\cos^2\theta, \quad 
    \rho := r + \ui a \cos \theta, \\
    \lambda &:= \Lambda/3, \quad 
    \chi := 1 + \lambda a^2 , \quad
    \sigma := \sin\!\theta, \\
    \Delta_r & := \varpi \left(1-\lambda  r^2\right)-2 M r + q^2 ,\quad
    \Delta_\theta := 1+ \lambda a^2 \cos^2\theta,
    \end{align}
\end{subequations}
all of which are real-valued except for $\rho\in\CC$. With these notations, the metric tensor $g_{ab}$ of the KNdS spacetime has components $g_{\alpha\beta}$ given (in matrix notation) by, e.g.,~\cite{Carter:1968ks,Gibbons:1977mu,wu2004entropy,Swearngin.24} 
\begin{equation} \label{comp_metric}
    g_{\alpha\beta}= \frac{1}{\Sigma\chi^2} \left(
\begin{array}{cccc}
 a^2 \sigma^2 \Delta_{\theta} -\Delta_r & 0 & 0 & -a \sigma^2 \left(\varpi \Delta_{\theta}-\Delta_r\right) \\
 0 & \Sigma^2\chi^2/\Delta_r & 0 & 0 \\
 0 & 0 & \Sigma^2\chi^2/\Delta_{\theta} & 0 \\
 -a \sigma^2 \left(\varpi \Delta_{\theta}-\Delta_r\right) & 0 & 0 & \sigma^2 \left(\varpi^2 \Delta_{\theta}-a^2 \sigma^2 \Delta_r\right) \\
\end{array}
\right).
\end{equation}

\subsection{Killing objects}

The KY tensor $f_{ab}$ has components that are independent of $q$, given by
\begin{equation} \label{comp_Z}
  f_{\alpha \beta} = \chi^{- 1} \left( \begin{array}{cccc}
  0 & 1 & - a r \sigma & 0\\
  - a \cos \theta & 0 & 0 & a^2 \sigma^2 \cos \theta\\
  a r \sigma & 0 & 0 & - \varpi \sigma r\\
  0 & - a^2 \sigma^2 \cos \theta & \varpi \sigma r & 0
\end{array} \right)
\end{equation}

Next, the scalar $\mathfrak{Z}$ built from the KY tensor \eqref{ffstar} involved in our analysis reads
\begin{equation} \label{comp_C}
    \mathfrak{Z} = a r \cos \theta,
\end{equation}
and the two Killing vectors $(\xi^a,\eta^a)$ (cf.\eqref{defxi}-\eqref{defeta}) have components $\xi^{\alpha} = \chi (1, 0, 0, 0)$ and $\eta^{\alpha} = \chi (a^2, 0, 0, a)$, such that, in a KNdS spacetime, 
\begin{equation}
    \xi^a = (\partial_t)^a \quand \eta^{\alpha} = \chi a \bigl( (\partial_\phi)^a + a (\partial_t)^a \bigr).
\end{equation}

\subsection{Curvature scalars}

Though they were not needed here, we complete the list of explicit coordinate expressions by giving the curvature scalars of the KNdS spacetime ($\bar{\rho}$ denotes the complex conjugate of $\rho\in\CC$):
\begin{equation}
  \Psi_2 = - \frac{M}{\bar{\rho}^3}, \quad R=4\Lambda \\ \quand \phi_2 = - \frac{q^2}{2\Sigma^2},
\end{equation}
justifying the relations in \eqref{justif}. The exact definition of these curvature scalars involves a complex null tetrad and goes beyond the scope of the present paper. These tools, however, are detailed in Paper III \cite{Ra.Iso.Dru.IntegO2.26}, and used in the Mathematica Notebook accompanying the present paper \cite{MMA_IntegO1}.

\end{document}